\documentclass[longauth]{aa}  
\usepackage{graphicx}
\usepackage[english]{babel}
\usepackage{placeins}
\usepackage{txfonts}
\usepackage{xspace}
\usepackage{natbib} 
\usepackage{amsmath}
\usepackage{array}
\usepackage{appendix}
\usepackage{mathabx}
\usepackage{comment}
\usepackage{rotating}
\usepackage{placeins}
\usepackage[table,dvipsnames]{xcolor}
\usepackage{hyperref}
\hypersetup{colorlinks=true,citecolor=blue}
\usepackage[capitalise]{cleveref}

\newcommand{\fastwind}{\textsc{Fastwind}\xspace} 
\newcommand{\kiwiGA}{\textsc{Kiwi}-GA\xspace}

\newcommand{\Msun}{$M_\odot$\xspace}
\newcommand{\teff}{$T_{\rm eff}$\xspace}
\newcommand{\logg}{$\log g$\xspace}
\newcommand{\yhe}{$y_{\rm He}$\xspace}

\newcommand{\vsini}{$\varv\sin i$\xspace}
\newcommand{\vmicro}{$\varv_{\rm micro}$\xspace}

\newcommand{\vwindturb}{$\varv_{\rm windturb}$\xspace}
\newcommand{\vclonset}{$\varv_{\rm cl, start}$\xspace}
\newcommand{\fcl}{$f_{\rm cl}$\xspace}
\newcommand{\fic}{$f_{\rm ic}$\xspace}
\newcommand{\fvel}{$f_{\rm vel}$\xspace}

\newcommand{\vinf}{$\varv_{\infty}$\xspace}

\newcommand{\mdot}{$\dot{M}$\xspace}

\newcommand{\logll}{$\log L / {L}_\odot$\xspace}

\newcommand{\hei}{He~\textsc{i}\xspace}
\newcommand{\heii}{He~\textsc{ii}\xspace}
\newcommand{\niii}{N~\textsc{iii}\xspace}
\newcommand{\ciii}{C~\textsc{iii}\xspace}

\newcommand{\niv}{N~\textsc{iv}\xspace}

\newcommand{\halpha}{H$\alpha$\xspace}

\newcommand{\heiiline}{He\textsc{~ii}~$\lambda$4686\xspace}

\newcommand{\cuvivline}{C\textsc{~iv}~$\lambda$1169\xspace}
\newcommand{\cuviiiline}{C\textsc{~iii}~$\lambda$1176\xspace}
\newcommand{\nvuvline}{N\textsc{~v}~$\lambda\lambda$1238-1242\xspace}
\newcommand{\siivline}{Si\textsc{~iv}~$\lambda\lambda$1394-1402\xspace}
\newcommand{\oivline}{O\textsc{~iv}~$\lambda$1340\xspace}
\newcommand{\ovline}{O\textsc{~v}~$\lambda$1371\xspace}
\newcommand{\oviline}{O\textsc{~vi}~$\lambda\lambda$1031-1038\xspace}
\newcommand{\CIVline}{C\textsc{~iv}~$\lambda\lambda$1548-1551\xspace}
\newcommand{\heiiuvline}{He\textsc{~ii}~$\lambda$1640\xspace}
\newcommand{\nivuvline}{N\textsc{~iv}~$\lambda$1718\xspace}
\newcommand{\nvopt}{N\textsc{~v}~$\lambda\lambda$4604-4620\xspace}
\newcommand{\pvline}{P\textsc{~v}~$\lambda\lambda$1118-1128\xspace}
\newcommand{\niiiuvline}{N\textsc{~iii}~$\lambda$1748-51-52\xspace}
\newcommand{\nivoptline}{N\textsc{~iv}~$\lambda$4058\xspace}
\newcommand{\ciiivoptline}{C\textsc{~iii}~$\lambda$5696\xspace}
\newcommand{\civvoptline}{C\textsc{~iv}~$\lambda\lambda$5801-5812\xspace}

\makeatletter
\newcommand\footnoteref[1]{\protected@xdef\@thefnmark{\ref{#1}}\@footnotemark}
\makeatother

\begin{document} 

    \title{X-Shooting ULLYSES: massive stars at low metallicity}

    \subtitle{XII. The clumped winds of O-type (super)giants in the Large Magellanic Cloud}

    \titlerunning{The clumped winds of O-type (super)giants in the Large Magellanic Cloud}

    \author{Sarah A. Brands \inst{\ref{inst:API}}
           \and 
           Frank Backs \inst{\ref{inst:API}}$^,$\inst{\ref{inst:leuven}}
           \and 
           Alex de Koter  \inst{\ref{inst:API}}$^,$\inst{\ref{inst:leuven}}
           \and 
           Joachim Puls  \inst{\ref{inst:LMU}}
           \and 
           Paul A. Crowther   \inst{\ref{inst:shef1}}
           \and           
           Hugues Sana \inst{\ref{inst:leuven}}
           \and 
           Frank Tramper \inst{\ref{inst:cab}}
           \and 
            Lex Kaper   \inst{\ref{inst:API}}
           \and    
           Jon O. Sundqvist \inst{\ref{inst:leuven}}
           \and 
           Joachim M. Bestenlehner \inst{\ref{inst:shef1}}$^,$\inst{\ref{inst:shef2}}
           \and 
           Florian A. Driessen  \inst{\ref{inst:API}}$^,$\inst{\ref{inst:leuven}}    
           \and
           Christiana Erba \inst{\ref{inst:STScI}}
           \and 
           Calum Hawcroft \inst{\ref{inst:STScI}}
           \and 
           Artemio Herrero \inst{\ref{inst:tenerife}}
           \and
           D. John Hillier \inst{\ref{inst:PITT}}
           \and
           Richard Ignace \inst{\ref{inst:ETSU}}
           \and 
           Roel R. Lefever \inst{\ref{inst:ari}}
           \and 
           N. Dylan Kee \inst{\ref{inst:goddard}}
           \and 
           Brankica Kub\'{a}tov\'{a} \inst{\ref{inst:cas}}
           \and
           Laurent Mahy \inst{\ref{inst:rob}}
           \and 
           Antony F.J. Moffat  \inst{\ref{inst:canada}}
           \and 
           Francisco Najarro \inst{\ref{inst:cab}}
           \and
           Raman K. Prinja  \inst{\ref{inst:london}}      
           \and 
           Varsha Ramachandran \inst{\ref{inst:ari}}
           \and
           Andreas A.C. Sander \inst{\ref{inst:ari}}
           \and 
           Jorick S. Vink \inst{\ref{inst:armagh}}
           \and 
           the XShootU collaboration
          }

\institute{
        {Astronomical Institute Anton Pannekoek, University of Amsterdam,  Science Park 904, 1098~XH, Amsterdam, The Netherlands\newline \email{s.a.brands@uva.nl}\label{inst:API}}
    \and 
        {Institute of Astronomy, KU Leuven, Celestijnenlaan 200D, 3001, Leuven, Belgium \label{inst:leuven}}
    \and 
        {LMU M\"unchen, Universit\"atssternwarte, Scheinerstr. 1, 81679 M\"unchen, Germany\label{inst:LMU}}
    \and
        {Astrophysics cluster, School of Mathematical and Physical Sciences, University of Sheffield, Hicks Building, Hounsfield Road, Sheffield S3 7RH, United Kingdom\label{inst:shef1}   } 
    \and 
        {Departamento de Astrof\'{\i}sica, Centro de Astrobiolog\'{\i}a, (CSIC-INTA), Ctra. Torrej\'on a Ajalvir, km 4,  28850 Torrej\'on de Ardoz, Madrid, Spain\label{inst:cab}}
      \and
        {School of Chemical, Materials and Biological Engineering, University of Sheffield, Sir Robert Hadfield Building, Mappin Street, Sheffield S1 3JD, United Kingdom\label{inst:shef2}}
    \and
        {Space Telescope Science Institute, 3700 San Martin Drive, Baltimore, MD 21218, USA\label{inst:STScI}}
    \and 
        {Universidad de La Laguna, Departamento de Astrof\'{i}sica, Avda. Astr. Francisco Sanchez, E-38206 La Laguna, Spain\label{inst:tenerife}}
    \and
        {Department of Physics and Astronomy \& Pittsburgh Particle Physics, Astrophysics, and Cosmology Center (PITT PACC), University of Pittsburgh, 3941 O’Hara Street, Pittsburgh, PA 15260, USA \label{inst:PITT}}
    \and
        {Department of Physics \& Astronomy, East Tennessee State University, Johnson City, TN 37615 USA\label{inst:ETSU}}
    \and 
        {{Zentrum f\"ur Astronomie der Universit\"at Heidelberg, Astronomisches Rechen-Institut, M\"onchhofstr. 12-14, 69120 Heidelberg, Germany\label{inst:ari}}}
    \and 
        {NASA Goddard Space Flight Center, 8800 Greenbelt Rd, Greenbelt, MD 20771, USA\label{inst:goddard}}
    \and 
        {Astronomical Institute, Czech Academy of Sciences, Fri\v cova 298, Ond\v rejov, 251 65, Czech Republic.\label{inst:cas}}
    \and 
        {Royal Observatory of Belgium, Avenue Circulaire/Ringlaan 3, B-1180 Brussels, Belgium \label{inst:rob}}
    \and 
        {D\'epartement de physique, Universit\'e de Montr\'eal, Complexe des Sciences, 1375 Avenue
Th\'er\`ese-Lavoie-Roux, Montr\'eal, Queb\'ec, H2V 0B3, Canada \label{inst:canada}}
    \and 
        {Department of Physics and Astronomy, University College London, Gower Street, London WC1E 6BT, UK\label{inst:london}}
    \and 
        {Armagh Observatory, College Hill, Armagh BT61 9DG, UK\label{inst:armagh}}
   }

\date{Received October 28, 2024; accepted March 10, 2025}

  \abstract
   {Mass loss governs the evolution of massive stars and shapes the stellar surroundings. To quantify the impact of the stellar winds we need to know the exact mass-loss rates; however, empirical constraints on the rates are hampered by limited knowledge of their small-scale wind structure, also referred to as `wind clumping'. }
   {We aim to improve empirical constraints on the mass loss of massive stars by investigating the clumping properties of their winds, in particular the relation between stellar parameters and wind structure. }
    {We analyse the optical and ultraviolet spectra of 25 O-type giants and supergiants in the Large Magellanic Cloud, using the model atmosphere code \fastwind and a genetic algorithm. We derive stellar and wind parameters including detailed clumping properties, such as the amount of clumping, the density of the interclump medium, velocity-porosity of the medium, and wind turbulence.  }
   {We obtain stellar and wind parameters for 24 of our sample stars and find that the winds are highly clumped, with an average clumping factor of $\langle f_\mathrm{cl}  \rangle= 33\pm14$, an interclump density factor of $\langle f_\mathrm{ic} \rangle = 0.2\pm0.1$, and moderate to strong velocity-porosity effects; $\langle f_\mathrm{vel} \rangle = 0.6\pm0.2$. The scatter around the average values of the wind-structure parameters is large. With the exception of a significant, positive correlation between the interclump density factor and mass loss, we find no dependence of clumping parameters on either mass-loss rate or stellar properties. }
   {In the luminosity range that we investigate, the empirical and theoretical mass-loss rates both have a scatter of about 0.5~dex, or a factor 3. Within this uncertainty, the empirical rates and theoretical predictions are in agreement. The origin of the scatter of the empirically inferred mass-loss rates requires further investigation. It is possible that our description of wind clumping is still not sufficient to capture effects of the structured wind; this could contribute to the scatter. }
   \authorrunning{Brands et al.}

   \maketitle
   
 
\section{Introduction}

Stars with an initial mass of $M_\mathrm{ini}\gtrsim 8$~\Msun (massive stars) have a large impact on their environment, in the early universe and in the present. 
They produce heavy elements, which they deposit into their surroundings through their strong stellar winds, and through the supernova explosions with which they end their lives \citep[][]{1957RvMP...29..547B,2006MNRAS.365..615S,2020ApJ...900..179K}. 
Moreover, both the stellar winds as well as the supernova explosions carry large amounts of mechanical energy into the environment \citep{Weaver1977}. In combination with their radiative feedback \citep[e.g.,][]{KahnF.D.1954,SpitzerLyman1978,2015MNRAS.448.3248G} this significantly affects the evolution of their host galaxies \citep[e.g.,][]{2011MNRAS.417..950H}. 

As the evolution of massive stars is dominated by mass loss \citep{2014ARA&A..52..487S}, it is key to have a thorough understanding of this phenomenon and to quantify the mass-loss rate as a function of stellar parameters. While impressive progress in this field has been made \citep[see, e.g.,][for a review]{2022ARA&A..60..203V}, uncertainties still remain. In particular, the inference of mass-loss rates from observations is hampered by small-scale structure present in the wind. 
Theoretically, radiation-driven winds are inherently unstable \citep[e.g.,][]{1970ApJ...159..879L}. The line-driven instability can naturally produce clumping in radiation-hydrodynamic simulations \citep[e.g.,][]{1988ApJ...335..914O,2018A&A...611A..17S,2022A&A...663A..40D}, but observational and structural considerations indicate that other mechanisms such as turbulent motions originating in deep sub-face layers could also play a significant role in triggering inhomogeneous winds \citep{2009A&A...499..279C}; in these simulations structure already exists in the photosphere \citep[][]{2015ApJ...813...74J,2022ApJ...924L..11S,2024A&A...684A.177D}. 
Wind structure, often referred to as `wind clumping' \citep{1988ApJ...334.1038M,1994ApJ...421..310M}, affects mass-loss diagnostics in a variety of ways (see, e.g., \citealt{2008A&ARv..16..209P}; \citealt{Hillier2020_UV_Review}, for a review). For example, recombination lines such as H$\alpha$ can become stronger due to the higher densities in the clumps  \citep[e.g.,][]{2004A&A...415..349R,2004A&A...413..693M,2008A&ARv..16..209P}, while the saturation of ultraviolet (UV) resonance lines can be affected by the low density material that is surrounding the clumps, the so-called interclump medium \citep[][]{2008ApJ...685L.149Z,2013A&A...559A.130S,2010A&A...510A..11S}. Furthermore, porosity in velocity space (velocity-porosity; sometimes also called `vorosity') seems to be crucial for reproducing the \pvline doublet \citep{2002ApJ...579..774C,2006ApJ...637.1025F,2007A&A...476.1331O,2008cihw.conf..121O,2010A&A...510A..11S,2013A&A...559A.130S,2021A&A...655A..67H}. Porosity effects occur when clumps become optically thick: in this case the clumps block part of the light, but as the medium is porous, some light slips through and can escape. 

Time-series spectroscopy of massive stars also reveals structure in the winds. In particular, large-scale structures in the stellar wind produce variable discrete absorption components (DACs) in the blue-shifted absorption troughs of P-Cygni type profiles \citep{1988MNRAS.231P..21P,1995ApJ...452L..53M,1996A&AS..116..257K,1999A&A...344..231K}. 
These DACs are likely the result of varying outflow conditions at the stellar surface caused by for example star spots  \citep{2017MNRAS.470.3672D}, non-radial pulsations \citep{2008ApJ...678..408L}, or stellar prominences \citep{2016A&A...594A..56S}. DACs have also been detected in UV spectra of O stars in the Large Magellanic Cloud (LMC; e.g., \citealt{1998MNRAS.300..828P}). 
This large-scale structure is prominent, but not addressed by the wind-clumping formalism as applied in this paper (see below). The clumping here refers to small-scale structure that manifests itself in the formation of the saturated black absorption troughs in UV resonance lines \citep{1982ApJ...255..286L,1993A&A...279..457P}, the production of X-rays \citep{1982ApJ...255..278L,1993A&A...276..117H,1997A&A...322..878F}, and the flaring observed in accreting X-ray sources in high-mass X-ray binaries \citep[e.g.][]{1993A&A...279..485K,2015A&A...576A.117G,2018MNRAS.475.3240E}. 

For the inference of mass-loss rates it is essential to take into account the small-scale wind clumping. If neglected,  the mass-loss rate of massive stars might be over- or underestimated. However, to date the wind-structure properties of massive-star winds remain poorly constrained. 
Therefore, the description of clumping in quantitative spectroscopy presently has multiple free parameters. The simultaneous analysis of optical and UV wind lines, which each respond differently to clumping effects, can help us to get constraints on these parameters. 

To date several authors have investigated the clumping properties with a clumping prescription allowing for optically thick clumps. \citet{2013A&A...559A.130S} analysed five Galactic O-type supergiants using a combination of the model atmosphere code PoWR \citep{2004A&A...427..697H} and 3D Monte Carlo simulations \citep{2012A&A...541A..37S}. They fit the  density of the interclump medium and the strength of velocity-porosity effects, but assume a fixed clumping factor, a quantity that defines the enhancement of the density within clumps compared to the density of a smooth wind with the same mass-loss rate. \citet{2021MNRAS.504..311F} investigate clumping in an O-type supergiant using a \mbox{CMFGEN} model \citep{1998ApJ...496..407H}  in which the clumps are described as dense spherical shells. Their model computes opacities from the density and velocity profile and can reproduce high ionization resonance transitions, but does not allow for porosity effects. 
\citet{2021A&A...655A..67H} were the first to use the parameterisation of \citet{2018A&A...619A..59S} for a detailed spectral analysis in which stellar and wind-clumping parameters, including velocity-porosity and wind turbulence, were derived simultaneously. 
They analyse a sample of 8 Galactic O-type supergiants and infer moderate velocity-porosity effects. 
Furthermore, they find that clumps are about 3-10 times as dense as the surrounding interclump medium. As their stars were of similar spectral type, \citet{2021A&A...655A..67H} were not able to investigate whether the clumping properties depend on stellar properties. 

\citet{brands2022} investigated a larger sample. They studied the wind structure of 3 WNh and 53 O-type stars in the cluster R136 in the LMC, using a similar method as \citet{2021A&A...655A..67H}, and found a tentative relation between stellar parameters and the clumping properties, where the winds of more luminous stars appeared less clumped. Both the clumping factors, as well as the other wind-structure properties, such as the density of the interclump medium and the strength of the velocity-porosity effects, pointed to this conclusion. However, the statistical significance of their findings was modest. 
\citet{2024AA...690A.126H} did a similar study for an LMC sample of 18 O-type stars of different spectral types. With data of higher signal-to-noise (S/N), and with a higher spectral resolution, they found significant correlations between the density of the interclump medium and temperature, as well as between the velocity-porosity and temperature. These authors could not find a trend in clumping factors. 

In this paper we aim to further investigate the clumping properties of massive-star winds, in particular the relation between stellar parameters and wind-structure properties. To this end, we analyse a subsample of 25 O-type giants and supergiants (luminosity class I, II and III) in the LMC; the sample contains stars of all spectral sub-types and thus spans a large range of temperatures.  
The data that we use for our analysis are part of the Ultraviolet Legacy Library of Young Stars as Essential Standards  (ULLYSES\footnote{\url{https://ullyses.stsci.edu/index.html}}) survey, a Hubble Space Telescope (HST) Director’s Discretionary programme comprising the UV spectra of about $250$ massive stars in the  Magellanic  Clouds, including about $150$~O-type stars (\citealt{2020RNAAS...4..205R}; see  \citealt{2024IAUS..361...15C}, for an overview of scientific goals and auxiliary datasets). The ULLYSES UV dataset is complemented by optical and NIR X-Shooter spectra of the X-Shooting ULLYSES (XshootU) programme\footnote{\url{https://massivestars.org/xshootu/}} \citep{Vinkinprep,Sanainprep}. 
A secondary goal of our paper -- reached in parallel with the wind-structure analysis -- is to derive accurate stellar parameters in order to contribute to the collective effort of creating an empirical spectral template database based on the ULLYSES and XshootU observations. 

The remainder of the paper is structured as follows. In \cref{sec:sample_and_data} we present the stellar sample, observations, and data reduction. In \cref{ullyses:methods} we present our method, that is, our fitting approach and the codes used in the process. In \cref{ullyses:results} we present the stellar and wind parameters. We put our results into context by comparing them with previous work and theoretical predictions in \cref{ullyses:discussion}. We conclude by summarising our results in \cref{ullyses:conclusionoutlook}.  

\section{Sample and data \label{sec:sample_and_data}}

\subsection{Observations and sample selection\label{sec:obs_sample}}

\renewcommand{\arraystretch}{1.1}

\begin{table*}[]
\centering
\newcolumntype{P}[1]{>{\centering\arraybackslash}p{#1}}
    \small 
    \caption{\centering Overview of our sample$^\star$: spectral types and UV observations used for spectral fitting$^{\star\star}$. \label{tab:obs_summary}}
    \begin{tabular}{p{2.5cm} l c l p{1.5cm} p{1.5cm} }
    \hline \hline
\\[-10.0pt]
Source & Spectral type & Ref~$\dagger$ & Revised spectral type&  \multicolumn{2}{c}{Instrument~$\ddagger$}  \\[-10.0pt] \\  
 &  \citep{Vinkinprep}$\dagger$ & & \citep{Bestenlehnerinprep}  &  {\footnotesize $\lambda<$1141~\AA} & {\footnotesize $\lambda>$1141~\AA} \\[-10.0pt] \\  \hline 
\\[-10.0pt]\\[-10.0pt]\\[-10.0pt]
\multicolumn{2}{p{0.2\textwidth}}{\textit{Presumed single}} \\[-10.0pt] \\ 
W61-28-23                 & O3.5 V((f$^{+}$))& a & O3.5 III(f*)      & COS-1096 & COS   \\
Sk~$-$67$^{\circ}$167       & O4 Inf$^+$       & b & O4 If          & FUSE/L   & STIS  \\
ST 92-4-18                & O5 If            & c & O4 Ifp         & -        & COS   \\
Sk~$-$67$^{\circ}$69        & O4 III(f)        & b & O4 III(f)      & -        & STIS  \\
Farina-88                 & O4 III(f)        & d & O4 III(f)      & -        & COS   \\
LMCe 078-1                & O6 Ifc           & e & O5 Ifpc        & -        & COS   \\
Sk~$-$67$^{\circ}$111       & O6 Ia(n)fpv       & f& O6 Ia(n)fpv    & FUSE/L    & STIS  \\
N11-018                   & O6 II(f$^+$)      & g& O6.5 II(f)     & FUSE/M    & COS   \\
Sk~$-$69$^{\circ}$50        & O7(n)(f)p         & f& O7 I(n)(f)p    & FUSE/L    & STIS  \\
Sk~$-$71$^{\circ}$50        & O6.5 III          & h& O7 III(f)p& FUSE/L    & STIS+ \\
Sk~$-$69$^{\circ}$104       & O6 Ib(f)          & i& O7 III(f)      & FUSE/L    & STIS+ \\
Sk~$-$68$^{\circ}$16        & O7 III            & c& O7 III(f)      &  -        & STIS  \\
    LH 9-34               & O8.5 Iaf          & j& O8-8.5 Ifp     & -         & STIS  \\
    Sk -66$^{\circ}$171   & O9 Ia            & k & O9 Ia          & FUSE/L &  STIS+  \\
    Sk -71$^{\circ}$41    & O9.7 Iab          & l& O9.5 Ib        & FUSE/M    & STIS  \\
Sk~$-$67$^{\circ}$5         & O9.7 Ib           & i& B0 Ib          & FUSE/L    & STIS+ \\
\\[-10.0pt]\\[-10.0pt]\\[-7.0pt]
\multicolumn{5}{p{0.40\textwidth}}{\textit{Signs of binarity}} \\[-10.0pt] \\ 
    LH~114-7             & O2 III(f*)+OB?    & m & O2 III(f*)     & -        & STIS  \\
    VFTS-267             & O3 III-I(n)f*     & n & O3.5 III(f*)   & -        & COS   \\
    Sk~$-$71$^{\circ}$46   & O4 If           & o   & O4 If           & -        & COS \\
    Sk~$-$67$^{\circ}$108  & O4-5 III          & k &  O5 III(f)     & FUSE/L   & STIS  \\
Sk~$-$71$^{\circ}$19       & O6 III             & h& O6 V((f))      & FUSE/L    & COS   \\
    Sk~$-$70$^{\circ}$115  & O6 If+O8:         & p & O6.5 Ifc       & FUSE/L   & STIS+ \\
    BI~173               & O8 II:             & q& O7.5 III((f))p  & FUSE/L   & STIS+ \\
    Sk~$-$68$^{\circ}$155  & B0.5 I             & r& O9 Ia          & FUSE/L   & COS $\diamond$ \\
BI~272                   & O7 II             & h & O7:V + B2          &  -       & STIS  \\
\hline 
\\[-8.0pt]
\multicolumn{6}{p{0.78\textwidth}}{\footnotesize $^\star$ The stars are ordered from early to late, adopting the spectral types of \citet{Bestenlehnerinprep}. }\\
\multicolumn{6}{p{0.78\textwidth}}{\footnotesize $^{\star\star}$ Optical observations are not specified here: all were obtained with VLT/X-shooter. }\\
\multicolumn{6}{p{0.78\textwidth}}{\footnotesize $\dagger$ \citet{Vinkinprep} list spectral types from literature; original references: 
\textbf{a}:~\citet{2005ApJ...627..477M}, 
\textbf{b}:~\citet{1987PASP...99..240G}, 
\textbf{c}:~\citet{2000AJ....119.2214M}, 
\textbf{d}:~\citet{2009AJ....138..510F}, 
\textbf{e}:~\citet{2017ApJ...837..122M}, 
\textbf{f}:~\citet{2010AJ....139.1283W}, 
\textbf{g}:~\citet{2006AA...456..623E}, 
\textbf{h}:~\citet{1986AJ.....92...48C}, 
\textbf{i}:~\citet{1977ApJ...215...53W}, 
\textbf{j}:~\citet{1992AJ....103.1205P}, 
\textbf{k}:~\citet{1988ApJ...335..703F}, 
\textbf{l}:~\citet{2018AA...615A..40R}, 
\textbf{m}:~\citet{2002AJ....123.2754W}, 
\textbf{n}:~\citet{2014AA...564A..40W}, 
\textbf{o}:~\citet{1995ApJ...438..188M}, 
\textbf{p}:~\citet{2004NewAR..48..727N},
\textbf{q}:~\citet{2002ApJS..141..443W}, 
\textbf{r}:~\citet{1978AAS...31..243R}. 
} \\
\multicolumn{6}{p{0.78\textwidth}}{\footnotesize $\ddagger$ `FUSE/L' and `FUSE/M' indicate that FUSE/LWRS and FUSE/MDRS data were used, respectively. `COS-1096' indicates COS/G130M/1096 data. `STIS' indicates STIS/E140M/1425 data, `STIS+' indicates STIS/E140M/1425 and STIS/E230M/1978 data. `COS' indicates COS/G130M/1291 and COS/G160M/1611 data. }\\
\multicolumn{6}{p{0.78\textwidth}}{\footnotesize $\diamond$ In the case of Sk -68$^{\circ}$155 the COS/G130M/1291 data do not reach lower than 1178~\AA, and we therefore use FUSE/LWRS data for the range 1141-1178~\AA\xspace. }\\
    \end{tabular}
\end{table*}

Our sample consists of all O-type supergiants, bright giants and giants (luminosity class I, II and III) in the LMC for which reduced ULLYSES \citep{2020AAS...23523203R} and XshootU \citep{Vinkinprep} data were available on 22 March 2022. This corresponds to Data Release 4 (DR4; June 2021) of the ULLYSES data and early Data Release 1 (eDR1; internal data release; March 2021) of the XshootU data \citep{Sanainprep}. We only include targets for which the available spectra cover at least the range $\lambda \lambda 1141-1708$~\AA\xspace. 
While our target selection was based on the spectral types as presented in the ULLYSES target list in March 2022 (which differs slightly from the spectral types in the overview \citealt{Vinkinprep}), in this paper we adopt spectral types from \citet[][]{Bestenlehnerinprep}; in several cases these sources list a different spectral type. We furthermore include one O-type supergiant not included in ULLYSES DR4. This source (Sk~$-66^{\circ}$171) is part of a benchmark study carried out by the XshootU collaboration (\citealt{Sanderinprep}; paper IV from this series) and meets all our sample requirements -- except for not being available on the date of our target selection. We use the ULLYSES DR5 data for this source and XshootU eDR1 data. 

An overview of our sample stars, spectral types, and data used can be found in \cref{tab:obs_summary}. The table contains two columns for spectral type: one column lists the type as in \citet{Vinkinprep}, who made an overview based on existing literature, the other column lists revised spectral types based on the XshootU spectroscopy \citep{Bestenlehnerinprep}; we note that for five sources the spectral type remains unchanged compared to the type listed in \citet{Vinkinprep}.  
In the end, adopting the spectral types from \citet{Bestenlehnerinprep}, our sample consists of one dwarf (Sk~$ -71^{\circ}$19), one early B-type supergiant (Sk~$-67^{\circ}5$), and 23 O-star supergiants, bright giants and giants.  

The used UV spectroscopy per source is summarised in \cref{tab:obs_summary}; a detailed log of the UV observations can be found on \href{https://zenodo.org/records/15013513}{Zenodo}. The UV data were obtained with either the Far Ultraviolet Spectroscopic Explorer (FUSE) or with the HST. The spectra covering the range $\lambda \lambda 1141-1708$~\AA\xspace are taken with either HST in combination with the Space Telescope Imaging Spectrograph (STIS) with grating E140M/1425 ($\lambda \lambda 1141-1708$~\AA, resolving power $R=45800$), or with HST in combination with the Cosmic Origins Spectrograph (COS) with gratings G130M/1291 and G160M/1611 ($\lambda \lambda 1141-1783$~\AA, $R=12000-17000$). These observations have a S/N in the range $8-21$. For seven STIS sources we also have observations in the E230M/1978 setting ($\lambda \lambda 1608-2366$~\AA, $R=30000$), with a S/N in the range $18-31$. We note that for two sources, Sk~$-70^{\circ}$115 and Sk~$-67^{\circ}$5, we have spectra taken with STIS/E230H and STIS/E140H; these gratings do not offer additional wavelength coverage compared to the E140M/1425 and E230M/1978 data that we also have for these sources, but they do have higher resolution. However, in order to keep the sample as homogeneous as possible, for our spectroscopic analysis we use only the data of the E140M/1425 and E230M/1978 gratings, also for Sk~$-70^{\circ}$115 and Sk~$-67^{\circ}$5. A similar case is that of Sk~$-68^{\circ}$155, for which we have E230M/1978 data, but use the data of the G130M/1291 and G160M/1611 gratings for the spectroscopic analysis, as we do for the other sources with COS data. 

Availability of spectroscopy in the far-UV range ($\lambda < 1141$~\AA) is not a selection criterion, but in case these data are available we include them in our analysis. This was the case for fifteen sources. For fourteen sources data in the far-UV were obtained with FUSE, using the LWRS or MDRS apertures ($\lambda \lambda 905-1180$~\AA, $R\simeq17500$); for one source the far-UV data came from HST/COS in combination with the G130M/1096 grating ($\lambda \lambda 940-1240$~\AA, $R =3000-12000$). For one additional source (Sk~$-71^{\circ}$46) FUSE data was available, but the S/N was very low ($< 3$) and therefore we did not include this data. The other FUSE observations have S/N in the range $4-18$; the COS/G130M1096 observation of W61-28-23 has S/N~$ =34$. 

All optical spectra were taken as part of the `X-shooting ULLYSES' or XshootU project (PI: Jorick Vink), an ESO large programme that was launched to complement the ULLYSES observations \citep{Vinkinprep}. 
The data were collected with the X-shooter spectrograph on the Very Large Telescope. X-shooter provides simultaneous coverage of three wavelength regions: near-UV and blue-optical (UVB arm; $3000\leq \lambda \leq 5000$~\AA), red-optical (VIS arm; $5000\leq \lambda \leq 10000$~\AA), and near-infrared \citep[NIR arm; $10000\leq \lambda \leq 25000$~\AA;][]{2011A&A...536A.105V}. In our analysis we only consider the UVB and VIS arms, with a resolving power of $R \approx 6700$ (at slit width 0.8") and $R \approx 11400$ (at slit width 0.7"), respectively. For most stars we have one epoch consisting of one exposure for each arm. Exceptions are Sk~$-66^\circ$171, for which we have two consecutive exposures in each arm, and ST~92-4-18 and BI~272, for which we have two epochs, each epoch one exposure per arm. A detailed account of the optical data reduction is presented in \citet{Sanainprep} and a log of the optical observations can be found in on \href{https://zenodo.org/records/15013513}{Zenodo}.

\subsection{Binarity}

Nine of our 25 stars show signs of binarity. We analyse these stars as if they were single stars because in all but one of these sources, one of the two stars dominates the spectrum. For one source (BI~272) we cannot obtain a good fit; it appears that both components of the binary contribute significantly and the single star approach cannot be employed (see \cref{ullyses:results}). 
We will now describe the signs of binarity of each of these sources. 

Three stars of our sample are explicitly marked as (possible) binaries in the \citet{Vinkinprep} overview. 
These are Sk~$-71^{\circ}$46, an eclipsing binary \citep{2011AcA....61..103G}, Sk~$-70^\circ$115, a system with a known period of 6.682~days and a mass ratio of $q = M_2/M_1 \sim 0.5$ \citep{2004NewAR..48..727N}, and LH~114-7 (listed as O2 III(f*)+OB? in \citealt{Vinkinprep}), 
which is suspected to have a companion: the He~\textsc{i}/He~\textsc{ii} ratio suggests a moderately early spectral type ($\sim$ O5), while the spectrum also contains nitrogen features that are more typical for very early-type stars \citep{2005ApJ...627..477M}. \citet{2010AJ....139.1283W} also classify this spectrum as composite. Upon inspecting the UV spectrum of LH~114-7, we see unusual shapes of the P-Cygni \CIVline and \ovline: both lines have absorption troughs that are very broad, but not saturated. This, too, hints at the presence of a companion star, that could be diluting the wind lines. 

In addition to the three binary stars listed by \citet{Vinkinprep}, we identify six other (potential) binaries in our sample by comparing the radial velocities of different epochs (of both optical and UV spectra; for details, see \cref{app:RVs}.), by inspecting the optical spectra for double sets of spectral lines, by identifying unusually shaped P-Cygni profiles in the UV \citep[e.g.,][]{1993ApJ...404..281S,2004A&A...423..267G} and by consulting the literature. For three stars we find radial-velocity variations with high significance ($\geq 4\sigma$); this concerns BI~173 and BI~272, as well as the eclipsing binary Sk~$-71^{\circ}$46 mentioned above. For this source we observe a secondary component in the red wings of the optical lines, about 250 km~s$^{-1}$ from the line centre of the primary component. Also for Sk~$-67^\circ$108 we find radial-velocity variations, albeit with a lower statistical significance ($1.9 \sigma$). Still, the binary nature of this source seems likely given the fact that the line profiles of this star look asymmetric: a secondary component seems present in the red wings of the optical lines. We find similar asymmetries in the spectrum of Sk~$-68^\circ$155, but in this case the secondary component is seen in the blue line wings. For this star we do not find significant radial-velocity variations. 
In the optical spectrum of VFTS-267 we see narrow \hei lines and broad \heii lines, suggesting that the spectrum might be composite. The radial velocities we derive from the two epochs we have for this star do not differ significantly from one another. \citet{2013AA...550A.107S} and \citet{2017A&A...598A..84A}, having a larger number of epochs, do report significant variations, but they cannot find a periodicity. Lastly, \citet{1986AJ.....92...48C} and \citet{1999AJ....117.2856S} mark Sk~$-71^{\circ}$19 as a spectroscopic binary. The latter (who inspect UV spectroscopy) remark that the star has inconsistent spectral features, namely strong N~\textsc{v}~$\lambda1240$ and O~\textsc{v}~$\lambda1371$ on the one hand pointing to a mid O-type star, and strong Al~\textsc{iii}/Fe~\textsc{iii} on the other hand, typical for B-type stars. Furthermore,  \CIVline is weak, atypical for an O6~III star. We indeed find these inconsistent features in our UV spectra.

\subsection{UV data reduction and preparation \label{sec:UVnorm}}

The UV spectra we used for the analysis are the High Level Science Products created by the ULLYSES team\footnote{\url{https://ullyses.stsci.edu/ullyses-download.html}}. These data are flux and wavelength calibrated, contain error spectra, and are available per grating, or in files in which the observations of different gratings have been merged. We use the files that contain spectra per individual grating, with the exception of the G130M/1291 and G160M/1611 observations, for which we use the merged files; we checked whether the wavelength calibration of the G130M/1291 and G160M/1611 gratings was consistent, which was the case for all sources. A detailed description of the data reduction of the UV spectra can be found in \citet{2020AAS...23523203R}.

For the normalisation we follow the method of \citet{brands2022}. In short, we find the position of the continuum by fitting the iron pseudo-continuum to \mbox{CMFGEN} \citep{1998ApJ...496..407H} models of \citet{2014AA...570A..38B} in five steps: \textit{1)} we mask strong wind lines and interstellar lines \textit{2)} we divide the observed UV flux by the normalised spectrum of the \mbox{CMFGEN} model, \textit{3)} we fit a polynomial to the ratio obtained in the previous step, \textit{4)} we use the polynomial as the normalisation model; that is, we obtain the normalised flux by dividing the observed UV flux by the polynomial, and \textit{5)} for each source we repeat steps \textit{2} to \textit{4} of this process for \mbox{CMFGEN} models with varying temperature, varying radial velocity, and, in some cases, varying rotational broadening (see below). For each broadened \mbox{CMFGEN} model we obtain a different normalised spectrum, of which we determine the goodness of fit of the iron pseudo-continuum compared to the assumed \mbox{CMFGEN} model. We adopt the best fitting normalised spectrum for our analysis. As a by-product, we also obtain the effective temperature\footnote{That is, the effective temperature belonging to the best fitting model; this is based on the fit to the iron-group lines only. }, the radial velocity, and in some cases the projected rotational velocity ($\varv \sin i$; see below). A more detailed description and demonstration of this normalisation method can be found in \citet{brands2022}. We note that all models used for the normalisation process have a surface gravity of $\log g = 4.0$. Since surface gravity has a similar effect on the iron forest as does temperature\footnote{An increase of 5000~K in temperature has approximately the same effect as a decrease of 0.5~dex in gravity.}, the temperatures derived from the iron forest will be systematically affected because temperature is used to `compensate' for gravity. 

We normalise all spectra in the manner described above. First, we normalise the spectra of the STIS/E140M or COS/G130M and COS/G160M gratings; in the process, the rotational broadening $\varv \sin i$ is a free parameter\footnote{We do not use these $\varv \sin i$ for any further analyses, but instead rely on the values found from the optical-only fits, see \cref{sec:fitsetup}.}. Next, we normalise the spectra obtained with other gratings, if available. For these settings, we do not fit $\varv \sin i$ but instead adopt the value found during the STIS/E140M or COS/G130M and COS/G160M normalisation procedure. For the FUSE and COS/G130M/1096 spectra we only normalise the region $\lambda >1100~$\AA, as we do not use diagnostics at shorter wavelengths for our analysis. 
For three stars  (Sk~$-71^\circ$50, Sk~$-68^\circ$155, and W61-28-23) the radial velocity we find from the FUSE data is poorly constrained or visibly deviating; in these cases we adopt for the FUSE data the radial velocity as found for the STIS/E140M or COS/G130M + COS/G160M gratings. 
For the STIS/E230M spectra we only normalise the region $\lambda <2000~$\AA, as we do not use diagnostics at longer wavelengths in our analysis. 

The normalisation method as described above is objective and works well when we obtain a good fit to the iron pseudo-continuum. 
This is usually the case, but sometimes the fit is not perfect and we see in the normalised spectrum `emission' where we do not expect this; that is, at wavelength ranges without wind lines the normalised flux lies several percent above unity. 
We inspect all diagnostics used for the fitting (\cref{tab:line_selection_UV}) and when the continuum clearly lies above unity, we renormalise the spectrum locally, that is, only around the diagnostic line that we are inspecting. We do this by estimating the correct location of the continuum by eye (in terms of normalised flux), and then divide the already normalised spectrum by this value. 

Now that we have normalised, radial-velocity corrected spectra, we correct for interstellar absorption lines. All spectra show a saturated interstellar Ly-$\alpha$ absorption line. 
The width of the profile varies per source, but in several cases \cuvivline and \cuviiiline are affected, and in all cases \nvuvline. 
We therefore correct for the interstellar absorption by fitting a Voigt-Hjerting function\footnote{Also called Voigt function (see e.g., \citealt{1978stat.book.....M}).} \citep{2006MNRAS.369.2025T, 2007MNRAS.382.1375T} to the Ly-$\alpha$ profile. For the damping factors of the Lorentzian component of the profile we use the radiative damping constants of the Ly-$\alpha$ transition. The hydrogen column density and central wavelength are free parameters that are fitted per source. We obtain good fits and can recover \cuvivline, \cuviiiline, and \nvuvline for all sources\footnote{While we do use \nvuvline for some experiments, we do not use it in our final analysis; see \cref{sec:dis:NITRO}.} and find an average central wavelength of $1215.2\pm0.4$~\AA, and hydrogen column densities that range from $\log~N$(H~\textsc{i}$~[\mathrm{cm}^{-2}]) = 20.7$ to $\log~N$(H~\textsc{i}$~[\mathrm{cm}^{-2}]) = 21.8$. 

Other interstellar lines in the UV that interfere with our stellar diagnostics are \siivline and \CIVline. We do not correct for these lines like we do for Ly-$\alpha$, but instead we clip (remove) the parts of the spectrum that are affected by this. This means that we lose part of our diagnostic, however, the loss of information is minimal, as the spectral resolution is sufficiently high and the interstellar lines are narrow compared to the stellar (wind) lines.

\subsection{Optical data reduction and preparation \label{sec:optical_dataprep}}

A detailed description of the data reduction of the optical spectra can be found in the paper concerning the XshootU Data Release 1 (DR1; \citealt{Sanainprep}). In brief, the reduction of the optical spectra (VIS and UVB arms\footnote{The near-infrared spectra will be available in a later data release. }) was performed using the ESO X-shooter pipeline v3.5.0 \citep{2011AN....332..227G}, and included bias, flat, and wavelength calibration, spectral rectification, sky subtraction, cosmic ray removal, flux normalisation, and extraction of a 1D spectrum. Co-added spectra are provided for stars with multiple exposures; in our sample we have three sources with multiple exposures. After inspecting radial-velocity differences between observations of different epochs, and finding none, we adopt these co-added spectra, rather than single epochs, in order to ensure S/N~$>100$ for all sources. We note that while XshootU DR1 contains telluric corrected spectra, this was not the case for the eDR1 spectra that we used for our analysis. If telluric lines were present in our spectra, we masked them during the fitting. 

In addition to flux calibrated spectra, all XshootU data releases contain normalised spectra. This normalisation was done automatically by fitting a modified Planck function to the flux-calibrated spectra. We adopt the normalised spectra, but before fitting we inspect all diagnostics by eye and if the continuum deviates from unity, we renormalise the spectrum locally. We do this by selecting continuum on both sides of the diagnostic that we are inspecting and carry out a linear fit through these points. We then divide the spectrum by the best linear fit to obtain the renormalised spectrum. 

\subsection{Photometry \label{sec:photometry_data}}

For assessing the luminosity of each source we adopt photometric values collected by \citet{Vinkinprep}, who present for most sources $U$, $B$, $V$, $J$, $H$, and $K_s$ magnitudes\footnote{Two sources lack $U$ magnitudes, one source lacks an $H$ magnitude.}. The adopted values and references per source are listed in \Cref{tab:photometry}. In \cref{sec:anchormag} we describe how these photometric values are used to derive the luminosity of each star. 

\section{Methods \label{ullyses:methods}}

For the analysis we rely on the model atmosphere code \fastwind. One \fastwind model can be computed in 15-45 minutes on a single CPU and without user intervention, which enables us to compute many models in an automated fashion. We exploit this by combining \fastwind with the genetic fitting algorithm Kiwi-GA\footnote{\url{https://github.com/sarahbrands/Kiwi-GA}}, facilitating the exploration of a large parameter space. In this section we discuss the two codes in more detail, followed by a description of our fitting approach. 

\subsection{Fastwind \label{sec:fastwind}}

\fastwind is a unified model atmosphere and radiative transfer code tailored to hot stars with winds \citep{1997A&A...323..488S,2005A&A...435..669P,2012A&A...537A..79R,2016A&A...590A..88C,2018A&A...619A..59S,2020A&A...642A.172P}. The atmosphere is described by a spherical quasi-hydrostatic photosphere which is linked to an expanding stellar wind at a velocity near the sonic point. The wind is specified by a pre-defined mass-loss rate $\dot{M}$, a terminal velocity $\varv_\infty$, and a classical $\beta$ velocity-law (e.g., Eq. (1) of \citealt{1997A&A...323..488S}). Furthermore, the user can provide parameters that describe the structure of the wind (wind-clumping) in a statistical manner (see below). For this work we use \fastwind V10.6.0. 

\fastwind incorporates non-LTE rate equations and takes into account the effects of line blocking and blanketing. In order to speed up the computation while maintaining precision, the atomic elements are split up in `explicit' and 'background' elements. The former are computed in the co-moving frame and for these elements spectral lines can be synthesised. The background elements on the other hand are computed in an approximate fashion; their radiation field is taken into account to ensure that the effects of line-blocking and blanketing are treated correctly, but individual transitions are not synthesised\footnote{In the latest version of \fastwind \citep[v11;][]{2020A&A...642A.172P} both explicit and background elements are treated in the co-moving frame. We do not use this version because the computation is slower and does not yet include an optically thick clumping prescription.}. For this work we include explicit elements H, He, C, N, O, Si, and P. 

\fastwind V10.6.0 includes a prescription for clumping which allows for optically thick clumps \citep{2018A&A...619A..59S}. This is implemented employing the formalism introduced in \citet{2014A&A...568A..59S}, where the wind consists of two components: over-dense clumps with a density $\rho_\mathrm{cl}$, and an under-dense interclump medium with a density $\rho_\mathrm{ic}$. In this prescription, the clumps are assumed to occupy a certain fraction of the total wind volume, $f_\mathrm{vol}$, such that a clumping factor can be defined:
\begin{equation}
    f_\mathrm{cl} \equiv \frac{\langle \rho^2 \rangle}{\langle \rho \rangle^2} = \frac{f_\mathrm{vol}\,\rho_\mathrm{cl}^2 + (1-f_\mathrm{vol})\rho_\mathrm{ic}^2}{[f_\mathrm{vol}\,\rho_\mathrm{cl} + (1-f_\mathrm{vol})\rho_\mathrm{ic}]^2},
\end{equation}
The clumping properties are simulated by adopting a single, `effective' opacity for the medium, which is rescaled as to account for clumps of arbitrary optical depths. The clumps are thus allowed to get optically thick, with porosity effects as a consequence. On the one hand, while optically thick clumps block some light, the porous medium allows some light to slip through. On the other hand, if the clumps do not follow the average velocity field, gaps between the clumps can effectively be closed because the Doppler shifted gas in the clumps spans a wider range of velocities. In that case, more light is blocked for line photons. 

The wind structure is described using six wind-structure parameters, of which we will treat five as free parameters in the fitting. 
The first three parameters describe the value of the clumping factor throughout the wind: clumping is assumed to start at a certain onset velocity ($\varv_\mathrm{cl,start}$), after which it increases linearly (in velocity space) to a maximum clumping factor (simply referred to as $f_\mathrm{cl}$), which is reached at a certain velocity $\varv_\mathrm{cl,max}$ (the third parameter) for which we assume $\varv_\mathrm{cl,max} = 2 \varv_\mathrm{cl,start}$ (following \citealt{2021A&A...655A..67H}). At $\varv < \varv_\mathrm{cl,start}$ the wind is assumed to be smooth; at $\varv > \varv_\mathrm{cl,max}$ the clumping factor stays constant at the maximum value. Because the value of $\varv_\mathrm{cl,max}$ is coupled to the value of $\varv_\mathrm{cl,start}$, we do not consider the former a free parameter. 

A fourth parameter, the interclump density factor $f_\mathrm{ic}$, describes the ratio between the interclump density and the mean density of the medium. It can vary independently from the clumping factor and is defined as: 
\begin{equation}
f_\mathrm{ic} = \rho_\mathrm{ic}/\langle\rho\rangle,
\end{equation}
with $\langle\rho\rangle$ the average density of the medium. The fifth parameter, which can also vary independently of the other parameters, concerns the velocity-porosity effects. These effects are expressed in terms of the normalised velocity-porosity clumping factor $f_\mathrm{vel}$: 
\begin{equation}
\label{eq:vel}
     f_\mathrm{vel} = \frac{f_\mathrm{vor}}{1+f_\mathrm{vor}}, \hspace{1.7cm} \mathrm{with} \hspace{0.2cm}  f_\mathrm{vor} = f_\mathrm{vol}\left|\frac{\delta\varv}{\delta\varv_\mathrm{sm}}\right|,
\end{equation}
where $f_\mathrm{vor}$ is the non-normalised velocity-porosity clumping factor; $\delta \varv$ the velocity span of the clumps, and $\delta \varv_\mathrm{sm}$ the velocity span of the underlying smooth velocity field. Spatial porosity is included as well, but does not play a significant role for the parameter range studied here. For the size of the clumps we assume a velocity stretch law \citep[see][]{2018A&A...619A..59S}. 

The last parameter describing the wind structure concerns the effects of non-monotonic velocity fields present in the wind, often associated with turbulent motions: while during the computation of the ionisation and excitation structure a fixed micro-turbulent velocity ($\varv_\mathrm{micro}$) is adopted, during the radiative transfer the turbulence increases from $\varv_\mathrm{micro}$ at the base of the wind to $\varv_\mathrm{windturb}$ at the point where the wind reaches its terminal velocity. For $\varv_\mathrm{micro}$ we adopt a fixed value of 15~km~s$^{-1}$, while $\varv_\mathrm{windturb}$ is a free parameter (as in, e.g., \citealt{2021A&A...655A..67H}). 
A more extensive, but simplified, description of the implementation of the formalism into \fastwind, as well as an illustration of the effects of wind clumping, porosity, and velocity-porosity, can be found in \citet[][their Fig. 5]{brands2022}. For all other details regarding the clumping implementation we refer the reader to \citet{2014A&A...568A..59S} and \citet{2018A&A...619A..59S}. 

We conclude this subsection by a note on wind-embedded shocks. Such shocks, caused by radiative instabilities \citep[e.g.,][]{1988ApJ...335..914O,1997A&A...322..878F}, can alter the ionisation balance of the wind and in this way affect spectral lines \citep[e.g.][]{2004ApJ...606..497G,2008ApJ...685L.149Z,2010ApJ...711L..30W,2016A&A...590A..88C,2021MNRAS.504..311F}. The shocks and the associated X-ray emission are implemented in \fastwind \citep{2016A&A...590A..88C,2020A&A...642A.172P}. The exact values of the parameters that describe the shocks can be tweaked. As we do not have observational constraints on the shock parameters, we adopt canonical values for all stars. For most parameters we follow the approach of \citet{brands2022}, but we introduce a new method to estimate the X-ray volume filling fraction (\cref{app:sec:X-rays}). 
 While the diagnostic lines considered in our fitting (\Cref{tab:line_selection,tab:line_selection_UV}) are not expected to be significantly affected by shocks, for other lines, such as \nvuvline and \oviline, the effect can be strong. We come back to this point in \cref{sec:dis:NITRO}. Details on the assumptions regarding shock and X-ray parameters can be found in \cref{app:sec:X-rays}.

\subsection{Genetic algorithm: Kiwi-GA \label{sec:kiwiGA}}

Given the large number of free parameters it is not feasible to compute a grid that covers our whole parameter space. Instead, we use a genetic algorithm in order to find the \fastwind model that matches best the observed spectra. Genetic algorithms can efficiently probe large parameter spaces by mimicking concepts of biological evolution (see below). They can be used for all kind of optimisation problems, and have in the past already been used successfully to fit the spectra of massive stars \citep[e.g.,][]{2005A&A...441..711M,2014A&A...572A..36T,2017AA...600A..81R,2021A&A...655A..67H,brands2022}. 

The genetic algorithm we use for this work is called Kiwi-GA \citep[][see also \citealt{2005A&A...441..711M} and \citealt{2021A&A...651A..96A}]{brands2022}. 
Kiwi-GA, like all genetic algorithms, starts out with an initial population of models of which the parameters are randomly sampled from a parameter space that is determined by the user. After all models are computed, the goodness of fit of each model to the data is assessed, by computing the chi-squared value: 
\begin{equation}\label{eq:chi2}
    \chi^{2} = \sum_{i=1}^N \left( \frac{\mathcal{F}_{\mathrm{obs},i} - \mathcal{F}_{\mathrm{mod},i}}{ \mathcal{E}_{\mathrm{obs},i}} \right)^2, 
\end{equation}
with $N$ the number of data points of the spectrum that is considered in the fit, $\mathcal{F}_{\mathrm{obs},i}$ the observed normalised flux,  $\mathcal{F}_{\mathrm{mod},i}$ the normalised flux of the model, and $\mathcal{E}_{\mathrm{obs},i}$ the uncertainty on the observed flux. 
Once the $\chi^{2}$ values are computed, parameters for the next generation of models are selected, by recombining the parameters of models of the previous generation. This is done in such a way, that the best fitting models (those with the lowest $\chi^2$) have the highest probability of being picked for the `reproduction process' (natural selection). After the recombination, small random changes are applied to part of the parameters (`mutations'). Once the new parameters are determined, the new set of models is computed. The models in this new `generation' will, typically, on average give slightly better fits to the data. The process of recombination and mutation, followed by model and fitness computation is repeated until the algorithm converges towards a specific set of parameters. In this work we fit at most 13 parameters simultaneously (see \cref{sec:fitsetup}), and for this we require the computation of 128 (number of models per generation) $\times$ 80 (number of generations) $\approx 10,000$ models for one fit. For more details about the workings of Kiwi-GA including a flowchart of the algorithm, we refer the reader to \citet{brands2022}.

\subsection{Best fit parameters and uncertainties\label{sec:Uncertainties}}

At the end of a Kiwi-GA run we inspect the fit by eye. In the case where the fit looks good, we adopt the parameters of the model with the lowest $\chi^2$ as best fit stellar and wind parameters.  Then, we derive uncertainties from the $\chi^2$-distributions that we find as a function of each parameter. In the past, this was done by normalising all $\chi^2$ values by the lowest $\chi^2$-value and by subsequently applying standard $\chi^2$ statistics and a cutoff value of $p = 0.05$ (see e.g., \citealt{2014A&A...572A..36T,brands2022}, for details). Initially, we applied this same method to the fits in this paper, but we noted that the uncertainties we found were clearly underestimated, in most cases being 0 for all parameters. The fact that this method underestimates uncertainties is due to the implicit assumption that a model with a perfect fit exists: a fit that does not differ significantly from the data. In reality, this is not the case: none of the models is `exactly true'. 
With a very large number of data points and/or a high S/N, as is the case for our high resolution spectra, small differences between models and data can easily be detected: a $\chi^2$ fit will nearly always find a significant difference (i.e., a bad fit). In practice, if we use the $\chi^2$ test in the case of a high number of data points, only the best fit model (which after the normalisation of $\chi^2$ values has $\chi^2_\mathrm{red} = 1$) will qualify as a statistically good fit, and the derived uncertainties will approach zero.

We are thus in need of a measure of goodness of fit that aims to quantify whether the model has an approximate or close fit to the data, rather than an exact fit. A statistical measure that allows for this is the root mean square error of approximation (RMSEA, \citealt{1980SteigerLind,1990Steiger}), frequently used in social and behavioural sciences. 
The RMSEA is computed for each model individually and is derived from the $\chi^2$-value: 
\begin{equation}
    \mathrm{RMSEA} = \sqrt{\mathrm{max}\left(\frac{\chi^2 - n_\mathrm{dof}}{n_\mathrm{dof}(N-1)},0\right)}, \label{RMSEA_eq1}
\end{equation}
with $n_\mathrm{dof}$ the degrees of freedom. 
Values closer to 0 represent a good or `close' fit. However, an absolute cutoff value to differentiate between `acceptable' or `unacceptable' does not exist. For this work, we adopt: 
\begin{equation}\label{eq:RMSEAcutoff}
    \alpha_\mathrm{RMSEA} = 1.04 \cdot \mathrm{min}( \mathrm{RMSEA}),
\end{equation}
where $\mathrm{min}( \mathrm{RMSEA})$ is the minimum RMSEA value of all models in the Kiwi-GA run and $\alpha_\mathrm{RMSEA}$ is the cutoff value. This means that all models with $\mathrm{RMSEA} < \alpha_\mathrm{RMSEA}$ are considered to have an acceptable fit; in other words, the uncertainty range of each parameter consists of all parameter values belonging to the models with $\mathrm{RMSEA} < \alpha_\mathrm{RMSEA}$. 
The value of 1.04 results in uncertainties that approximately resemble $1 \sigma$ uncertainties. The value is calibrated using Kiwi-GA runs of the low-resolution data of the R136 study of \citet{brands2022}, where the $\chi^2$ method still works well, since we have fewer data points: if we adopt the RMSEA measure and $\alpha_\mathrm{RMSEA}$ as in \cref{eq:RMSEAcutoff} we obtain similar uncertainties as we do with the $\chi^2$ method. For obtaining uncertainties that resemble approximately $2\sigma$ equivalent errors, we use a factor 1.09; also this value is calibrated using the runs of \citet{brands2022}. 
We note that using a relative cutoff value (dependent on the best fit) has the effect that for stars for which we obtain a relatively poor fit, we adopt larger uncertainties than we do for stars where the fit is good; this is what we would realistically expect to be the case. That being said, if for a majority of the diagnostics the model and the observed spectrum do not match, we cannot trust the resulting best fit values and uncertainties at all, and reject the fit altogether. 

We implement the RMSEA-method as described above into Kiwi-GA and use it for obtaining uncertainty margins on our derived parameters. We stress that, while we are confident that for high resolution spectra this method is an improvement over the original $\chi^2$ method, it is by no means perfect. The provided uncertainties that we quote throughout this paper should thus be regarded as estimated values. 
We stress furthermore, that the new RMSEA-method has no influence on the determination of the relative fitness between the models, i.e., on the order of the models from fittest to least fit. Thus, even if it would be used during the run (and not only during the post-processing), it would not influence the result, as it is essentially a scaled $\chi^2$ value, as can be seen when we substitute $\chi^2_\mathrm{red} = \chi^2/n_\mathrm{dof}$ in \cref{RMSEA_eq1}: 
\begin{equation}
    \mathrm{RMSEA} = \sqrt{\mathrm{max}\left(\frac{\chi^2_\mathrm{red} - 1}{N-1},0\right)},   \label{RMSEA_eq2}
\end{equation}
and consider the fact that we employ the RMSEA method only when the best fit is not an exact fit (i.e., when the $\chi^2$-value is not only affected noise, but also by intrinsic differences between model and data), in all cases $\chi^2_\mathrm{red} > 1.0$ and thus $\frac{\chi^2_\mathrm{red} - 1}{N-1} > 0$, reducing \cref{RMSEA_eq2} to: 
\begin{equation}
    \mathrm{RMSEA} = \sqrt{\frac{\chi^2_\mathrm{red} - 1}{N-1}} \label{RMSEA_eq3}. 
\end{equation}
Keeping in mind that $N$ is constant for all models in a run it is clear from \cref{RMSEA_eq3} that the order of the models by fitness is same whether their fitness is assessed by $\chi^2$ or RMSEA. A Kiwi-GA output that is analysed with the $\chi^2$ measure compared to with the RSMEA measure will thus always result in the same best fit parameters; the only aspect that is changing is the size of the uncertainties. When applying the RMSEA method to the current study, we obtain larger uncertainties than we would have, had we used the $\chi^2$ method. 

\subsubsection{Luminosity \label{sec:anchormag}}

Kiwi-GA derives the stellar luminosity by using a de-reddened absolute magnitude as an anchor. That is, the stellar radius $R_*$ of the model is chosen such that the spectral energy distribution of the model matches the anchor magnitude (see \citealt{brands2022}, for details). 
In this work we use the absolute magnitude in the $K_s$ band, $M_K$, as our anchor. 
The $K_s$-band is the optimal choice for a luminosity anchor because at these wavelengths ($2.2 \mu$m) the extinction is low, while thermal radiation of dust is not yet an issue.
We obtain the $K_s$-band magnitude for each source using photometry in different bands available from literature (see \cref{sec:photometry_data}), and using this to estimate the extinction. 
We assess the reddening towards each source by applying the ``extinction without standards" technique  \citep[e.g.,][]{1966ApJ...144..305W,2005AJ....130.1127F}. For this, we adopt the best fitting \mbox{CMFGEN} models from the normalisation process of the STIS/E140M and COS/G130M + COS/G160M gratings for the intrinsic spectral energy distribution. We fit the extinction law of \citet{1999PASP..111...63F} with updated values for the spline anchor points from E. Fitzpatrick as in the Goddard IDL Astrolib routine \texttt{FM\_UNRED}\footnote{{\tiny \url{https://idlastro.gsfc.nasa.gov/ftp/pro/astro/fm_unred.pro}}}. We adopt $R_V = 3.1$ and a distance of $d = 49.59$~kpc \citep{2019Natur.567..200P}. By varying $A_V$ and the absolute flux of the adopted model, we find for each source the absolute magnitude and extinction as a function of wavelength. We obtain good SED fits for all stars: the residuals of the fits are typically $\lesssim5\%$; in a few cases the largest residuals are around $\sim10\%$. The adopted magnitudes and derived values for $A_{V}$ and $A_{K_s}$ are listed in \Cref{tab:photometry}.

\begin{table}
    \centering
    \small 
    \caption{Free parameters in the optical-only and optical + UV fits.  \label{tab:fitsetup_summary} }
    \renewcommand{\arraystretch}{1.3}
    \begin{tabular}{p{1.7cm}>{\raggedright\arraybackslash}p{4.4cm}}
    \hline \hline 
    Fit & Free parameters \\ \hline 
    Optical-only & \teff, $g$, \vsini, \mdot, \yhe, $x_{\rm C}$, $x_{\rm N}$, $x_{\rm O}$ \\ 
    Optical~+~UV & \teff, $g$, \mdot, $x_{\rm C}$, $x_{\rm N}$, $x_{\rm O}$, $\beta$,  \vinf,  \vwindturb, \fcl, \vclonset, \fic, \fvel \\ \hline 
    \end{tabular}
\end{table}

\subsection{Fitting strategy\label{sec:fitsetup}}

We analyse the full sample two times with Kiwi-GA. First, we carry out for each star a run using only the optical spectroscopy, with the goal of constraining the helium abundance 
 $y_\mathrm{He}$ ($= n_\mathrm{He}/n_\mathrm{H}$, with $n_\mathrm{He}$ and $n_\mathrm{H}$ the number density of helium and hydrogen, respectively) and projected surface rotation velocity ($\varv \sin i$). Next, we carry out for each star an optical~+~UV run, in which we constrain 13 stellar and wind parameters simultaneously, but do not leave  $y_\mathrm{He}$ and \vsini free; we fix these parameters to the values obtained in the optical only run. 
 
The reason for this two-step approach is that we do not expect that the UV spectroscopy can improve our measurements of $y_\mathrm{He}$ and \vsini. On the contrary, if we leave \vsini free when fitting UV lines, we find high values of \vsini that are clearly too high for a good fit with the optical photospheric lines. Apparently, a higher \vsini leads to better UV line fits\footnote{This is somewhat curious, as from conservation of angular momentum one would expect rotation to vary $\propto~r^{-1}$, implying that the rotation is actually slower in the region where the wind lines are formed (unless the co-rotation is forced by for example a magnetic field); with the simple convolution that we use for modelling \vsini, this `slow down' towards larger radii is not taken into account.  }. We do not consider these higher \vsini values to represent the true rotational broadening well and therefore adopt the \vsini value of the optical-only fit for the optical~+~UV runs. We also adopt the helium abundance from the optical fits as the UV spectra would only add noise to the abundance measurement: the only additional helium line in the UV, \heiiuvline, is usually not very strong, and moreover blended with iron-group lines. 

For the optical-only run we have eight free parameters: effective temperature (\teff), gravitational acceleration ($g$), mass-loss rate (\mdot), 
\vsini, $y_\mathrm{He}$, and carbon, nitrogen and oxygen abundances ($x_{j} = \log(n_{j}/n_\mathrm{H}) + 12$, with $n_\mathrm{H}$ the number density of hydrogen and $n_j$ the number density of element $j$, with $j~\epsilon~\mathrm{C,N,O}$). We do not separate \vsini and macroturbulence, so that our \vsini values are effectively upper limits.  In these runs, we adopt a fixed clumping factor $f_\mathrm{cl} = 10$, interclump density factor of $f_\mathrm{ic} = 0.1$, velocity-porosity $f_\mathrm{vel} = 0.5$, wind acceleration parameter $\beta=1.0$, and clumping onset velocity $\varv_\mathrm{cl,start} = 0.05$. For silicon we adopt an abundance of $x_\mathrm{Si} = \log(n_\mathrm{Si}/n_\mathrm{H}) + 12 = 7.06$ \citep{2022MNRAS.515.4130C}, for other elements that have a fixed abundance we adopt $0.5Z_{\odot}$, with $Z_{\odot}$ from \citet{2009ARA&A..47..481A}. 
The terminal velocity, \vinf, we fix to an estimated value that we obtain by reading off the wavelength of the blue edge of \CIVline, or in a few cases where this line is weak, from the blue edge of \nvuvline. We note that we want reasonable assumptions for the wind parameters to ensure that the wind lines fit well, but that the exact values are not important for our final results, as we only use those runs to obtain \vsini and $y_\mathrm{He}$. 
For the optical~+~UV run we fix \vsini and $y_\mathrm{He}$ to the best fit values of the optical only run, and then fit 13 free parameters: \teff, $g$, $x_\mathrm{C}$, $x_\mathrm{N}$, $x_\mathrm{O}$ and \mdot as in the optical-only run, and in addition seven more wind parameters: $\beta$,  \vinf,  \vwindturb, \fcl, \vclonset, \fic, and \fvel. After completing both runs, we thus have two sets of values for \teff, $g$, $x_\mathrm{C}$, $x_\mathrm{N}$, $x_\mathrm{O}$ and \mdot, one from the optical-only, and one from the optical~+~UV run. Unless noted explicitly otherwise, we will adopt the values of the optical~+~UV run for further analyses. The run setups are summarised in \Cref{tab:fitsetup_summary}.  The range in which each parameter was allowed to vary during the optical~+~UV Kiwi-GA run of each star is shown in the fitness distribution plots that can be found on \href{https://zenodo.org/records/15013513}{Zenodo}. 

\subsection{Diagnostic line selection \label{tab:line_selection}}

\begin{table}[]
    \centering 
    \small
    \caption{Optical diagnostics.\label{tab:optlines}}
    \begin{tabular}{>{\raggedright\arraybackslash}p{1.2cm} >{\raggedright\arraybackslash}p{1.5cm}  >{\raggedright\arraybackslash}p{4.7cm}}
        \hline \hline 
        Main ion & Diagnostic & Transitions \\ \hline 
        H~\textsc{ i} & H$\epsilon$ & H~\textsc{ i}~$\lambda$3970.1, He~\textsc{ ii}~$\lambda$3968.3,       O~\textsc{ iii}~$\lambda$3961.6\\
            & H$\delta$ & H~\textsc{ i}~$\lambda$4101.7, He~\textsc{ ii}~$\lambda$4100.0, N~\textsc{ iii}~$\lambda$4097.4, $\lambda$4103.4, Si~\textsc{ iv}~$\lambda$4088.9, $\lambda$4116.1\\
            & H$\gamma$ & H~\textsc{ i}~$\lambda$4340.4, He~\textsc{ ii}~$\lambda$4338.7\\
            & H$\beta$ & H~\textsc{ i}~$\lambda$4861.4, He~\textsc{ ii}~$\lambda$4859.3, N~\textsc{ iii}~$\lambda$4858.7, $\lambda$4859.0, $\lambda$4861.3, $\lambda$4867.1, $\lambda$4867.2, $\lambda$4873.6 \\          
            & H$\alpha$ & H~\textsc{ i}~$\lambda$6562.8, He~\textsc{ ii}~$\lambda$6559.8\\
        He~\textsc{ i} & He~\textsc{ i}~$\lambda$4026 & He~\textsc{ i}~$\lambda$4026.2, He~\textsc{ ii}~$\lambda$4025.6 \\
         &  He~\textsc{ i}~$\lambda$4387 & He~\textsc{ i}~$\lambda$4387.9, N~\textsc{ iii}~$\lambda$4379.0, $\lambda$4379.1\\ 
         & He~\textsc{ i}~$\lambda$4471 & He~\textsc{ i}~$\lambda$4471.5 \\
         & He~\textsc{ i}~$\lambda$4922 & He~\textsc{ i}~$\lambda$4921.9 \\
         & He~\textsc{ i}~$\lambda$5875 & He~\textsc{ i}~$\lambda$5875.6 \\
         & He~\textsc{ i}~$\lambda$7065 & He~\textsc{ i}~$\lambda$7065.2 \\
He~\textsc{ ii} & He~\textsc{ ii}~$\lambda$4200 & He~\textsc{ ii}~$\lambda$4200.0, N~\textsc{ iii}~$\lambda$4195.8, $\lambda$4200.1, $\lambda$4215.8 \\
         & He~\textsc{ ii}~$\lambda$4541 & He~\textsc{ ii}~$\lambda$4541.6, N~\textsc{ iii}~$\lambda$4534.6 \\ 
         & He~\textsc{ ii}~$\lambda$4686 & He~\textsc{ ii}~$\lambda$4685.7, He~\textsc{ i}~$\lambda$4713.1\\ 
         & He~\textsc{ ii}~$\lambda$5411 & He~\textsc{ ii}~$\lambda$5411.3\\ 
         & He~\textsc{ ii}~$\lambda$6406 & He~\textsc{ ii}~$\lambda$6406.4\\
         & He~\textsc{ ii}~$\lambda$6527 & He~\textsc{ ii}~$\lambda$6527.1, H~\textsc{ i}~$\lambda$6562.8\\
         & He~\textsc{ ii}~$\lambda$6683 & He~\textsc{ ii}~$\lambda$6682.8, He~\textsc{ i}~$\lambda$6678.2\\
 C~\textsc{iii}          & C~\textsc{ iii}~$\lambda$4650 & C~\textsc{ iii}~$\lambda$4647.4, $\lambda$4650.2, $\lambda$4651.5, $(\circ)$ \\
           & C~\textsc{ iii}~$\lambda$5695 & C~\textsc{ iii}~$\lambda$5695.9 \\
 C~\textsc{ iv} & C~\textsc{ iv}~$\lambda$5801 & C~\textsc{ iv}~$\lambda$5801.3, $\lambda$5812.0\\

N~\textsc{ iii}~$\dagger$   & N~\textsc{ iii}~$\lambda$4640 & N~\textsc{ iii}~$\lambda$4634.1, $\lambda$4640.6, $\lambda$4641.9, $(\circ)$\\
       N~\textsc{ iv}   & N~\textsc{ iv}~$\lambda$3480 & N~\textsc{ iv}~$\lambda$3478.7, $\lambda$3483.0, $\lambda$3485.0\\
                   & N~\textsc{ iv}~$\lambda$4058& N~\textsc{ iv}~$\lambda$4057.8 \\
                   & N~\textsc{ iv}~$\lambda$6380 & N~\textsc{ iv}~$\lambda$6380.8 \\
  N~\textsc{ v} & N~\textsc{ v}~$\lambda$4603& N~\textsc{ v}~$\lambda$4603.7\\
                  & N~\textsc{ v}~$\lambda$4620& N~\textsc{ v}~$\lambda$4620.0\\
        O~\textsc{ iii}~$\dagger$ & O~\textsc{ iii}~$\lambda$5592 & O~\textsc{ iii}~$\lambda$5592.4\\
         \hline 
        \multicolumn{3}{p{8.4cm}}{\small  $(\circ)$ The N~\textsc{ iii}~4640 and C~\textsc{ iii}~4650 complexes are typically blended.}\\
        \multicolumn{3}{p{8.4cm}}{\small $\dagger$  We stress that several transitions of N~\textsc{iii} are included as blends with H$\delta$, H$\beta$, He~\textsc{ i}~$\lambda$4387, He~\textsc{ ii}~$\lambda$4200, and He~\textsc{ ii}~$\lambda$4541, and the O~\textsc{ iii}~$\lambda$3961.6 transition is included as a blend with H$\epsilon$.}\\
        
    \end{tabular}
    
\end{table}

We include the same set of optical lines for all stars. The selection contains the Balmer lines, important for constraining surface gravity, as well as many \hei and \heii lines, important for helium abundance, temperature, and rotation constraints. The selection also includes several lines that are in most cases (partially) formed in the wind, namely \halpha, \heiiline, \niv~$\lambda$4058 and the \niii and C~\textsc{iii} triplets around $\lambda\lambda4640-4650~$\AA. Furthermore, we include as many carbon (C), nitrogen (N) and oxygen (O) lines as possible, in order to constrain the CNO-abundances. For C and N we have fair coverage, with multiple ions and at least two atmospheric lines per atom, for O we only have the relatively weak O~\textsc{iii}~5592 line, and the O~\textsc{iii}~$\lambda3962$ feature in the wing of H$\epsilon$. The optical diagnostic lines that we include in the fitting are listed in \cref{tab:optlines}. All transitions that we consider are listed per diagnostic. 
Note that while silicon is not listed explicitly in the ions used for optical diagnostics, Si~\textsc{ iv}~$\lambda4089$ and Si~\textsc{ iv}~$\lambda4116$ are included as part of H$\delta$. These transitions are especially important for the cooler stars. 

In the optical line selection three helium singlets are included (\hei~$\lambda 4387$, \hei~$\lambda 4922$, and \hei~$\lambda 6678$). These lines are known to sometimes have modelling issues related to uncertainties in the oscillator strengths associated with two Fe~\textsc{iv} transitions  \citep[see][]{2006A&A...456..659N}. In our sample we do not experience significant problems with these lines: for the cooler stars ($T_\mathrm{eff} < 40000$~K) the lines are strongly in absorption and well reproduced by the data. For the hotter stars ($T_\mathrm{eff} > 40000$~K) we do see a discrepancy between the models and the data, where in the data the lines are in absorption or not visible, and in the models the lines are in emission. However, for these hot stars these lines are extremely weak (their line centres being not deeper than 1\% of continuum, in both the models and the data) and therefore they will not affect the fit significantly. To check this we carried out additional runs without singlets for one hot star and one cool star, Farina-88 (O5 If) and Sk~$-67^\circ$5 (O9.7 Ib). These stars were selected as `extreme' cases, that is, where the fit to the \hei lines was poor compared to other stars in our sample. For both stars we find the same results for the fits with and without the singlets, with similar line profiles and stellar parameters that agree within errors. We therefore conclude that, for this particular sample, we can safely include the singlets as a diagnostic. 

The UV diagnostics that we include in the fitting are listed in \cref{tab:line_selection_UV}. Again, all transitions that we consider are listed per diagnostic. The exact line selection for each source is dictated by both the strength of the diagnostic lines relative to the iron pseudo-continuum, as well as the data availability. 
The iron pseudo-continuum that is present in the UV observations cannot (or only in a very approximate way) be modelled by \textsc{Fastwind} V10, forcing us to avoid spectral regions where the iron-group lines dominate the spectrum. 
Data availability is discussed in \cref{sec:obs_sample}; for all stars the data covers $1141~$\AA~$< \lambda < 1708$~\AA, for a subsample we have also other UV ranges. 
In practice this means that \cuvivline, \cuviiiline, \CIVline, and \oivline are included for all stars, \pvline is included always when data is available\footnote{Other diagnostics are available in the FUSE range \citep[see, e.g.,][]{2002ApJS..141..443W}, but these are not included in our fitting as they are strongly affected by X-rays,  blended with $H_2$ Lyman-Werner absorption bands, or the relevant ions are not included as explicit ions in \fastwind.}, \nivuvline and N~\textsc{iii}~$\lambda$1751 are included if they can be distinguished from the pseudo-continuum and data is available, and \heiiuvline and \ovline are included when these lines show clear wind signatures (i.e., if they are seen in emission, or have broad, blue shifted absorption). We considered to include the X-ray sensitive \nvuvline doublet as a diagnostic, but our lack of knowledge about the shock induced X-rays for these stars led us to decide against it. In \cref{sec:dis:NITRO} we discuss this in more detail.  
 Lastly, \siivline is included only when it is dominating both the iron pseudo-continuum as well as the interstellar absorption; that is, when the line has a P-Cygni profile. The run summaries (\cref{fig:examplefit:farina88} and on \href{https://zenodo.org/records/15013513}{Zenodo}) show for each star the  exact line selection.  

\begin{table}[]
    \centering 
    \small
    \caption{UV diagnostics. \label{tab:line_selection_UV}}
    \begin{tabular}{l >{\raggedright\arraybackslash}p{1.7cm}  >{\raggedright\arraybackslash}p{4.3cm}}
        \hline \hline 
        Main ion & Diagnostic & Transitions \\ \hline 
        He~\textsc{ ii} & He~\textsc{ ii}~$\lambda$1640 & He~\textsc{ ii}~$\lambda$1640.4 C~\textsc{ iii}~$\lambda$1640.0 \\
        C~\textsc{ iii} & C~\textsc{ iii}~$\lambda$1176 & C~\textsc{ iii}~$\lambda$1174.9, $\lambda$1175.3, $\lambda$1175.6, $\lambda$1175.7, $\lambda$1176.7, $\lambda$1177.0,  $(\bullet)$\\
        C~\textsc{ iv} & C~\textsc{ iv}~$\lambda$1169 & C~\textsc{ iv}~$\lambda$1168.9, $\lambda$1169.0, $(\bullet)$\\
          & C~\textsc{ iv}~$\lambda$1550 & C~\textsc{ iv}~$\lambda$1548.2, $\lambda$1550.8\\
        N~\textsc{ iii} & N~\textsc{ iii}~$\lambda$1751$\dagger\ddagger$ & N~\textsc{ iii}~$\lambda$1747.9, $\lambda$1751.2, $\lambda$1751.7 \\        
        N~\textsc{ iv} & N~\textsc{ iv}~$\lambda$1718$\dagger$ & N~\textsc{ iv}~$\lambda$1718.6 \\
        O~\textsc{ iv}  & O~\textsc{ iv}~$\lambda$1340 & O~\textsc{ iv}~$\lambda$1338.6, $\lambda$1343.0, $\lambda$1343.5 \\
        O~\textsc{ v}   & O~\textsc{ v}~$\lambda$1371$\ddagger$ & O~\textsc{ v}~$\lambda$1371.3 \\
        Si~\textsc{ iv} & Si~\textsc{ iv}~$\lambda$1400$\ddagger$ & Si~\textsc{ iv}~$\lambda$1393.8, $\lambda$1402.8 \\
        P~\textsc{ v} & P~\textsc{ v}~$\lambda$1118$\dagger$ & P~\textsc{ v}~$\lambda$1118.0, $\lambda$1128.0, Si~\textsc{ iii}~$\lambda$1113.2, Si~\textsc{ iv}~$\lambda$1128.3\\ 
        \hline 
        \multicolumn{3}{p{8.1cm}}{\small $\dagger$ Covered in our data only for part of the sample. } \\
        \multicolumn{3}{p{8.1cm}}{\small $\ddagger$ For several sources the line was weaker or at similar strength as the iron-group lines in this part of the spectrum, and the line was excluded. } \\
        \multicolumn{3}{p{8.1cm}}{\small $(\bullet)$ C~\textsc{ iii}~1176 and C~\textsc{ iv}~1169 are blended for stars with strong winds.} \\
    \end{tabular}

\end{table}

\subsection{Derived parameters}

Apart from the free fitting parameters we derive several quantities from the best fit parameters of the optical~+~UV fits, including 
the spectroscopic mass $M_\mathrm{spec}$, the Eddington factor for electron scattering $\Gamma_{\mathrm{Edd},e}$, and H and \hei and ionising fluxes $Q_0$ and $Q_1$\footnote{Here, by convention, $Q_x = q_x 4 \pi R_*^2$, with $q_x$ the ionising radiation (number of photons) per unit surface area per second and $x \in \{0,1\}$.}. 
Furthermore, we derive the initial mass $M_\mathrm{ini}$, the current evolutionary mass $M_\mathrm{evol}$, and the age $\tau$ using the \textsc{Bonnsai} tool\footnote{The BONNSAI web-service is available at \url{https://www.astro.uni-bonn.de/stars/bonnsai/}.} \citep{2014A&A...570A..66S,2017A&A...598A..60S},  in combination with the grids of \citet{2011A&A...530A.115B} and \citet{kohler2015}. \textsc{Bonnsai} is a Bayesian framework that allows us to compare observed stellar parameters to stellar evolution models in order to infer full posterior distributions of model parameters. Our input parameters are luminosity, temperature, and an upper limit on the value we derive for \vsini. We use default settings, with the exception of the prior for the initial rotational velocity, for which we assume the distribution of \citet{2013A&A...560A..29R} instead of a flat distribution (\Cref{tab:stellarparams}). 

\section{Results \label{ullyses:results}}

With \kiwiGA we obtain stellar and, in most cases, wind parameters for all single stars in the sample, as well as for all (suspected) binaries, except for BI~272 (see below). 
The best fit parameters and associated $1 \sigma$ uncertainties can be found in \Cref{tab:stellarparams} and \ref{tab:bestfitswind}. These tables include the helium abundance and projected surface rotation as from the optical-only fits\footnote{The complete set of optical-only best fit values can be found on \href{https://zenodo.org/records/15013513}{Zenodo}. }; all other fit parameters come from the optical~+~UV fits, or are derived from the optical~+~UV fits\footnote{For LH~114-7 we were not able to derive evolutionary parameters. Likely this star is evolved beyond the main sequence (see \cref{sec:HRD}) and therefore its parameters are not covered by the grids of \citet{2011A&A...530A.115B} and \citet{kohler2015}. }. For some stars we were not able to constrain one or more clumping parameters; while \kiwiGA does output best fit values (and these values are used for obtaining the best fit spectrum), the $2\sigma$ uncertainties on the derived values of these sources are so large that they span the full parameter space. Therefore, the actual values are meaningless and so they are not included in the analysis of the wind structure parameters (\cref{section:wind_structure}). However, these stars are included in the mass-loss rate analysis (\cref{section:masslossrates}) as their inferred mass-loss rates are not affected: the uncertainty on the unconstrained wind structure parameters is captured in the uncertainty of the mass-loss rates. For completeness and reproducibility purposes all clumping values are listed in \Cref{tab:bestfitswind}; in case a value was essentially unconstrained the value is displayed in brackets. 

For one star, BI~272, we could not obtain a satisfying fit, it appears that the spectrum of this source contains of two components of similar strength: a single broad component is visible in He~\textsc{ii}~4541 and two components for He~\textsc{i}~4471 (one broad, one narrow). This is confirmed by a higher resolution Magellan MIKE spectrum; the source appears to be a mid O-type star (He~\textsc{ii}, Si~\textsc{iv}, part of He~\textsc{i}) + early B-type star (remainder of He~\textsc{i}, Si~\textsc{iii}, Mg~\textsc{ii}; \citealt{Bestenlehnerinprep}).
We do  not present our best fit parameters for this source. 
In the spectra of the other binaries one component clearly dominates the spectrum, allowing us to obtain a good fit to at least a subset of lines. 
For example, for VFTS-267 we obtain good fits with all lines except for several \hei lines, which seem to have a contribution from a cooler secondary star. 
Therefore, we are confident that stellar parameters, such as the effective temperature, that we find from this fit, provide a reasonable estimate of the parameters of the primary (dominating) component, although the formal uncertainties presented are likely underestimated. 
We include these binaries in several plots concerning the bulk properties of our sample stars, but in all cases mark them clearly, so that they can be distinguished from the presumably single stars. 
We are more cautious with the wind properties we derive for these stars, in particular the wind-structure parameters. As these often have subtle effects on the lines, we deem the wind-structure values we derive for the binaries less reliable, and will not consider them in our wind-structure analysis. 

\begin{figure*}
    \centering
    \includegraphics[width=0.92\textwidth]{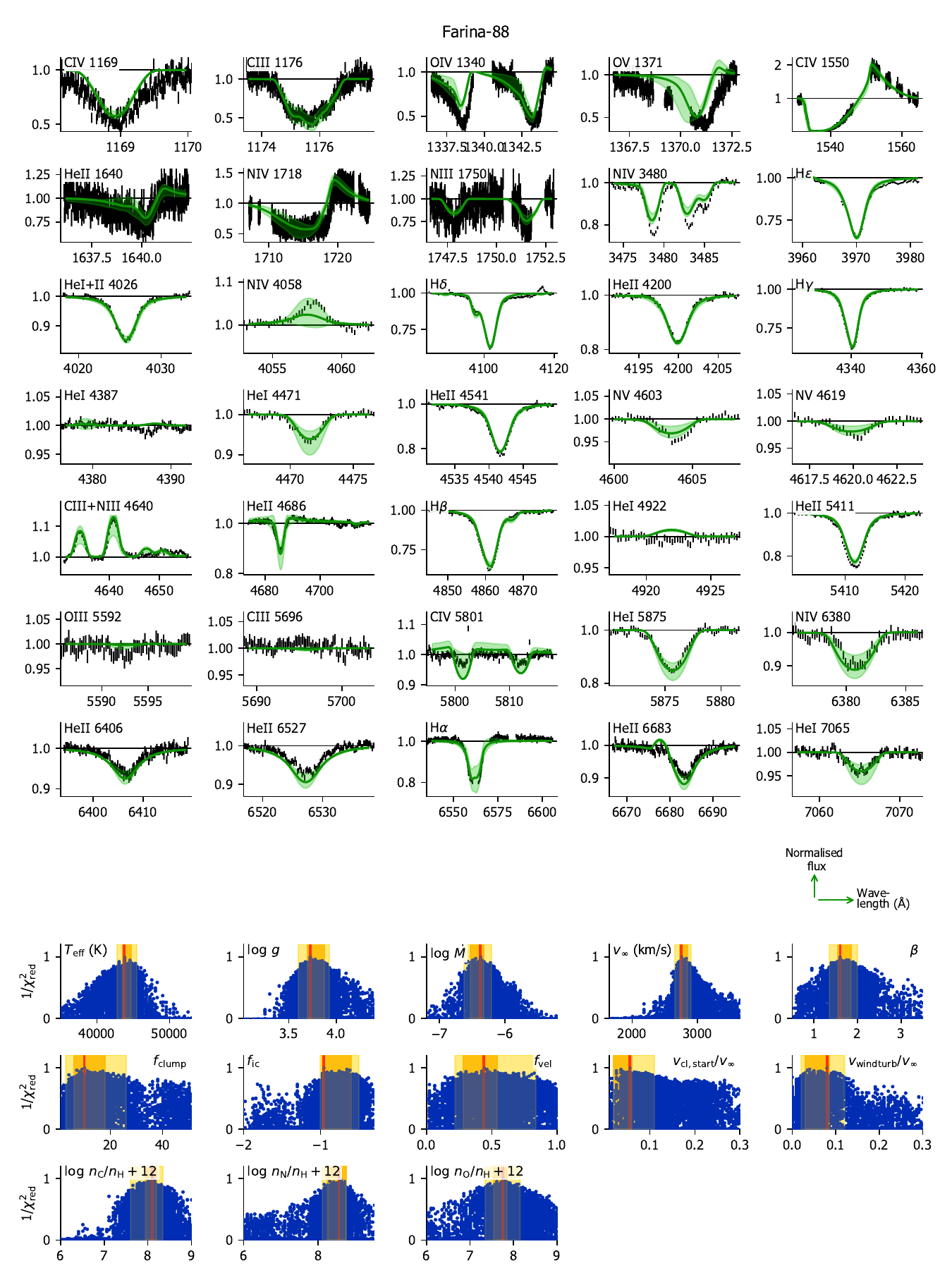}
    \caption{Output summary of the optical and UV Kiwi-GA  run of Farina-88 (O4 III(f)). The top part of the figure shows all used diagnostics: the observed spectra (black; vertical bars show uncertainty on each observed flux), the best fit model (dark green solid line), and the uncertainty region (light green shaded area; this area covers all model spectra of which the parameters lie within the $2\sigma$ uncertainties). The bottom part of the figure shows the fitness distribution (blue dots) of all parameters that were fitted in the optical and UV run; red vertical lines indicate the best fit value, the shaded regions indicate $1\sigma$ (orange) and $2\sigma$ (yellow) uncertainty margins.  Output summaries of the other stars can be found on \href{https://zenodo.org/records/15013513}{Zenodo}. 
    \label{fig:examplefit:farina88}}
\end{figure*}

A comparison of observed spectra and best fit models for all stars is presented in 
\cref{fig:windlines}, which includes the UV diagnostics and the main wind-sensitive optical lines (the \ciii-\niii complex at 4640, \heiiline and \halpha), and \cref{fig:otherlines}, showing the remainder of the optical lines. In these figures the stars are ordered by temperature; we remark in this context that the temperatures we derive are generally in good agreement with the adopted spectral types (\cref{app:spectraltype}), as well as with literature values for stars that were previously analysed (\cref{app:literature}). 
We are aware that the plotted spectra in Figs. \ref{fig:windlines} and \ref{fig:otherlines} are rather small; this was done to ensure that the spectra of all stars can be viewed simultaneously, allowing us to see trends as a function of temperature and to assess the general fit quality of individual lines. 
A more detailed display per star can be found on  \href{https://zenodo.org/records/15013513}{Zenodo}; the figures presented there consist of larger plots of each diagnostic that are clearly labelled. Furthermore, fitness distribution diagrams for all free parameters are included. An example of such an overview is shown in \cref{fig:examplefit:farina88} for the star Farina-88. We briefly discuss the fits, per line.

\paragraph{\pvline}
This line displays a broad P-Cygni profile for the hotter stars (\teff~$\gtrsim 35000$~K)\footnote{With the exception of W61-28-23.}, while for the cooler stars the P-Cygni profile is narrower or the line is simply in absorption. 
The strength and shape of \pvline is matched by the models in most cases. 
In the fitting process we adopted for all stars a fixed phosphorus abundance of half Solar \citep{2009ARA&A..47..481A}, given the metallicity of the LMC (i.e., $\log n_\mathrm{P}/n_\mathrm{H} + 12 = 5.12$). This is consistent with limits from the analysis of the interstellar P~\textsc{ii} lines \citep{2015ApJ...811...78T}, and does not seem to give problems for the fitting. Only for N11-018 and Sk~$-68^\circ$155 (binary) the model predicts lines that are clearly too strong.

\paragraph{\cuvivline-\cuviiiline} 
This group of lines is well reproduced for cooler stars (\teff~$\lesssim 40000$~K), where it displays a P-Cygni shape. For the hotter stars, the \cuvivline component is underpredicted. Possibly, the observed spectra show not only the \cuvivline line here, but also components of iron, which are not included in our models. 

\paragraph{\oivline and \ovline} The \oivline complex is only mildly affected by the wind for most stars; this line shows the strongest wind signature in the spectrum of W61-28-23. In many cases the absorption in the models of this line is too weak. We tested whether our adopted value of  \vmicro could be the reason for the poor fit by redoing the fits of three stars with \vmicro kept free. While the best fit values obtained for  \vmicro were slightly higher than the value adopted by us (17-21~km~s$^{-1}$ instead of 15~km~s$^{-1}$) the fits to \oivline did not improve, and also the inferred oxygen abundance remained unaffected (see also \cref{sec:dis:CO}). The other oxygen line, \ovline, was included only when it was prominently visible, which is the case for hotter stars ($T_\mathrm{eff}\gtrsim 40000$~K). Generally, it is somewhat too weak in the models; the cause of this is unidentified\footnote{\citet{brands2022} notice something similar in the fits of their WNh stars, and name possible causes in their Appendix E.1.}. 

\paragraph{\siivline} This doublet was only included in the fitting when it displays a P-Cygni profile, which is typically the case for stars with \teff~$\lesssim 40000$~K. The line is strongest for stars with lower temperatures (\teff~$\lesssim 36000$~K) and is generally well reproduced; only the shape of the $\lambda1402$ absorption component is sometimes too `round' compared to the observations, which show a more `linear' decrease of flux in the absorption trough. For the hottest stars, \siivline is weak and it is hard too distinguish iron components from absorption caused by the \siivline line. Nonetheless, the strength of the model spectra is in line with the observations. 

\paragraph{\CIVline} This line displays a strong, saturated P-Cygni profile for nearly all stars in the sample, and is reproduced well by the models in all cases. Even for LH~114-7, which is likely a binary and has an unusually shaped \CIVline, the single star models find a relatively good fit. Clearly visible in \cref{fig:windlines} is the trend in decreasing terminal velocity as a function of temperature. This is further discussed in \cref{sec:ull:terminalvelocity}.

\paragraph{\heiiuvline} This line was only fitted for the hotter stars, where it shows a (sometimes very broad) P-Cygni profile. It is well reproduced by the models. 

\paragraph{\nivuvline and \niiiuvline} We see something similar as for \siivline: the strength is generally well reproduced by the models, but there is a mismatch in the exact shape. This line displays a strong and broad P-Cygni profile for the earlier type stars (\teff~$\gtrsim 35000$~K). For the later type stars the line is hard to distinguish from iron-group lines present at similar wavelengths. The nearby \niiiuvline lines are visible at all temperatures in the range that we consider, but do not show a clear wind signature. For the cooler stars, the models typically underpredict the strength of these lines somewhat.  

\paragraph{Optical wind lines}
The main wind diagnostics in the optical are the \niii-\ciii complex around $\lambda$4640, \heiiline, and \halpha. The \niii-triplet is in emission for most stars, and with a few exceptions well reproduced. The \ciii lines are in weak emission or absent, or in strong absorption for the cooler stars ($\lesssim 32000$~K). For \heiiline we see a wide variety of shapes, usually a combination of emission and absorption -- with the exception of Sk~$-67^\circ167$, where the line is broad and strongly in emission. It is challenging to reproduce this line: though generally the strength of the line is matched by the models, in about half of the single star fits the exact shape cannot be reproduced. We do not know the cause of this, but note that it is mostly the strength of the line that is of importance for $\dot{M}$ determinations rather than the exact shape. 
For \halpha, which is (partially) in emission for almost all stars, the models fit better, reproducing strength and shape well in nearly all cases. 
In \cref{fig:otherlines} we find several more wind lines: \nivoptline, \ciiivoptline, and \civvoptline. \nivoptline is in moderate emission for the hottest stars ($\gtrsim 40000$~K), and in strong emission for the very hot (binary) star LH~114-7. It is challenging to reproduce this line. For cooler stars the line is in weak absorption or absent. \ciiivoptline appears in emission for the cooler stars ($\lesssim 38000$~K), whereas \civvoptline is in emission for the hottest stars ($\gtrsim 38000$~K). 

\paragraph{Optical photospheric lines}
The photospheric lines in the optical (\cref{fig:otherlines}) are generally well reproduced with a few exceptions. The Si~\textsc{iv} lines in the wings of H$\delta$ are problematic for a significant number of stars, being either strong in absorption in the observations, but too weak in the models (for stars with \teff~$\lesssim 32000$~K) or being in emission, while the models show a flat profile (for stars with \teff~$\gtrsim 37000$~K). The \niv lines around $\lambda$3480 are too weak in the models for the hottest stars, but fit well at lower temperatures (\teff~$\lesssim 38000$~K). \heii~$\lambda5411$ is in most cases slightly too weak in the models; 
the same is the case for \hei~$\lambda5875$ in the cooler stars. For \heii~$\lambda5411$, a possible cause can be contamination by diffuse interstellar bands, which are not included in our models. Also the strength of \hei~$\lambda4471$ is a bit underpredicted for the cooler stars. The \hei singlets (\hei~$\lambda 4387$, \hei~$\lambda 4922$, and \hei~$\lambda 6678$) are generally reproduced well; the largest deviations between models and observations appear at \teff~$\approx34000-35000$~K for \hei~$\lambda 6678$, which is modelled slightly too deep in this regime. 
Lastly, we remark that upon comparing the fitness of the helium lines of the optical-only, and the combined optical+UV fit, their fits are typically equally good; only 
for Sk~$-67^{\circ}$167 we see that the fit gets worse when including the UV but keeping the helium abundance fixed at the optical-only derived value \footnote{For this star \hei+\heii~$\lambda 4026$, \hei~$\lambda4471$, \hei~$\lambda5875$, and \hei~$\lambda7065$ fit significantly better in the optical-only analysis.}. 

\section{Discussion \label{ullyses:discussion}}

\subsection{ Hertzsprung-Russell diagram \label{sec:HRD}}

\begin{figure}
    \centering
    \includegraphics[width=0.49\textwidth]{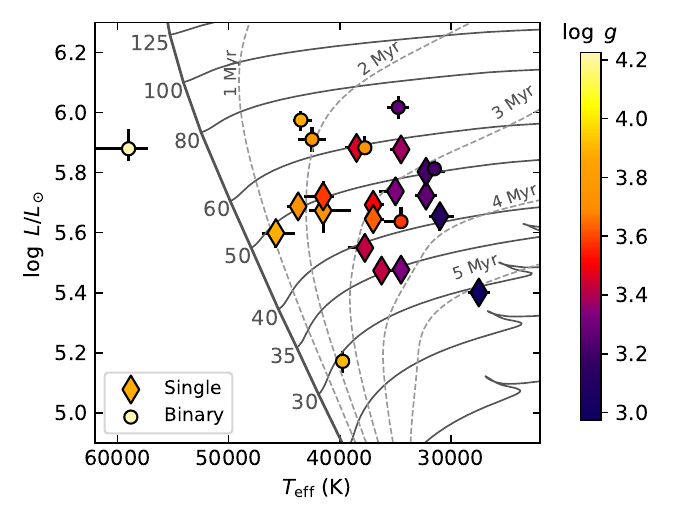}
    \caption{Our sample stars in the HRD. Presumed single stars are indicated with diamonds; suspected binaries with circles. The colour of each marker corresponds to its surface gravity, where lighter colours correspond to higher values. As expected, a trend in surface gravity is seen as a function of temperature and age, which are correlated: more evolved stars are cooler and have lower surface gravity, implying a larger radius.   
    In the background, evolutionary tracks (solid lines) and isochrones (dashed lines) of the grids of \citet{2011A&A...530A.115B} and \citet{kohler2015} are shown (initial rotation rates: $170-200$~km~s$^{-1}$). The vertical solid line indicates the position of the ZAMS; numbers on the left of the ZAMS refer to the initial mass of each model. The figure shows that our sample stars have an initial mass in the range $\approx 25-75~$\Msun, and an age of $1-5$~Myr. One source, LH~114-7, lies on the blue side of the ZAMS; we cannot estimate its age and mass from this HRD (but see \cref{fig:lh114_hrd}). 
    \label{fig:hrd_logg}}
\end{figure}

\begin{figure}
    \centering
    \includegraphics[width=0.49\textwidth]{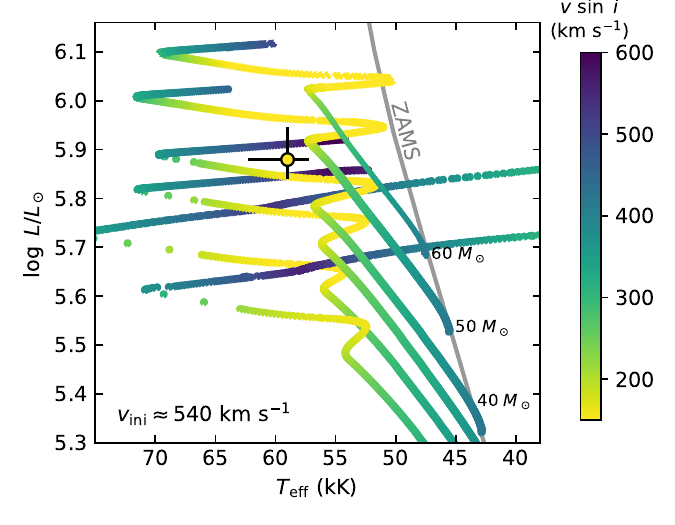}
    \caption{The HRD position of LH~114-7 compared to evolutionary tracks of the grid of \citet{2011A&A...530A.115B} with an initial rotation of $\approx 540$~km~s$^{-1}$\label{fig:lh114_hrd}. Initial masses are indicated next to the ZAMS (grey solid line); the ZAMS of the track with the lower luminosity tracks fall off the scale of this plot; the initial masses of these models are, in order of decreasing luminosity, 35, 30 and 25~$\mathrm{M}_\odot$. Colour coding indicates $\varv\ \sin \ i$, both for the tracks and the observation. The surface rotational velocities of the tracks have been multiplied by $\pi/4$, to obtain an (average) indication for $\varv\ \sin \ i$. We find the best match of the observed \teff, $L$ and $\varv\ \sin \ i$ of LH~114-7 to be with the $M_\mathrm{ini} = 40$~\Msun track (but see text). \label{fig:hrd_lh1147}}
\end{figure}

\Cref{fig:hrd_logg} shows our sample stars in the Hertzsprung-Russell diagram (HRD). All sources have moved away from the zero-age main-sequence (ZAMS); at an age of $1-5$~Myr, the stars are, on average, at 40-60\% of their main sequence lifetime. Effective temperature and surface gravity are correlated as can be seen from the colour gradient, implying furthermore that surface gravity is correlated with age, given that stars move to lower temperatures during the main sequence. 

One source, LH~114-7 (O2 III(f*)), is very hot (\teff~$\approx59000$~K) and lies on the blue side of the ZAMS. 
With \textsc{Bonnsai} we did not obtain evolutionary parameters for this star. While we can reproduce the combination of observed \teff, $L$, and $\varv~\sin~i$ well with the \citet{2011A&A...530A.115B} track of $M_\mathrm{ini} = 40$~\Msun and $\varv_\mathrm{ini} = 538$~km~s$^{-1}$ (see \cref{fig:hrd_lh1147}), the observed helium abundance and surface gravity do not match. However, 
we suspect that this star is a binary given its unusual combination of spectral lines: with this high temperature we obtain a good fit for \heii and the highly ionised metal lines, but \hei lines, absent at such high temperatures, are clearly visible in the observed spectrum. Possibly, this would affect our inference of the surface properties. 
Assuming that both \ovline and \CIVline are saturated for one of the stars and absent for the other, we estimate the light ratio of the two stars to be approximately 1:1, which would translate in a luminosity of \logll~$\approx~5.6$ for either star, positioning the star near the $M_\mathrm{ini} = 25$~\Msun track in \cref{fig:hrd_lh1147}, assuming that the temperature we inferred is approximately correct.  
Although beyond the scope of this paper, it might be interesting to investigate further the evolutionary history of this system.

\subsection{Mass-loss rates\label{section:masslossrates}}

\begin{figure}[h]
    \centering
    \includegraphics[width=0.49\textwidth]{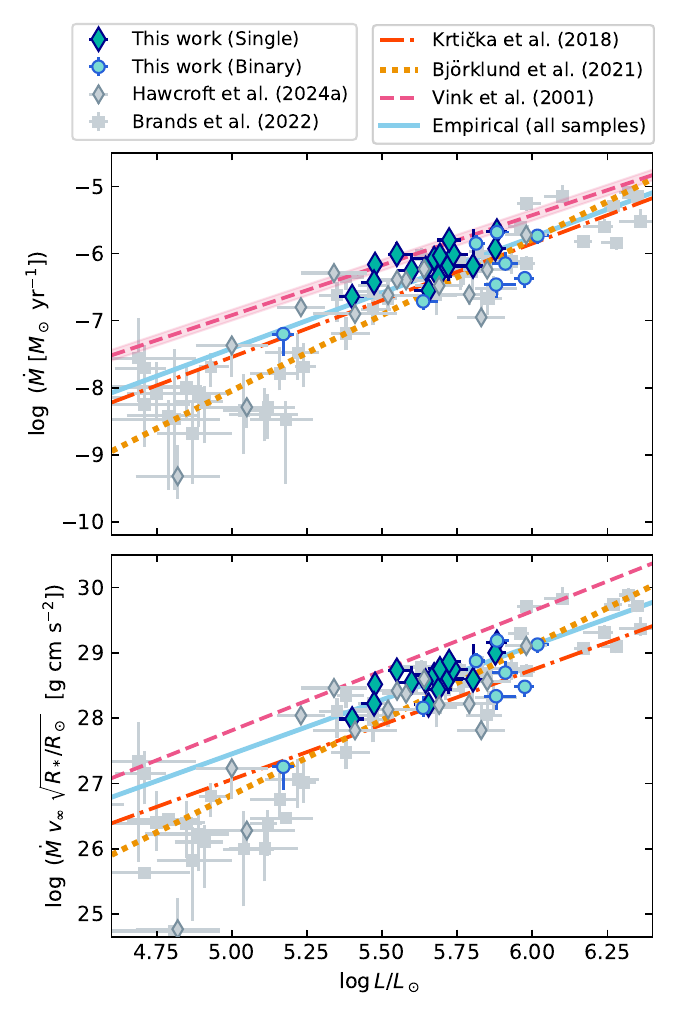}
    \caption{Mass-loss rates (top panel) and modified wind momentum (bottom panel) derived from the optical+UV analysis of the single stars (green diamonds) and binaries (blue circles), compared to the mass-loss predictions of \citet[][pink dashed lines]{2001AA...369..574V}, \citet[][dark orange dashed-dotted lines]{2018AA...612A..20K}, and \citet[][light orange dotted lines]{2021AA...648A..36B}. The pink shaded region around the \citet[][]{2001AA...369..574V} relation in the top panel indicates spread of the predicted rates due to the inclusion of stellar parameters other than luminosity (\teff and $M_\mathrm{evol}$). 
    In grey we show the LMC observations of \citet[][]{brands2022} and \citet[][]{2024AA...690A.126H}. 
    Most stars of our sample have mass-loss rates that are lower than the predictions of  \citet[][]{2001AA...369..574V}, and higher than those of \citet[][]{2018AA...612A..20K} and \citet[][]{2021AA...648A..36B}. The light blue solid lines shows the empirical relation obtained by fitting a linear relation through all points of the three  samples combined; for the mass-loss rate, it shows a good agreement with the relation of \citet{2018AA...612A..20K}, having nearly the same slope and an average offset of only $\sim0.1$~dex. 
    \label{fig:mdot_vs_theory}}
\end{figure}

\begin{figure}[h]
    \centering
    \includegraphics[width=0.49\textwidth]{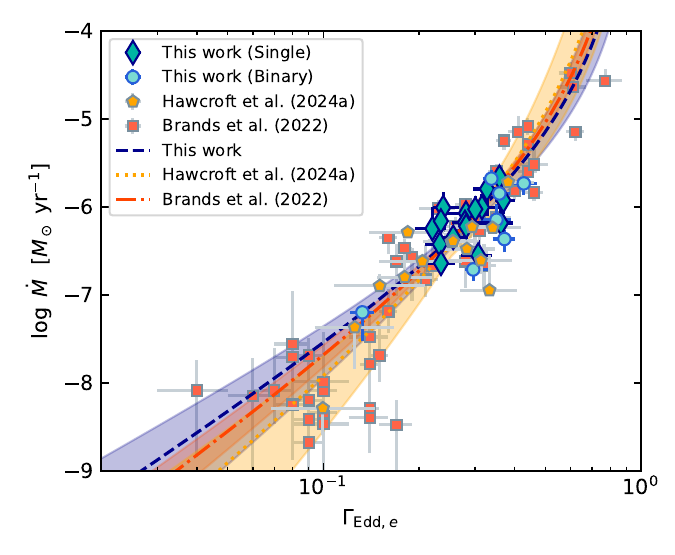}
    \caption{Mass-loss rate versus Eddington factor for electron scattering for three samples of LMC O-type stars: \citet[][red squares]{brands2022} \citet[][yellow pentagons]{2024AA...690A.126H}, and the sample presented in this work (green diamonds for the presumed single stars, and blue circles for the binaries). To each sample we fitted the mass-loss prescription of \citet[][their equation (12); see \Cref{tab:cak} for our best fit values]{bestenlehner2020_CAK}. Within uncertainties the fits to the three different samples are in agreement. }
    \label{fig:gamma_mdot}
\end{figure}

In this section we discuss the observed mass-loss rates in the context of theoretical predictions and other observational studies of LMC stars that use the same methodology for the spectroscopic analysis. We do not address the metallicity dependence of mass loss; this topic is discussed in detail in \citet{Backsinprep}, who include the results of this study in their analysis. 

\Cref{fig:mdot_vs_theory} shows the observed mass-loss rates and modified wind momentum ($D_\mathrm{mom} = \log (\dot{M} \varv_{\infty} \sqrt{R_*/R_\odot}$)) as a function of luminosity, as well as theoretical predictions of \citet{2001AA...369..574V}, \citet{2018AA...612A..20K} and \citet{2021AA...648A..36B}\footnote{\label{bjorklundtypo}We note that Eqs. (19) and (20) of \citet{2021AA...648A..36B} contain a typo: the double minus sign in the second terms of these equations should be read as a single minus sign (private communication with R. Bj\"{o}rklund). In the \href{https://arxiv.org/abs/2008.06066}{arXiv pre-print} the equations are correct. }$^,$\footnote{The relation for wind momentum of \citet[][their Eq. 15]{2000A&A...362..295V} is semi-empirical as these authors do not predict terminal velocities but use observed values for \vinf and $R_*$. The relation plotted in \cref{fig:mdot_vs_theory} is corrected for lowered \mdot and \vinf as a result of the lower metallicity in the LMC compared to the Milky Way (\citealt{1992ApJ...401..596L}). 
 \citet{2018AA...612A..20K} do not provide an explicit relation for the wind momentum; for \cref{fig:mdot_vs_theory} we computed $D_{\rm mom}$ for all the models in their LMC grid, and obtain a linear fit through these points as a function of luminosity. \citet{2021AA...648A..36B} provide a metallicity dependent relation for the wind momentum. }. 
In the background we show the LMC mass-loss rates and wind momentum of \citet[][stars in R136, 30 Doradus]{brands2022} and \citet[][stars in other regions of 30 Doradus]{2024AA...690A.126H}, which are derived using the same method\footnote{All analyses concern the simultaneous fitting of optical and UV spectra using \fastwind and a genetic algorithm, adopting the clumping prescription of \citet{2018A&A...619A..59S}.}. 
We carry out orthogonal distance regression (ODR) fits through the observations of all three LMC samples (\citealt{brands2022}, \citealt{2024AA...690A.126H}, and this work, as shown in \cref{fig:mdot_vs_theory}) and derive the following empirical relations:
\begin{equation}
\label{eq:meanmdot}
    \log \dot{M} \ [M_\odot \mathrm{yr}^{-1}] = (-15.70\pm 0.46) + (1.65\pm 0.08) \log L/L_\odot
\end{equation}
and 
\begin{equation}
    \log D_\mathrm{mom} \ [\mathrm{g \ cm \ s^{-2}}] = (17.48 \pm 0.53) + (1.94 \pm 0.09) \log L/L_\odot,
\end{equation}
for LMC mass loss and modified wind momentum, respectively. 
We stress that all mass-loss rates presented here, and thus the above equations, rely on the simultaneous fit of the wind-structure (clumping) parameters, and therefore need not to be `corrected' for clumping, e.g., scaled by a factor $1/\sqrt{f_\mathrm{cl}}$ as one could do when working with models with a smooth wind. 

The observed mass-loss rates of our sample stars show a spread of  $\approx 0.5$~dex for a given luminosity (\cref{fig:mdot_vs_theory}). The origin of this scatter lies partially in statistical uncertainties, which for the mass-loss rate typically are $0.1-0.2$~dex (\cref{tab:bestfitswind}). 
We do not know the origin of the remainder of the scatter. A mass dependence does not seem to be the cause, as the scatter is similarly present when we look at \logll versus modified wind momentum (\cref{fig:mdot_vs_theory}), or at $ \Gamma_{\mathrm{e}}$ versus \mdot (\cref{fig:gamma_mdot}). Furthermore, we find no correlation between \teff and the residuals of the best empirical fit (light blue solid line in \cref{fig:mdot_vs_theory}), implying that also a temperature dependence cannot explain the scatter. 
Likely, the the scatter is caused by a physical process that we do not consider in our analysis, or that we do not consider correctly. The treatment of the wind structure is a candidate for such a process; we discuss this further in \cref{section:wind_structure}. 

\begin{table}[]
    \centering
    \small
     \caption{Best fit values of the mass-loss prescription of \citet{bestenlehner2020_CAK} fitted to three different samples. \label{tab:cak}}
    \begin{tabular}{l l l}
    \hline \hline
    \\[-10.0pt]
         &  $\log \dot{M}_{e,\mathrm{trans}}$ & $\alpha_\textrm{eff}$ \\
         \hline
        This work               & $-5.29\pm0.19$ & $0.53\pm0.13$\\
        \citet{2024AA...690A.126H} & $-5.07\pm0.39$ & $0.40\pm0.12$\\
        \citet{brands2022}      & $-5.19\pm0.09$ & $0.46\pm0.04$\\
       All samples combined & $-5.19\pm 0.07$ & $0.47 \pm 0.03$ \\
    \hline
    \end{tabular}
   
\end{table}

\begin{table}[]
    \centering
    \small 
    \caption{\label{tab:quantify_mdot_dmom_fits} Mean ratios of theoretical (subscript `t') to observed (subscript `o') mass-loss rates and modified wind momenta. }
    \begin{tabular}{l c c}
      \hline \hline 
     \\[-10.0pt]
       & $\langle\dot{M}_\mathrm{t}/\dot{M}_\mathrm{o}\rangle$ & $\langle D_\mathrm{mom,t}/D_\mathrm{mom,o}\rangle$ \\ 
      \hline 
\citet{2001AA...369..574V} & 2.47 & 4.13 \\
\citet{2018AA...612A..20K} &0.87 & 0.57 \\
\citet{2021AA...648A..36B} & 0.76 & 0.95\\ 
     \hline 
    \end{tabular}

\end{table}

Comparing our mass-loss rates and wind momentum to theoretical predictions\footnote{See \citet{brands2022} for a summary of the differences and similarities between the theoretical prescriptions considered here.}, we find that \citet{2001AA...369..574V} overestimate both the mass-loss rate and wind momentum, while \citet[][]{2018AA...612A..20K} and \citet{2021AA...648A..36B} underestimate these quantities. The ratio between the observed and theoretical quantities for the stars in our sample is given in \Cref{tab:quantify_mdot_dmom_fits}. The ratios show that the predictions of \citet[][]{2018AA...612A..20K} best match the observed mass-loss rates, while those of \citet{2021AA...648A..36B} provide only a slightly poorer fit. The observed modified wind momenta are best reproduced by the prescription of \citet{2021AA...648A..36B}. 

Upon comparing the observed rates to those of \citet{brands2022} and \citet{2024AA...690A.126H}, we find for our single stars rates that they are on average 0.3~dex higher in the luminosity range where these samples overlap; the rates of these authors agree well with \citet[][]{2018AA...612A..20K} and \citet{2021AA...648A..36B} in this luminosity range. The reason for the discrepancy between the different samples seems to be related to the typical stellar masses of the different samples. Upon comparing the mass-loss rates of the samples as a function of $\Gamma_{\mathrm{Edd},e}$ rather than \logll, we find a good agreement between the samples; see \cref{fig:gamma_mdot}. The figure includes a fit to the mass-loss prescription of \citet{bestenlehner2020_CAK} for each of the samples. \citet{bestenlehner2020_CAK} describes mass loss as a function of $\Gamma_{\mathrm{Edd},e}$, the transition mass-loss rate $\dot{M}_{e, \mathrm{trans}}$, and the force multiplier $\alpha_\mathrm{eff}$\footnote{\citet{brands2022} include a brief explanation of \citet{bestenlehner2020_CAK} and the physical meaning of $\alpha_\mathrm{eff}$.}. 
In the  $\Gamma_{\mathrm{Edd},e}$-range where our samples overlap, the best fits are nearly identical. At low luminosities the three fits differ, but for both our sample as well as that of \citet{2024AA...690A.126H} in this $\Gamma_{\mathrm{Edd},e}$-range the best fit is an extrapolation rather than a true fit due to the lack of data. The best fit values for $\dot{M}_{e, \mathrm{trans}}$ and  $\alpha_\mathrm{eff}$, for each individual sample as well as those of all samples combined, can be found in \cref{tab:cak}. 

\subsection{Wind structure parameters\label{section:wind_structure}}

\begin{figure}
    \centering
    \includegraphics[width=0.47\textwidth]{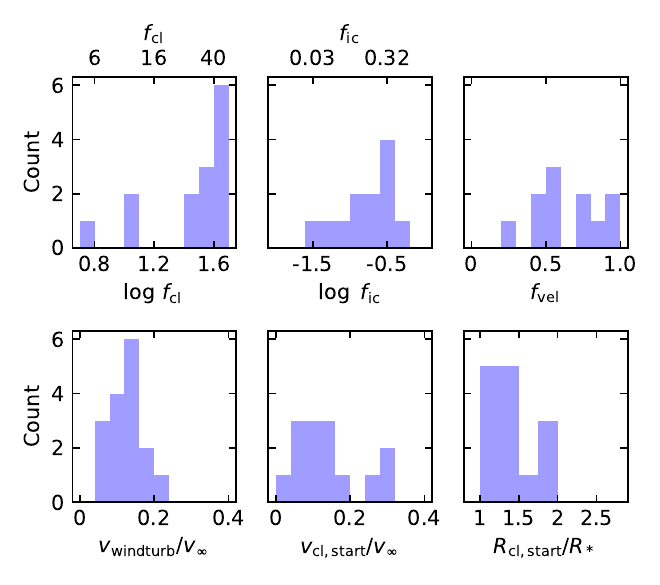}
    \caption{Distribution of observed values of wind parameters for the presumed single stars in our sample for which we obtain good fits (see text). From left to right, top to bottom, we show the clumping factor, the interclump density factor, the velocity-porosity, the wind turbulence, and the onset velocity and radius of clumping. Each distribution peaks around a certain value, except for the distribution of the velocity-porosity, which is more or less flat for $f_\mathrm{vel}\gtrsim0.4$. See also \cref{tab:typical_windstructure}. 
    \label{fig:windpar_dist}}
\end{figure} 

Assuming that the winds of all our stars are governed by the same wind physics, and given the inferred uncertainties on our measurements, we hypothesise to see a certain degree of structure in our results. Either, we would expect that the parameter space is only partially occupied (i.e., all parameters have a certain value), or else we would expect to find trends as a function of stellar (\teff,  $L$) or wind ($\dot{M}$) parameters. 
In this section we investigate trends in the wind-structure properties derived for the (presumed) single stars of our sample. 

\Cref{fig:windpar_dist} shows the distributions of the best fit values of wind-structure parameters, and weighted mean values are listed in \Cref{tab:typical_windstructure}. 
The distribution of the parameters cluster around certain values: clumping factors at $f_\mathrm{cl} \approx 25-50$; the interclump density at $f_\mathrm{ic} \approx 0.1-0.3$; the onset velocity of clumping and the wind turbulence cluster around $0.05-0.15~\varv_\infty$. We find the flattest distribution for the velocity-porosity: the values range from $0.4-1.0$, with the exception of one star (Sk $-69^\circ$104\footnote{This star is the only presumed single star that does not follow the relation between terminal velocity and temperature of \citet{2024A&A...688A.105H}; see \cref{sec:ull:terminalvelocity}. }) for which we find $f_\mathrm{vel} = 0.28\pm0.23$. 
It thus appears that the winds of O-type giants, bright giants and supergiants are highly clumped, with moderate to strong velocity-porosity effects. 
A mean interclump density factor of $f_\mathrm{ic} = 0.2$ and a clumping factor of $f_\mathrm{cl} = 33$ implies that on average (for this sample) the density of the clumps is way higher than that of the surrounding medium ($\rho_\mathrm{cl}/\rho_\mathrm{ic} \approx 200$; Eqs. (4), (8) and (17) of \citealt{2018A&A...619A..59S}), but at the same time the clumps take up only a fraction of the space  ($f_\mathrm{vol} \approx 0.03$; Eq. (17) of \citealt{2018A&A...619A..59S}), such that $20\%$ of the wind mass is contained in the interclump medium ($(1-f_\mathrm{vol})
f_\mathrm{ic} = 0.2$). 

The best fit values for the clumping onset peak around $0.05-0.15\varv_\infty$;  expressed in stellar radii this is around $1R_*-1.5R_*$. This is higher than what is found by \citet{2006A&A...454..625P}, who study the radial stratification of clumping in O-stars and infer a typical onset of clumping at $1.05R_*$. \citet{2013A&A...559A.130S} find that a non-void interclump medium can hide the signature of the onset of clumping; \citet{2006A&A...454..625P} assume a void interclump medium, while we assume that it is non-void. However, we do not find such degeneracy from our correlation analysis (\cref{sect:windcorr}), implying that this does not explain the differences between our findings and those of \citet{2006A&A...454..625P}. 

\begin{figure*}
    \centering
    \includegraphics[width=1.0\textwidth]{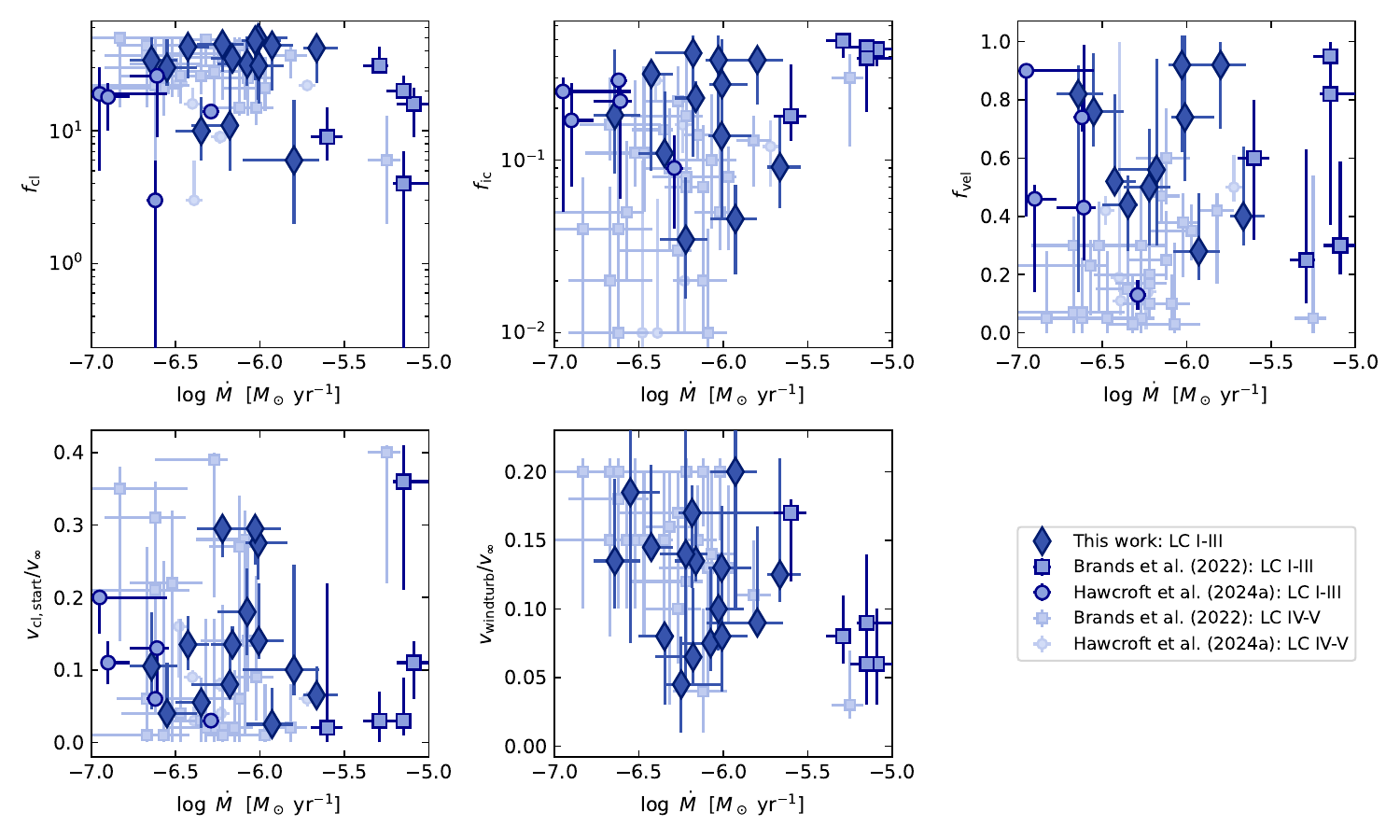}
    \caption{Wind structure parameters versus mass-loss rate. Diamonds correspond to stars of our sample. We also show the samples of \citet[][squares]{brands2022} and \citet[][circles]{2024AA...690A.126H}. The luminosity classes of the latter two samples are indicated with different shades. 
    We find a positive correlation between the interclump density (\fic) and the mass-loss rate, but find no correlations with any stellar parameter or mass-loss rate for the other wind-clumping parameters. 
    \label{fig:ull:windpar_correlation}}
\end{figure*}

\Cref{fig:ull:windpar_correlation} shows the inferred wind-structure parameters as a function of mass-loss, and also includes the values that are obtained by \citet{brands2022} and \citet{2024AA...690A.126H}; these authors also analyse optical and UV data of LMC stars using a genetic algorithm and \fastwind, fitting the same wind-structure parameters\footnote{With the exception that \citet{2024AA...690A.126H} do not fit wind turbulence, but assume a fixed value of $\varv_\mathrm{windturb} = 0.1\varv_\infty$ for all stars. }. \citet{brands2022} find a tentative correlation between the mass-loss rate and several clumping parameters ($f_\mathrm{cl}$, $f_\mathrm{ic}$, $f_\mathrm{vel}$, $\varv_\mathrm{windturb}$), where they infer smoother winds for stars with high mass-loss rates ($\log \ \dot{M} > -6.0$) than for stars with lower mass-loss rates. \citet{2024AA...690A.126H} cannot confirm this trend, but do infer a statistically significant correlation between effective temperature and  $f_\mathrm{ic}$ and $f_\mathrm{vel}$. We now combine their and our data sets and compute the Kendall correlation coefficient ($\tau_{k}$; \citealt{Kendall1938}; see also our \cref{sect:windcorr}) to quantify possible correlations between stellar and wind parameters (\teff, $L$, $\dot{M}$) and the wind-structure parameters ($f_\mathrm{cl}$, $f_\mathrm{ic}$, $f_\mathrm{vel}$, $\varv_\mathrm{cl,start}$, and $\varv_\mathrm{windturb}$). In these tests we only consider the stars with luminosity classes I-III. We find no significant trends, with the exception of a correlation between $\dot{M}$ and $f_\mathrm{ic}$, where higher values of $f_\mathrm{ic}$ generally correspond to higher $\dot{M}$. We find $\tau_k = 0.34$ with $p=0.02$, indicating a positive correlation.  

\begin{table}[]
    \centering
    \caption{Mean wind-structure parameters of our O-type (super)giants. \label{tab:typical_windstructure}}
    \small 
    \begin{tabular}{l l@{$\pm$}l}
         \hline\hline 
         Parameter & \multicolumn{2}{l}{Value} \\\hline 
         $f_\mathrm{cl}$ &  33&14 \\
         $f_\mathrm{ic}$ & 0.2&0.1 \\
         $f_\mathrm{vel}$ & 0.6&0.2 \\
         $\varv_\mathrm{cl,start}$ & 0.12&0.04 $\varv_\infty$ \\
         $\varv_\mathrm{windturb}$ & 0.14&0.09 $\varv_\infty$ \\
         \hline 
    \end{tabular}
    
\end{table}

Despite the trend that we recover for $\dot{M}$ and $f_\mathrm{ic}$, and the fact that the wind-structure parameters are somewhat clustered in certain parts of the parameter space (\cref{fig:windpar_dist}), there is still a large scatter in the obtained values, significantly exceeding the uncertainties. First of all, this raises the question whether our diagnostic lines are sensitive enough for these properties. Can we distinguish between them, in other words, are we able to break possible degeneracies between them? 
The fact that the fitness distributions from our \kiwiGA runs indicate that specific combinations fit significantly better than others, suggests that we do: it implies that different diagnostics respond in a distinct manner to the different wind parameters; otherwise the distributions would not have been peaked. 
We analyse the \kiwiGA output in more detail by probing degeneracies between the different wind parameters in \cref{sect:windcorr}. We  note in this context that we find no systematic differences between the wind structure parameters of the stars with higher resolution STIS data and those for which we have lower resolution COS data; neither do we find such differences between the sources for which we do or do not include the far-UV. 

Assuming that our diagnostics are adequate, a possible reason for the scatter and the lack of trends could be that physical processes are manifesting that are not included in our models. For example, in addition to the small-scale, incoherent structure that is the focus of this paper, large-scale, coherent structures could affect the line profiles \citep[e.g.,][]{1984ApJ...283..303M,2015ApJ...809...12M}, for example related to magnetic fields \citep{2016A&A...594A..56S,2021A&A...656A.131D}. Such variations in the line profiles could affect the inferred mass-loss rates. =\citet{2024ApJ...971..166M} quantify this effect and report that the rms intrinsic error on the mass-loss rate inferred from a single observation is 15\% for O-type stars. This is an average value, \citet{2024ApJ...971..166M} further report that the effect is larger for lower \teff. 
In addition, it is also possible that the properties of the small-scale clumps are intrinsically variable. In that case, the scatter we observe could be (partially) intrinsic: depending on the time when we observe a star, the (small-scale) clumping properties observed could be different. In the context of variability it is important to realise that the optical and UV spectra considered in this study are not taken at the same time. The optical and UV spectra could therefore correspond to different states of the wind, which could make it impossible to obtain a good fit for optical and UV lines simultaneously, and possibly lead to erroneous results  \citep{2024ApJ...971..166M}.

The scatter could also be related to the adopted parameterisation of the wind structure. It could be that this parameterisation is inadequate, i.e., does not represent well enough the clumpy structure of the wind  -- regardless of additional physical processes that may or may not be there. 
If this would be the case, the result could be that the wind-structure parameters are being reduced to merely `fitting parameters', which, while they may improve the fits, do not necessarily represent real physical quantities. Apart from the lack of structure in the values of the clumping parameters, also the high values of $\beta$ that we derive ($\langle\beta\rangle = 1.6 \pm 0.6$) could point to this phenomenon: possibly $\beta$ is used to somehow compensate for wind-structure effects not accounted for in our models. In addition, the relatively high values we obtain for the interclump density in combination with the derived clumping factors suggest that on average a significant fraction of the material ($20\%$) is not contained in clumps. If this is indeed true, then the current \fastwind approach of using one NLTE solution for all the material could be difficult to justify. \citet{2024arXiv241014937V}, who study the mass loss and wind structure of a sample of B-type (super)giants, come to a similar conclusion (see their Sect. 4.5). 

If the effect that the possibly inadequate wind-structure prescription has on the inferred mass-loss rates is random rather than systematic, it should be averaged out with a large enough sample size, and we would, despite large scatter, still be able to capture the overall trend in mass loss versus luminosity or $\Gamma_{\mathrm{Edd},e}$. In order to investigate the validity of our clumping parameterisation further, we would need to extend our theoretical knowledge of the wind structure: ideally, we would have 3D hydrodynamical models of the wind structure, and use these to design and calibrate our 1D parameterisation\footnote{Such tests have been carried out in a simplified manner by \citet{2010A&A...510A..11S}, who used 1D models where the structures have been randomly re-shuffled to 3D space, to check a couple of effects.}.  
In this context we note that the 2D radiation-hydrodynamic O-star wind models of \citet[][their Fig. 8.3.]{DriessenPhDthesis} and \citet{2024A&A...684A.177D} tentatively suggest that the density distribution of wind parcels resembles a Gaussian-like rather than a bi-modal distribution. A similar distribution is found from the 3D radiation-hydrodynamic simulations of \citet{2022A&A...665A..42M}, although these authors simulate a Wolf-Rayet star wind rather than an O-star wind. These recent findings, just as the relatively high inferred interclump medium density discussed above, contradict the assumption of a two-component medium as adopted in \fastwind, and warrant further investigation. 

\subsection{Comparison to `optically-thin clumping'}

\begin{figure}
    \centering
    \includegraphics[width=1.0\linewidth]{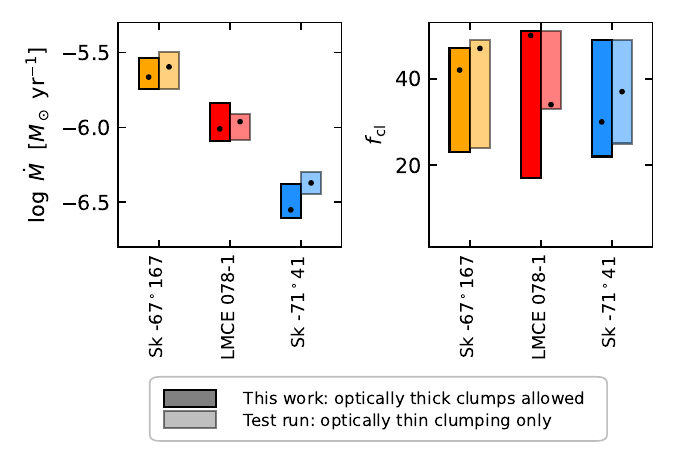}
    \caption{Comparison of mass-loss rates (left) and clumping factors (right) obtained with our default optical+UV fitting method using the clumping parameterisation of \citet{2018A&A...619A..59S} allowing for optically thick clumps (darker shades), versus those obtained using the optically-thin clumping approach. Yellow, red, and blue bars correspond to the $1\sigma$ uncertainties of Sk~$-67^\circ167$, LMCe~078-1, and Sk~$-71^\circ41$, respectively, and the best fit values are indicated with a black dot. The values derived with the two parameterisations are in good agreement for both mass-loss rates and clumping factors.  \label{fig:thinclumps}}
    
\end{figure}

In this brief section we compare our wind structure parameterisation with an alternative, frequently used wind-structure parameterisation, namely the so called `microclumping' or `optically-thin clumping' approach. Here, the interclump medium is assumed to be void, while all matter is assumed to be contained in clumps that are small and rarefied enough that they stay optically thin. We test how this parameterisation would affect the derived mass-loss rates and clumping factors by re-fitting three semi-randomly picked stars\footnote{We use the same stars as we use for our micro-turbulence/abundance tests; \cref{sec:abundance} details on how these stars were selected. We note that all three stars have a significant interclump density in the best fit optically thick models ($f_\mathrm{ic} = 0.09$ for Sk~$-67^\circ167$, $f_\mathrm{ic} = 0.14$ for LMCe~078-1, and $f_\mathrm{ic} = 0.25$ for Sk~$-71^\circ41$).} in the same manner as our optical+UV runs, except for changing the clumping parameterisation to the optically-thin clumping approach; \fic and \fvel are no longer considered. The results are compared to our default optical+UV runs in \cref{fig:thinclumps}. We see that the mass-loss rates that are derived with the different clumping prescriptions are in good agreement, as are the clumping factors. The similarity between the derived values suggests that if the parameterisation adopted in this paper may not be adequately describing the physics in the wind, and this causes systematic errors, the optically-thin approach suffers from similar errors. 

\subsection{Diagnostics and degeneracy between wind parameters \label{sect:windcorr}}

While a detailed study of the behaviour of different diagnostics is beyond the scope of this paper, 
it is possible to address potential degeneracies between wind parameters given our set of diagnostics by analysing the correlation plots of the best fitting models.
In order to estimate the correlation between two parameters $X$ and $Y$ we carry out the following procedure. First, we select of each run the best fitting models (the approximately $2\sigma$ equivalent models, see \cref{sec:Uncertainties}) and make a scatter plot. 
The shape of this plot represents the correlation between $X$ and $Y$ in terms of fitness. The second step in our procedure is to quantify this shape with a correlation measure. For this we use the Kendall correlation coefficient, $\tau_k$, which can assume a value between $-1$ and $+1$, where $-1$ is a strong negative, and $+1$ a strong positive correlation. The corresponding $p$-value, $p_k$, reflects the significance of the obtained $\tau_k$. We compute $\tau_k$ and $p_k$ for each parameter combination $X-Y$, for each star (\kiwiGA run). The third step is to count (per $X-Y$ pair) for how many stars a significant ($p_k < 0.05$) correlation is found  per ``direction'' (either positive, $\tau_k > 0$, or negative, $\tau_k < 0$) and compute the average $\tau_k$ per direction. For $X-Y$ pairs where the correlation is found for most stars and mostly in a certain direction, we conclude that there likely is a correlation between those parameters. 

We apply this method to all combinations of wind parameters and, as a check on whether the above method works for finding correlations, also to \teff$-\log~g$, which we know are positively correlated. Our findings are summarised in  \Cref{tab:correlations}, and scatter plots as well as the $\tau_k$ value per star can be found on \href{https://zenodo.org/records/15013513}{Zenodo}. \Cref{tab:correlations} shows that indeed, we recover the positive correlation between \teff and $\log~g$. Furthermore, we find a nearly equally strong (but negative) correlation between \mdot and $\sqrt{f_{\rm cl}}$, and between \mdot and $\beta$. The correlation between \mdot and $\beta$ has been noticed previously \citep[e.g.][]{2006A&A...454..625P,2021A&A...655A..67H} and originates likely in optical wind lines that are formed close to the star: here both an increase in \mdot and a decrease in $\beta$ would lead to stronger mass-loss signatures, although a decrease in $\beta$ would affect the shape of the line differently. With somewhat less certainty (i.e., only for 50-70\% of our stars) we find correlations between \mdot and $\log f_\mathrm{ic}$ (negative), \mdot and $\log f_\mathrm{vel}$ (negative), \mdot and $\varv_\mathrm{cl,start}$ (positive), and between $\log f_\mathrm{ic}$ and $f_\mathrm{vel}$ (negative). 
We stress that the fact that we find correlations between the wind parameters does not mean we are not able to break the degeneracies between these parameters. We can, up to a certain point: this is reflected in the shape of the \kiwiGA distributions (which are generally peaked rather than flat for the wind parameters), together with the uncertainties on the derived parameters, which have become larger due to the correlations (compared to the case where, e.g., only noise in the data contributed to the uncertainty). What the correlations tell us is that part of the diagnostics considered respond similarly to changes in different parameters. 
For example, both a higher \teff and a higher \logg affect the Balmer lines similarly, and therefore we recover a correlation between the two; this is reflected by the elongated shape of the correlation plot. On the other hand, we also fit other lines (e.g., He~\textsc{i} and He~\textsc{ii} and lines) that respond differently to an increase in \teff versus an increase in \logg, and therefore can break the degeneracy; this is reflected by the fact that within the elongated correlation plot there is an `island' where the fittest models reside (the global minimum of the fit), i.e. the best fitting parameter combinations. 
We conclude our discussion on correlations by warning that the correlations presented in \Cref{tab:correlations} apply specifically to the adopted set of diagnostics, in combination with the parameter space covered by our sample, and cannot necessarily be generalised. 

\begin{table}[]
    \centering \small 
    \caption{\centering Correlations between between fitting parameters. \label{tab:correlations}}
    \begin{tabular}{lrcrc}

\hline \hline
           & \multicolumn{2}{c}{Positive correlation} & \multicolumn{2}{c}{Negative correlation} \\ 
Parameters & \%  & $\langle \tau_{k} \rangle$ &  \% & $\langle \tau_{k} \rangle$ \\ \hline
\cellcolor{NavyBlue!35}$T_{\rm eff}~\mathrm{(kK)}$~-~$\log \ g$ &  \cellcolor{NavyBlue!35} $95 $ &\cellcolor{NavyBlue!35}$0.35 \pm 0.09$ &  &  \\
\cellcolor{NavyBlue!35}$\log \ \dot{M}$~-~$\sqrt{f_{\rm cl}}$ &  &  &  \cellcolor{NavyBlue!35} $83 $ &\cellcolor{NavyBlue!35}$-0.29 \pm 0.11$ \\
\cellcolor{NavyBlue!15}$\log \ \dot{M}$~-~$\log \ f_{\rm ic}$ &   $12 $ &$0.16 \pm 0.11$ &  \cellcolor{NavyBlue!15} $54 $ &\cellcolor{NavyBlue!15}$-0.13 \pm 0.05$ \\
\cellcolor{NavyBlue!15}$\log \ \dot{M}$~-~$f_{\rm vel}$ &  &  &  \cellcolor{NavyBlue!15} $54 $ &\cellcolor{NavyBlue!15}$-0.16 \pm 0.09$ \\
\cellcolor{NavyBlue!35}$\log \ \dot{M}$~-~$\beta$ &   $8 $ &$0.12 \pm 0.00$ &  \cellcolor{NavyBlue!35} $79 $ &\cellcolor{NavyBlue!35}$-0.29 \pm 0.08$ \\
\cellcolor{NavyBlue!15}$\log \ \dot{M}$~-~$v_{\rm cl,start}$ &  \cellcolor{NavyBlue!15} $70 $ &\cellcolor{NavyBlue!15}$0.27 \pm 0.15$ &  &  \\
$\sqrt{f_{\rm cl}}$~-~$\log \ f_{\rm ic}$ &   $25 $ &$0.11 \pm 0.04$ &   $20 $ &$-0.11 \pm 0.05$ \\
$\sqrt{f_{\rm cl}}$~-~$f_{\rm vel}$ &   $16 $ &$0.09 \pm 0.03$ &   $12 $ &$-0.13 \pm 0.05$ \\
$\sqrt{f_{\rm cl}}$~-~$\beta$ &   $41 $ &$0.13 \pm 0.06$ &   $20 $ &$-0.16 \pm 0.06$ \\
$\sqrt{f_{\rm cl}}$~-~$v_{\rm cl,start}$ &   $37 $ &$0.13 \pm 0.06$ &   $8 $ &$-0.16 \pm 0.06$ \\
\cellcolor{NavyBlue!15}$\log \ f_{\rm ic}$~-~$f_{\rm vel}$ &  &  &  \cellcolor{NavyBlue!15} $70 $ &\cellcolor{NavyBlue!15}$-0.18 \pm 0.09$ \\
$\log \ f_{\rm ic}$~-~$\beta$ &   $20 $ &$0.13 \pm 0.04$ &   $8 $ &$-0.14 \pm 0.01$ \\
$\log \ f_{\rm ic}$~-~$v_{\rm cl,start}$ &   $12 $ &$0.09 \pm 0.00$ &   $33 $ &$-0.12 \pm 0.06$ \\
$f_{\rm vel}$~-~$\beta$ &   $37 $ &$0.13 \pm 0.04$ &   $8 $ &$-0.16 \pm 0.10$ \\
$f_{\rm vel}$~-~$v_{\rm cl,start}$ &   $12 $ &$0.13 \pm 0.08$ &   $20 $ &$-0.15 \pm 0.13$ \\
$v_{\rm cl,start}$~-~$\beta$ &   $29 $ &$0.25 \pm 0.11$ &   $4 $ &$-0.21 \pm 0.00$ \\
\hline 
\multicolumn{5}{p{8.0cm}}{\tiny \textbf{Notes.} \% refers to the percentage of Kiwi-GA runs for which we found a significant positive or negative correlation for the relevant parameter combination. The mean value of $\tau_k$ is computed separately for the significant positive, and significant negative values of $\tau_k$. In dark and light blue we mark parameter combinations for which we find a significant correlation for at least 75\% or at least 50\% of our runs, respectively. The results presented here should be regarded as an indication of possible correlations given the adopted set of  diagnostics and the parameter space covered by our sample, and cannot necessarily be generalised.
} \\

    \end{tabular}
\end{table}

\subsection{Terminal velocities \label{sec:ull:terminalvelocity}}

\begin{figure}
    \centering
    \includegraphics[width=0.43\textwidth]{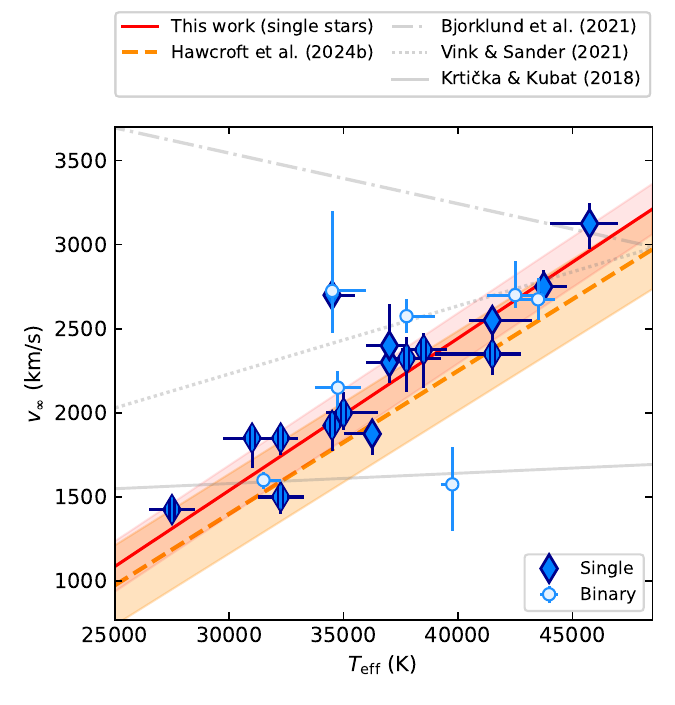}
    \caption{Terminal velocities of our sample stars as a function of \teff (blue diamonds indicate single stars, with hatched symbols corresponding to supergiants; white circles indicate binaries). In the background we show the best linear ODR fit to the single star observations of our sample (red solid line, $\varv_\infty = 0.090T_\mathrm{eff} - 1175$), as well as the LMC relation derived by \citet[][orange dashed line, $\varv_\infty = 0.085T_\mathrm{eff} - 1150$]{2024A&A...688A.105H}. 
    In the background we show in grey the trends of the theoretical predictions of \citet[][dash-dotted line]{2021AA...648A..36B}, \citet[][dotted line]{2021MNRAS.504.2051V}, and \citet[][solid line]{2018AA...612A..20K}.  We note that one source, the binary LH~114-7 ($\varv_\infty = 3425~$km~s$^{-1}$), falls outside the range of this plot due to its high effective temperature ($T_\mathrm{eff} = 59000$~K). 
 \label{fig:vinf_vs_teff}}
\end{figure}

\Cref{fig:vinf_vs_teff} shows our terminal velocities as a function of effective temperature, together with the empirical LMC relation of \citet{2024A&A...688A.105H}, and the theoretical predictions\footnote{See \citet{brands2022} for a summary of the differences and similarities between the theoretical prescriptions considered here.} by \citet[][]{2021AA...648A..36B}, \citet[][]{2021MNRAS.504.2051V}, and \citet[][]{2018AA...612A..20K}.  Concerning the theoretical predictions it is clear that none matches the empirical results. Only the prediction of \citealt{2021MNRAS.504.2051V} is similar to some extent; this relation somewhat matches the slope and absolute value of the observations, although the fit is far from perfect. 

\citet{2024A&A...688A.105H} derive their relation based on 37 O-type stars of the ULLYSES sample, 12 of which overlap with our sample of presumed single stars. They measured the terminal velocities using the 
Sobolev with Exact Integration method \citep[SEI;][]{1987ApJ...314..726L}. Investigating the dependence of \vinf on both escape velocity and temperature, they found that the latter provides a tighter relation. 
Within uncertainties the terminal velocities of our stars are in agreement with the relation of \citet{2024A&A...688A.105H}, with the exception of one outlier (Sk $-69^\circ$104), of which the velocity exceeds the predicted value of \citet{2024A&A...688A.105H} with $\sim$900~km~s$^{-1}$ (nearly four times the uncertainty on their relation), and three binary stars. However, the values we find are systematically higher, on average 200~km~s$^{-1}$ (considering only the presumed single stars, and excluding Sk $-69^\circ$104). We find that the difference between both results lies in the wind turbulence velocity. This velocity is fitted both in our full spectral fit, as well as in the SEI analysis, but with the latter typically higher values are obtained. Upon comparing $\varv_\mathrm{max} = \varv_\infty + \varv_\mathrm{windturb}$ of our analysis and the SEI analysis of \citet{2024A&A...688A.105H}, we find a good agreement without any systematic offsets. Possibly, the difference in the derived values for $\varv_\mathrm{windturb}$ is related to a different approach for computing the radial dependence of the line optical depth: in SEI this dependence is parametrised by a power law \citep{1987ApJ...314..726L,1995A&A...295..136H,2014A&A...568A..59S}, while in \fastwind this is explicitly treated in non-LTE. It could be the case that, if the optical depth structure adopted in SEI is off, a higher turbulent velocity is required in order to reproduce the observed profile. 

 \citet{2024A&A...688A.105H} also infer the terminal velocities of this sample by `direct measurement', that is, by obtaining the wavelength of the bluest edge of the P-Cygni absorption showing zero flux \citep[e.g.][]{1990ApJ...361..607P}. For the 37 stars on which  \citet{2024A&A...688A.105H} base their relation (stars with saturated profiles, labelled `rank i' stars in their paper; nearly all have luminosity class I, II, or III), they find a systematic offset between the two methods, where the SEI method results in velocities that are about 200~km~s$^{-1}$ lower than those of the direct measurement method (green points in their Fig. 3). This implies that their direct measurements are in agreement with our detailed spectral fitting. Indeed, when comparing the values that \citet{2024A&A...688A.105H} obtain with this method for their class i stars, to our values, we find no systematic offset. 
 
 Overall, we confirm the tight relation of \vinf with temperature of \citet{2024A&A...688A.105H}, but suggest a slightly modified relation for LMC stars with temperatures in the range $30000-45000$~K:
 \begin{equation}\label{eq:vinffit}
     \varv_\infty = (0.090 \pm 0.008) \ T_\mathrm{eff} - (1175\pm293),
 \end{equation}
 with \vinf in $\mathrm{km}\ \mathrm{s}^{-1}$, and \teff in K. Here we assumed that the detailed \fastwind modelling yields more reliable optical depths in the outer wind, and thus more reliable turbulent velocities than does SEI. We note that the inclusion or exclusion of \nvuvline (see \cref{sec:dis:NITRO}) does not affect our obtained terminal velocities significantly. Only for two of the suspected binaries we find terminal velocities that are around 450~km~s$^{-1}$ higher (Sk~$-71^\circ$19) or lower (BI~173) when \nvuvline is included. This would not affect the fit leading to \cref{eq:vinffit}, for which we only considered presumed single stars. We furthermore note that our fits to the blue edge of \CIVline are usually excellent, increasing the reliability of the results derived here. 

\begin{figure}
    \centering
    \includegraphics[width=0.4\textwidth]{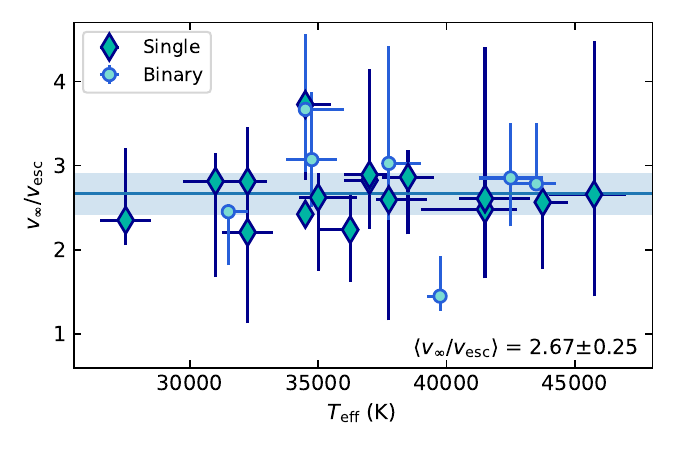}
    \caption{\label{fig:vinf_vesc} The ratio of terminal velocity and effective escape velocity as a function of effective temperature. For our presumed single stars this ratio has a mean value of $2.67\pm0.25$ (indicated by the horizontal line and shaded area).  }
    
\end{figure}

We conclude this section by inspecting the ratio $\varv_\infty/\varv_\mathrm{esc}$, where $\varv_\mathrm{esc} = \sqrt{2GM_\mathrm{evol}(1-\Gamma_{\mathrm{eff},e})/R_*}$ is the effective escape velocity. For our presumed single stars we find an average ratio of $\langle \varv_\infty/\varv_\mathrm{esc} \rangle = 2.67\pm0.25$ (\cref{fig:vinf_vesc}). This ratio is consistent with the typical ratio  of $2.58\pm0.20$ for Galactic O-stars \citep{1995ApJ...455..269L}, and with value of $\langle \varv_\infty/\varv_\mathrm{esc} \rangle = 2.4 \pm 0.4$ as found by \citet{2024A&A...688A.105H} for LMC O-stars.

\subsection{CNO abundances and their uncertainties\label{sec:abundance}}

\begin{figure}
    \centering
    \includegraphics[width=0.49\textwidth]{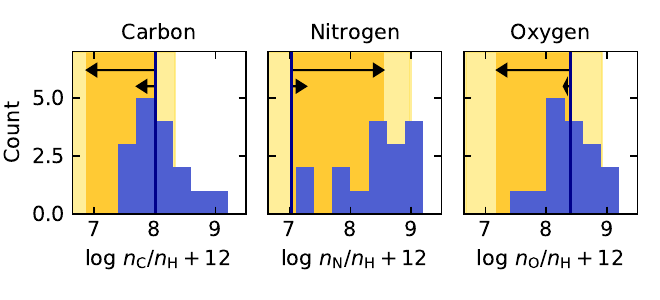}
    \caption{Distribution of observed CNO abundances for our (presumed) single stars. The vertical line indicates the baseline abundances (taken from \citealt{2022MNRAS.515.4130C}); the black arrows denotes the expected change in abundance when the star evolves (nitrogen enrichment, oxygen and carbon depletion), where the size of the top arrow indicates a relatively large change in abundance (corresponding to a $M_\mathrm{ini} = 80$~\Msun, $\varv_\mathrm{ini} = 300$~km~s$^{-1}$ model of the \citet{kohler2015} grid at 2.5~Myr), and the bottom arrow a relatively small change (corresponding to a $M_\mathrm{ini} = 40$~\Msun, $\varv_\mathrm{ini} = 200$~km~s$^{-1}$ model of the \citet{kohler2015} grid at 2.5~Myr). The orange shaded region indicates the values that are in agreement with the expected enrichment/depletion; this range of `allowed values' is based on the change in the $M_\mathrm{ini} = 80$~\Msun, $\varv_\mathrm{ini} = 300$~km~s$^{-1}$ model, and is extended with the average uncertainty on the observationally inferred stellar abundances (light yellow shaded region). For both carbon and oxygen part of the stars fall outside the region of expected values. \label{fig:cno_baseline_dist}}
   
\end{figure}

As a by-product of our wind study we determine C, N, and O abundances for the stars in our sample. Most of the diagnostic lines from which these abundances are derived are UV wind lines. Since these lines are sensitive for many stellar and wind (structure) parameters, the derived abundances should be treated with caution. 
Given that our sample consists of somewhat evolved stars (\cref{fig:hrd_logg}), we expect from the CNO-cycle to see nitrogen enrichment and oxygen and carbon depletion, compared to the LMC baseline abundances. How much enrichment or depletion we expect depends on the stellar mass, rotation rate and age. \Cref{fig:cno_baseline_dist} shows the range of possible expected values per element: changes as large as 1-1.5~dex can be expected, but values close to baseline are possible too.  

\subsubsection{Nitrogen\label{sec:dis:NITRO}} 

\begin{figure}
    \centering
    \includegraphics[width=0.45\textwidth]{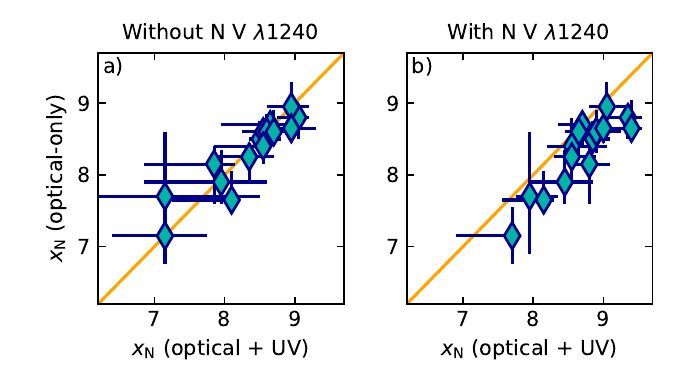}
    \includegraphics[width=0.45\textwidth]{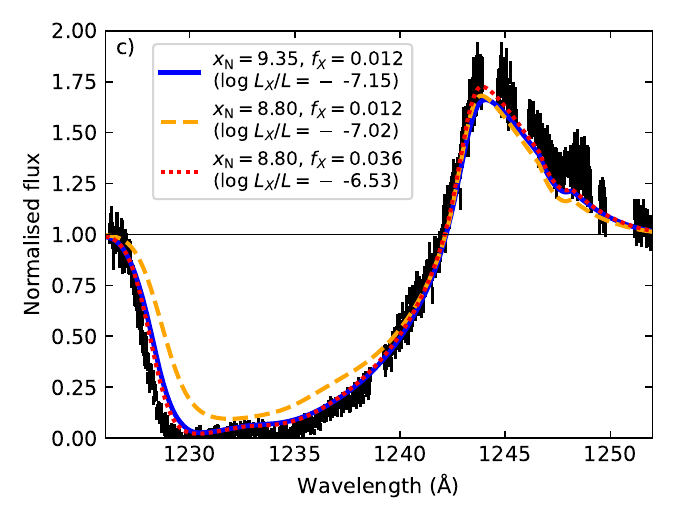}
    \caption{Effect of \nvuvline on the derived nitrogen abundance. Panels $a$ and $b$ compare the nitrogen abundance of the presumed single stars, derived from the optical-only runs, to that derived from the optical~+~UV fits in case of excluding  \nvuvline (panel $a$) and in case of including \nvuvline (panel $b$). In the fits where \nvuvline is included, the nitrogen abundance is systematically higher than the values found from the optical-only runs. In panel $c)$ we show, as an example, the \nvuvline doublet of Sk~-67$^{\circ}$167, with the best fit model of the optical~+~UV run including \nvuvline (blue solid line), a model where we altered the nitrogen abundance to the optical-only best fit value (orange dashed line), and a model where, in addition, the X-ray volume filling fraction was increased (red dotted line). The models with a high nitrogen abundance and low X-ray volume filling fraction, and a lower nitrogen abundance and higher X-ray volume filling fraction are very similar and both match the observed spectrum.  
    \label{fig:nitrogen_xrays}}   
\end{figure}

For all sources we find indeed nitrogen enrichment compared to the LMC baseline abundance ($x_\mathrm{N} = 7.03$; \citealt{2022MNRAS.515.4130C,Vinkinprep}), with a sample mean of $\langle x_N\rangle = 8.25\pm 0.56$ (see also \cref{fig:cno_baseline_dist}). Typical uncertainties on individual values are large: on average the $1 \sigma$ uncertainty range is $\Delta x_N =0.93$ dex. For nitrogen (contrary to carbon and oxygen) we have a substantial set of optical lines available that allows us to check the results of our optical + UV analysis and make the measurement less dependent on UV wind lines. Our optical + UV values agree with the results of the optical-only runs (\cref{fig:nitrogen_xrays}, panel \textit{a}). In combination with the fact that we find enrichment for all sources we deem our inferred nitrogen abundances reliable within the derived uncertainties. 

Nonetheless, one should remain cautious with the interpretation of wind line derived abundances, as a systematic error is easily introduced when one or more other parameters regarding the wind properties are not modelled properly. For example, we did not include \nvuvline as a diagnostic because this line is strongly affected by X-rays. Although we model the X-rays to the best of our ability (see \cref{sec:fastwind,app:sec:X-rays}), we have indications that the implementation is not optimal. When including \nvuvline as a diagnostic in our fits, we note that the derived nitrogen abundance is consistently higher than the nitrogen abundance derived from the optical-only fits, on average 0.33~dex (\cref{fig:nitrogen_xrays}, panel $a$ and $b$). 

In order to investigate the cause of this discrepancy we recompute the best fit models from several optical~+~UV runs that include \nvuvline as a diagnostic, adopting the nitrogen abundance of the best fit value of the optical-only run. With this lower abundance, the \nvuvline is too weak compared to the data. It thus appears that \nvuvline pushes towards a higher nitrogen abundance. On the other hand, a higher amount of X-rays can have the same effect on the line as an increased nitrogen abundance: experiments on three of our sample stars showed that for obtaining a good fit of \nvuvline with the nitrogen abundance as from the optical-only fit, we need to increase the volume filling factor $f_X$ such that the corresponding X-ray luminosity equals $\log L_X / L \approx -6.5$ (\cref{fig:nitrogen_xrays}, panel $c$). While such values are observed at times, typically the X-ray luminosities of O-type stars are lower \citep[e.g.,][]{2004ApJ...606..497G,2006MNRAS.372..661S,2009A&A...506.1055N,2022MNRAS.515.4130C}.  

The fact that we need higher X-ray luminosities than are typically observed in order to model this line, could mean that our X-ray assumptions are somehow inadequate. 
For a better understanding of X-rays in massive-star winds and their proper implementation, it might be interesting to systematically study the effect of the different X-ray parameters by modelling \nvuvline simultaneously with other X-ray sensitive wind lines, such as \oviline. For this to be meaningful, one would need an observed (upper limit on) the X-ray flux\footnote{Apart from the X-ray prescription that we adopted for this work \citep{2016A&A...590A..88C}, another prescription is recently implemented in \fastwind \citep[][partly based on an analysis by \citealt{2013MNRAS.429.3379O}]{2020A&A...642A.172P}; this new parameterisation should be included in the study.}. The high required X-ray luminosities could also be related to the way the wind structure is implemented in \fastwind: the same ionisation structure is adopted for both the clumps and the interclump medium, while differences between them can be expected. The models of \citet[][see also \citealt{2023MNRAS.518.5001F}]{2021MNRAS.504..311F} allow for different ionisation structures,  which enables them to reproduce high ionization resonance transitions with an X-ray luminosity that is consistent with typical observed values. 

We conclude this section by pointing out that \citet{brands2022} do include \nvuvline in their modelling, which could affect their inferred nitrogen abundances. However, Fig. 8 of \citet{2016A&A...590A..88C} shows that the effect might be smaller in their parameter range compared to that of our sample: for $T_\mathrm{eff} \gtrsim 43000$~K the effect of X-rays on the N~\textsc{v} ionisation balance is small. Most stars in the sample of \citet{brands2022} fall in this temperature range. 

\subsubsection{Carbon and oxygen\label{sec:dis:CO}} 

\begin{figure}
    \centering
    \includegraphics[width=0.47\textwidth]{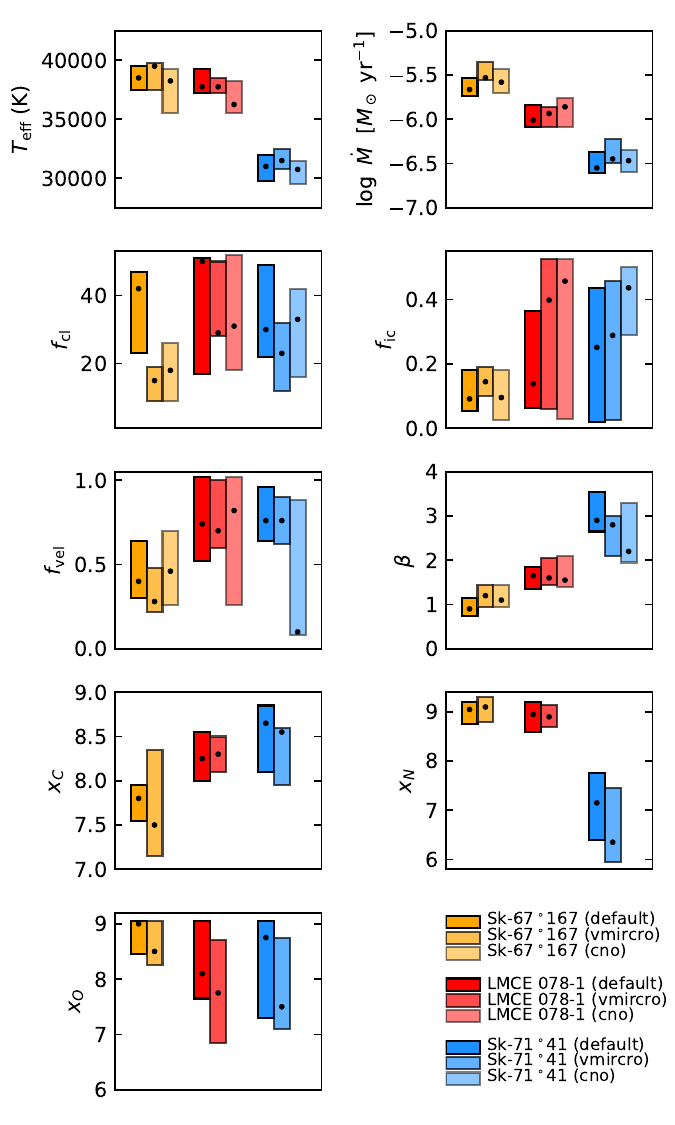}
    \caption{For three stars (displayed in different colors: orange for Sk$~-67^\circ167$, red for LMCe 078-1, and blue for Sk$~-71^\circ41$) we compare the results of our default runs (darkest shade, \Cref{tab:stellarparams,tab:bestfitswind}) with two different test runs: in one case we set \vmicro as a free parameter (middle shade; labeled 'vmicro'),  in the other case we fix the CNO abundances to the LMC baseline values (lightest shade; labeled 'cno'). Each bar indicates the $1\sigma$ range of parameter values, the dot indicates the best fitting model. A free \vmicro does not result in significant changes in abundance. Furthermore, we see that both \teff as well as wind parameters are robust against changes in \vmicro and abundance. \label{fig:CNOvmicrotest}}
\end{figure}

For carbon and oxygen we do not always find the expected depletion (\cref{fig:cno_baseline_dist}). The sample means, $\langle x_C \rangle = 8.03\pm 0.36$ and $\langle x_O \rangle = 8.33\pm 0.48$, are close to the baseline abundances ($x_\mathrm{C} = 8.01$, $x_\mathrm{O} = 8.40$; \citealt{2022MNRAS.515.4130C,Vinkinprep}), while on average we expect them to be somewhat lower. 
Uncertainties on the C and O abundances are typically large, spanning on average $\Delta x_C = 0.68$ and $\Delta x_O =1.05$ dex, respectively, and considering this, all but three sources\footnote{W61-28-23 and Sk~$-71^\circ41$ have  an anomalously high carbon abundance, and Sk~$-67^\circ167$ has an anomalously high oxygen abundance.} are consistent with C and O baseline abundance or lower. Nonetheless, one would still expect the mean value to be below the baseline for these elements. 
Unless the evolution of the carbon and oxygen enriched sources is unusual (e.g., affected by binary interaction), the somewhat high mean values and the three anomalously high abundances could imply that our inference of the C and O abundances is of poor quality. 
Possibly, the relatively high C and O abundances reflect limitations of our wind-structure model (see \cref{section:wind_structure}): this would affect the shape and strength of the wind lines, which in turn could affect the abundance determination. There is currently no way to test this hypothesis. Another possible cause for the relatively high abundances could lie in the adopted model atoms for C and O, which are not tested as extensively as the N model atom. The oxygen model in particular needs to be more carefully tested \citep{2019A&A...623A...3C}. 
Upon inspecting the line profiles (\cref{ullyses:results}) we notice that the fits to \oivline are often poor; this could affect the abundance determination and possibly finds its origin in the model atom. 
Micro turbulence, for which we adopt a fixed value of \vmicro$~= 15~$km~s$^{-1}$ for our study, could also affect the abundance determinations. We test the effect of our assumption by re-fitting three stars with \vmicro as a free parameter: of the presumed single stars with a higher than baseline oxygen and/or carbon abundance, we pick three stars at random (Sk$~-67^\circ167$, LMCe 078-1, and Sk$~-71^\circ41$). We find best fit values in the range $\varv_\mathrm{micro} = 17-21$~km~s$^{-1}$, not too far from our original assumption, and no significant changes in abundance when varying \vmicro (\cref{fig:CNOvmicrotest}).  
Lastly, we point out the possibility that the high abundances that we derive reflect the limitations of our 1D models. For the abundance determination of the Sun, the use of 3D, time-dependent hydrodynamical models resulted in downward revisions of in particular C, N, O and Ne abundances \citep{2009ARA&A..47..481A}. Although the first steps toward 3D treatment of massive star atmospheres have been made (e.g., \citealt{2022A&A...665A..42M,2023ApJ...945...58S}), it remains to be seen whether this will result in similar changes in abundance as for Solar type stars.

While at this point we cannot be certain what are the true values of the abundances, we can check whether our results, in particular the derived wind parameters, are robust against changes in abundance, in particular for stars with a high oxygen or carbon abundance. 
For this we look again at the results of our test where we vary \vmicro, and in addition fit the same three stars, but now with fixed values of CNO abundances, where we adopt LMC baseline values. 
The tests shows that both \teff and the wind parameters are robust against these changes (\cref{fig:CNOvmicrotest}). The only parameter that seems affected significantly, for one source, is \fcl. 
Apparently in some cases, a small change in micro turbulence and/or the corresponding abundances can significantly affect the value of the clumping factor.  
This could explain part of the scatter we see for the clumping factor in \cref{fig:ull:windpar_correlation}. We stress that we see this only for the clumping factor (and only for one star) and that the mass-loss rate and the other wind-structure parameters have not changed significantly. 

\section{Summary and outlook \label{ullyses:conclusionoutlook}}

Through the analysis of optical and UV spectroscopy we retrieved stellar and wind parameters of 17 (presumed) single, and 7 (likely) binaries in the LMC. All stars are part of the ULLYSES and XshootU surveys. This analysis contributes to the ultimate goal of characterising the full ULLYSES sample, aiding the construction of an empirical library of spectral templates for stars at sub-solar metallicity. 

Our sample consists of O-type giants, bright giants and supergiants and we find that these are massive ($20-60$~\Msun), somewhat evolved stars; with ages of $1-5$~Myr they are about half way through their main-sequence lifetime, and the nitrogen surface abundance of most stars is enhanced. Furthermore, all sample stars have strong stellar winds, 
with mass-loss rates ranging from $\log \ \dot{M} \ [\mathrm{M}_\odot~\mathrm{yr}^{-1}] = -7.2  $ to $-5.6$, and terminal velocities  of $\varv_\infty = 1400 -3400$~km~s$^{-1}$. We study in detail the properties of the stellar winds, in particular, we scrutinised the wind-structure parameters adopting a clumping prescription that allows for clumps to become optically thick.  
Our approach relies on the simultaneous fit of the wind-structure (clumping) parameters, and therefore our mass-loss rates need not to be `corrected' for clumping, e.g., scaled by a factor $1/\sqrt{f_\mathrm{cl}}$ as one could do when working with models with a smooth wind. 
Our main findings are:
\begin{itemize}
    \item The mass-loss rates that we infer for the O-type supergiants in our sample are lower than the predicted rates of \citet{2001AA...369..574V}, but higher than those of \citet{2018AA...612A..20K} and \citet{2021AA...648A..36B}. However, in the luminosity regime covered by our sample (\logll=$~5.3-6.0$) the theoretical predictions are not drastically different, and the empirical and theoretical mass-loss rates have converged within 0.5~dex, or a factor of three. 
    \item We combine our sample with the LMC samples of \citet{2024AA...690A.126H} and \citet{brands2022}, both obtained using the same method as we do, and present empirical \mdot-$L$ and $D_\mathrm{mom}$-$L$ relations for the LMC. 
    \item We find high clumping factors of $\langle f_\mathrm{cl}  \rangle= 33\pm14$, moderate to strong velocity-porosity effects, $\langle f_\mathrm{vel} \rangle = 0.6\pm0.2$, and an interclump density factor of $\langle f_\mathrm{ic} \rangle = 0.2\pm0.1$. The latter, in combination with a clumping factor of $f_\mathrm{cl} = 33$, implies that for the winds of our sample stars, 20\% of the wind material is contained in the interclump medium. 
    \item We find a positive correlation between $f_\mathrm{ic}$ and mass loss, but no dependence of other wind-structure parameters on either mass-loss rate or stellar properties. 
    \item The scatter on empirical mass-loss rates inferred from the optical and UV spectra is larger than our typical error bars on the mass-loss rates. This also holds for the scatter on the wind-structure properties (clumping parameters). We do not know whether this scatter has a physical origin or is the consequence of systematic errors. Possibly, the parameterisation of the wind structure that we adopt is inadequate, i.e., does not represent well enough the clumpy structure of the wind. This warrants further investigation. 
    \item We confirm the empirical relation between \teff and \vinf inferred by \citet{2024A&A...688A.105H}, but notice a systematic offset between our \vinf values (derived with \fastwind), compared to theirs (derived with SEI). On average using SEI results in values that are 200~km s$^{-1}$ lower. 
    \item We consider the nitrogen abundances that we infer to be reliable, but oxygen and carbon abundances are typically higher than expected. This might be related to the model atoms of carbon and oxygen, or find its origin in 3D effects currently not treated in our 1D models; this warrants further investigation. 
   The uncertain abundances do not alter our conclusions about the wind parameters, as we find that these are robust against changes in CNO abundance.
\end{itemize}
This paper comprises only a small subset of the ULLYSES stars, and while we included all spectral sub-types, an expansion of the sample might give us better handles on typical mass-loss values. 
To this end, a future study with a larger sample, possibly also including (early) O-type dwarfs, would be valuable. 
In parallel to this work, \citet{Backsinprep} analysed ULLYSES O-type (super)giants in the Small Magellanic Cloud \citep[SMC][]{Backsinprep} and combined their results with Milky Way and LMC samples (including the sample of this work) in order to investigate the dependence of mass-loss rates on metallicity. At \logll$ = 5.75$ they find $\dot{M}~\propto~Z^m$, with $m\sim1$, while at lower luminosities a steeper scaling of $m>2$ is inferred. Such studies of the metallicity dependence of mass loss would also benefit from expanding the analysis to larger samples, in particular towards lower luminosities, and if possible including galaxies with an even lower metallicity than the SMC (e.g., \citealt{2024ApJ...974...85T}; \citealt{Fureyinprep}).  

In parallel to these sample studies it is essential that the nature of the scatter on the observed mass-loss rates and wind properties is investigated. What is the effect of line variability on the inferred wind parameters? What is the minimum set of diagnostics required to break degeneracies between mass-loss rate, clumping, interclump density and velocity-porosity? Lastly, it is absolutely crucial to investigate the validity of the clumping parameterisation of the \fastwind models, for example with the use of 2D and 3D simulations of massive-star winds \citep[e.g.,][]{2022A&A...663A..40D,2022A&A...665A..42M,2024A&A...684A.177D}. 

\section*{Data availability}

Additional figures, tables, and data can be found on \href{https://zenodo.org/records/15013513}{Zenodo}: \url{https://zenodo.org/records/15013513}. 

\begin{acknowledgements}
This publication is part of the project `Massive stars in low-metallicity environments: the progenitors of massive black holes' with project number OND1362707 of the research TOP-programme, which is (partly) financed by the Dutch Research Council (NWO). 
SB is furthermore supported by a grant from the Dutch Research School for Astronomy (NOVA). 
This work used the Dutch national e-infrastructure with the support of the SURF Cooperative using grant no. EINF-3257. 
AACS, VR, and RRL are supported by the Deutsche Forschungsgemeinschaft (DFG - German Research Foundation) in the form of an Emmy Noether Research Group -- Project-ID 445674056 (SA4064/1-1, PI Sander). AACS is further supported by funding from the Federal Ministry of Education and Research (BMBF) and the Baden-Württemberg Ministry of Science as part of the Excellence Strategy of the German Federal and State Governments. 
BK gratefully acknowledges support from the Grant Agency of the Czech Republic (GAČR 22-34467S). The Astronomical Institute in Ond\v{r}ejov is supported by the project RVO: 67985815.
FT and FN acknowledge grant PID2022-137779OB-C41 funded by the Spanish MCIN/AEI/ 10.13039/501100011033. 
DJH acknowledges support from the STScI through grant HST-AR-16131.001-A. 
JS and FB acknowledge the support of the European Research Council (ERC) Horizon Europe under grant agreement number 101044048. JS further acknowledges support from Belgian Research Foundation Flanders (FWO) Odysseus program under grant number G0H9218N and FWO grant G077822N. 
PAC and JMB acknowledge financial support from the Science and Technology Facilities Council via research grant ST/V/000853/1 (P.I. Vik Dhillon). 
This research is based on observations obtained with the NASA/ESA Hubble Space Telescope, retrieved from the Mikulski Archive for Space Telescopes (MAST) at the Space Telescope Science Institute (STScI). STScI is operated by the Association of Universities for Research in Astronomy, Inc. under NASA contract NAS 5-26555. 
\end{acknowledgements}

\bibliographystyle{aa}
\bibliography{references}

\begin{appendix}

\onecolumn

\section{Photometry}

The used photometry, as well as the best fitting extinction values and the resulting absolute magnitudes can be found in \Cref{tab:photometry}. 

\begin{table*}[bh]

     \caption{Photometry and extinction properties of the stars in our sample. \label{tab:photometry}}
     \small
\begin{tabular}{l c c c c c c c c c c c}
\hline \hline
Source  & $U$  & $B$  & $V$  & $J$  & $H$  & $K$ & $A_V$ & $M_V$ & $A_K$ & $M_K$ & Ref\\ 
  & [mag]  & [mag]  & [mag]  & [mag]  & [mag]  & [mag] & [mag] & [mag] & [mag] & [mag] & \\ \hline
\\[-10.0pt]
\multicolumn{12}{p{0.2\textwidth}}{\textit{Presumed single}} \\ 
W61-28-23 & 12.52 & 13.65 & 13.81 & 14.16 & 14.20 & 14.24 & 0.55 & --5.22 & 0.06 & --4.30 & \tiny{c, l}\\
Sk -67$^\circ$167 & 11.24 & 12.31 & 12.53 & 12.90 & 12.88 & 12.94 & 0.54 & --6.48 & 0.06 & --5.60 & \tiny{b, l}\\
ST 92-4-18 & 12.63 & 13.64 & 13.66 & 13.77 & 13.77 & 13.78 & 0.91 & --5.73 & 0.10 & --4.80 & \tiny{c, l}\\
Sk -67$^\circ$69 & 11.88 & 12.93 & 13.09 & 13.52 & 13.64 & 13.60 & 0.49 & --5.88 & 0.06 & --4.93 & \tiny{i, l}\\
Farina-88 & 12.87 & 13.56 & 13.63 & 13.86 & 13.82 & 13.92 & 0.93 & --5.78 & 0.11 & --4.66 & \tiny{d, l}\\
LMCE 078-1 & 12.49 & 13.46 & 13.49 & 13.68 & $\cdots$ & 13.75 & 0.83 & --5.82 & 0.09 & --4.82 & \tiny{d, l}\\
Sk -67$^\circ$111 & 11.33 & 12.39 & 12.60 & 12.95 & 12.95 & 13.02 & 0.53 & --6.41 & 0.06 & --5.52 & \tiny{b, l}\\
N11-018 & $\cdots$ & 13.04 & 13.13 & 13.33 & 13.31 & 13.35 & 0.79 & --6.13 & 0.09 & --5.22 & \tiny{h, l}\\
Sk -69$^\circ$50 & 12.16 & 13.15 & 13.31 & 13.60 & 13.67 & 13.66 & 0.63 & --5.80 & 0.07 & --4.89 & \tiny{b, l}\\
Sk -71$^\circ$50 & 12.31 & 13.32 & 13.44 & 13.73 & 13.76 & 13.82 & 0.65 & --5.69 & 0.07 & --4.73 & \tiny{i, l}\\
Sk -69$^\circ$104 & 10.86 & 11.89 & 12.10 & 12.53 & 12.62 & 12.63 & 0.45 & --6.83 & 0.05 & --5.90 & \tiny{f, l}\\
Sk -68$^\circ$16 & 11.68 & 12.66 & 12.85 & 13.36 & 13.43 & 13.45 & 0.44 & --6.07 & 0.05 & --5.08 & \tiny{b, l}\\
LH 9-34 & 11.66 & 12.61 & 12.69 & 12.88 & 12.86 & 12.86 & 0.82 & --6.61 & 0.09 & --5.71 & \tiny{g, l}\\
Sk -66$^\circ$171 & 11.02 & 12.04 & 12.19 & 12.62 & 12.57 & 12.62 & 0.55 & --6.84 & 0.06 & --5.92 & \tiny{e, l}\\
Sk -71$^\circ$41 & 11.78 & 12.75 & 12.82 & 12.94 & 12.98 & 12.91 & 0.86 & --6.52 & 0.10 & --5.67 & \tiny{b, l}\\
Sk -67$^\circ$5 & 11.53 & 12.48 & 12.68 & 13.03 & 13.07 & 13.16 & 0.56 & --6.35 & 0.06 & --5.38 & \tiny{d, l}\\
\\[-10.0pt]
\multicolumn{12}{p{0.2\textwidth}}{\textit{Signs of binarity}} \\ 
LH 114-7 & 12.27 & 13.41 & 13.66 & 14.12 & 14.26 & 14.38 & 0.32 & --5.13 & 0.04 & --4.13 & \tiny{b, m}\\
VFTS-267 & $\cdots$ & 13.44 & 13.49 & 13.34 & 13.27 & 13.30 & 1.19 & --6.18 & 0.14 & --5.31 & \tiny{j, l}\\
Sk $-71^\circ$46 & 12.37 & 13.26 & 13.27 & 13.02 & 12.93 & 12.99 & 1.30 & --6.51 & 0.15 & --5.64 & \tiny{a, l}\\
Sk -67$^\circ$108 & 11.33 & 12.37 & 12.57 & 13.07 & 13.06 & 13.15 & 0.44 & --6.34 & 0.05 & --5.38 & \tiny{b, l}\\
Sk -71$^\circ$19 & 13.15 & 14.03 & 14.23 & 14.77 & 14.87 & 14.85 & 0.45 & --4.70 & 0.05 & --3.68 & \tiny{d, l}\\
Sk -70$^\circ$115 & 11.16 & 12.14 & 12.24 & 12.37 & 12.30 & 12.33 & 0.88 & --7.12 & 0.10 & --6.25 & \tiny{e, l}\\
BI 173 & 11.78 & 12.80 & 12.96 & 13.24 & 13.27 & 13.30 & 0.64 & --6.15 & 0.07 & --5.25 & \tiny{b, l}\\
Sk -68$^\circ$155 & 11.84 & 12.74 & 12.75 & 12.63 & 12.64 & 12.60 & 1.14 & --6.87 & 0.13 & --6.01 & \tiny{b, l}\\
BI 272 & 12.05 & 13.06 & 13.28 & 13.72 & 13.79 & 13.86 & 0.45 & --5.64 & 0.05 & --4.67 & \tiny{k, l}\\
\hline
\multicolumn{12}{p{14.6cm}}{\tiny \textbf{Notes.} $A_V$, $M_V$, $A_K$ and $M_K$ are derived in this work by fitting the \citet{1999PASP..111...63F} extinction curve, adopting $R_V = 3.1$ (see main text). Observed magnitudes are taken from the literature overview of \citet{Vinkinprep}; 
references are indicated in the last column, where the first letter refers to the $U$, $B$, and $V$ photometry, and the second to the $J$, $H$, and $K_s$ photometry:  \textbf{a}: \citet{1995ApJ...438..188M},
\textbf{b}: \citet{2002ApJS..141...81M},
\textbf{c}: \citet{2000AJ....119.2214M},
\textbf{d}: \citet{2004AJ....128.1606Z},
\textbf{e}: \citet{1975AAS...19..259I},
\textbf{f}: \citet{1972AAS....6..249A},
\textbf{g}: \citet{1999MNRAS.306..279S},
\textbf{h}: \citet{2006AA...456..623E},
\textbf{i}: \citet{1979AAS...38..239I},
\textbf{j}: \citet{2011AA...530A.108E},
\textbf{k}: \citet{1982AAS...50....7I},
\textbf{l}: $J$ and $K_s$ from \citet{2011AA...527A.116C}, $H$ from \citet{2003yCat.2246....0C},
\textbf{m}: $J$, $H$, and $K_s$ from \citet{2012yCat.2281....0C}.
}
\end{tabular}
\end{table*}

\section{Results of the combined optical and UV fits}

\begin{sidewaysfigure*}
    \centering
    \includegraphics[width=1.0\textwidth]{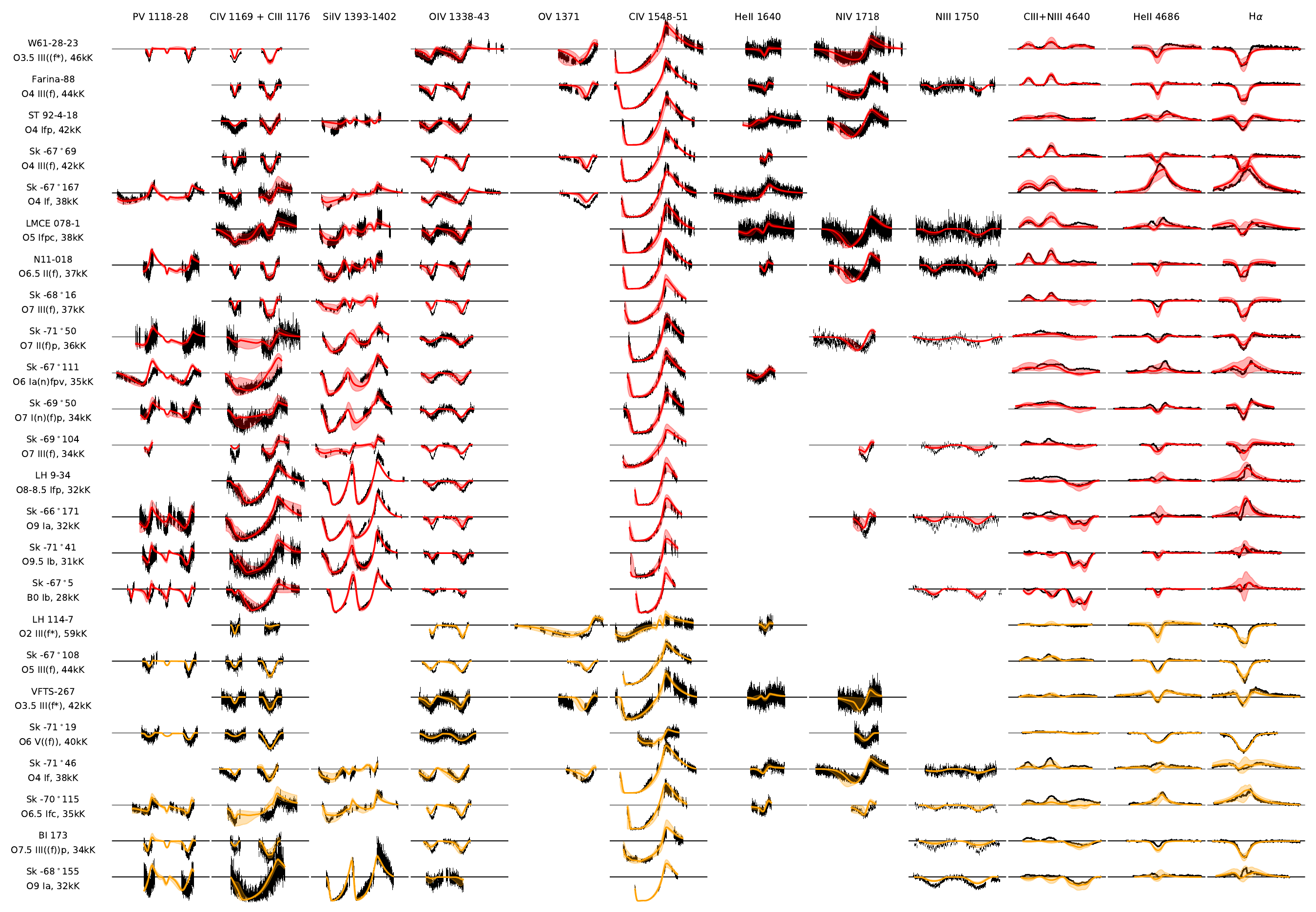}
    \caption{Normalised spectra (black) and best fit models of the UV lines, as well as the \niii-\ciii complex, \heiiline, and \halpha for all presumed single stars (red) and binaries (orange), ordered by increasing \teff. The shaded areas indicate $2\sigma$ uncertainties. Black horizontal lines indicate the continuum; the continua of the different stars are plotted 1.5 normalised flux units apart for the UV lines, and 0.45 for the optical lines. \cref{fig:examplefit:farina88} and similar plots for the other sources that can be found on \href{https://zenodo.org/records/15013513}{Zenodo} provide a more comprehensive overview per star.  \label{fig:windlines}}
\end{sidewaysfigure*}

\begin{sidewaysfigure*}
    \centering
    \includegraphics[width=1.0\textwidth]{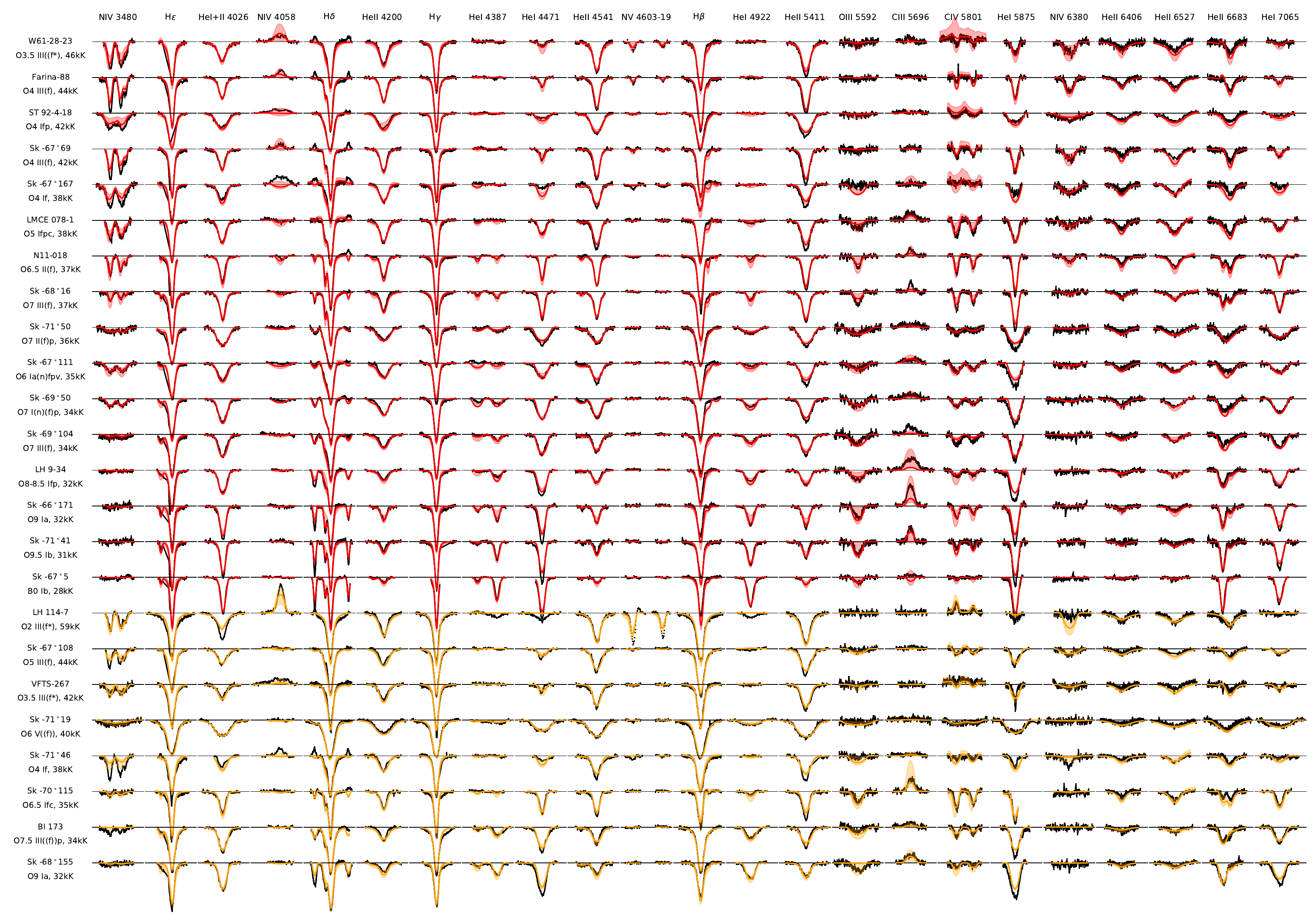}
    \caption{As \cref{fig:windlines}, but for the remainder of the optical lines. The continua of the different stars are plotted 0.25 normalised flux units apart. \label{fig:otherlines}}
\end{sidewaysfigure*}

A comparison of the observed spectra and best fit models for all stars is presented in 
\cref{fig:windlines} (UV diagnostics and the main wind-sensitive optical lines; the \ciii-\niii complex at 4640, \heiiline and \halpha), and \cref{fig:otherlines} (remainder of the optical lines). The spectra are ordered by temperature. For more details, see \cref{ullyses:results}. 
A detailed fit overview per star, including fitness distributions of the fitted parameters can be found on \href{https://zenodo.org/records/15013513}{Zenodo}. 

The best fit parameters and associated $1 \sigma$ uncertainties of the optical + UV fits can be found in \Cref{tab:stellarparams} and \ref{tab:bestfitswind}. \Cref{tab:stellarparams} includes the helium abundance and projected surface rotation as from the optical-only fits. The complete set of parameters derived from the optical-only fit can be found on \href{https://zenodo.org/records/15013513}{Zenodo}.

\renewcommand{\arraystretch}{1.4}
\begin{sidewaystable*}[]
    \centering
    \small 
    \parbox{22cm}{\caption{Best fit values of stellar parameters, abundances, masses, and age. 
    \label{tab:stellarparams}}}
\begin{tabular}{l r@{$\pm$}l  r@{$\pm$}l  r@{$\pm$}l  r@{$\pm$}l  r@{$\pm$}l  r@{$\pm$}l  r@{$\pm$}l  r@{$\pm$}l  r@{$\pm$}l  r@{$\pm$}l  r@{$\pm$}l  r@{$\pm$}l  r@{$\pm$}l }
\hline \hline
Source  &  \multicolumn{2}{c}{$\log L/\mathrm{L}_\odot$} &  \multicolumn{2}{c}{$T_\mathrm{eff}$} &  \multicolumn{2}{c}{$\log g$} &  \multicolumn{2}{c}{$R_*$} &  \multicolumn{2}{c}{$x_\mathrm{C}$} &  \multicolumn{2}{c}{$x_\mathrm{N}$} &  \multicolumn{2}{c}{$x_\mathrm{O}$} &  \multicolumn{2}{c}{$y_\mathrm{He}$} &  \multicolumn{2}{c}{$\varv \sin i$} &  \multicolumn{2}{c}{$M_\mathrm{spec}$} &  \multicolumn{2}{c}{$M_\mathrm{evol}$} &  \multicolumn{2}{c}{$M_\mathrm{ini}$} &  \multicolumn{2}{c}{Age}\\
 &  \multicolumn{2}{c}{} &  \multicolumn{2}{c}{[K]} &  \multicolumn{2}{c}{[cm~s$^{-2}$]} &  \multicolumn{2}{c}{[$\mathrm{R}_\odot$]} &  \multicolumn{2}{c}{} &  \multicolumn{2}{c}{} &  \multicolumn{2}{c}{} &  \multicolumn{2}{c}{} &  \multicolumn{2}{c}{[km~s$^{-1}$]} &  \multicolumn{2}{c}{[$\mathrm{M}_\odot$]} &  \multicolumn{2}{c}{[$\mathrm{M}_\odot$]} &  \multicolumn{2}{c}{[$\mathrm{M}_\odot$]} &  \multicolumn{2}{c}{[Myr]}\\
\hline
\\[-10.0pt]
\multicolumn{8}{p{0.2\textwidth}}{\textit{Presumed single}} \\ 
W61-28-23 & 5.60 & $^{0.04}_{0.05}$  & 45750 & $^{1250}_{1750}$  & 3.88 & $^{0.25}_{0.23}$  & 10.1 & $^{0.3}_{0.3}$  & 9.00 & $^{0.05}_{0.60}$  & 8.65 & $^{0.25}_{0.70}$  & 8.45 & $^{0.55}_{0.40}$  & 0.235 & $^{0.025}_{0.080}$  & 130 & $^{25}_{20}$  & 29 & $^{19}_{10}$  & 47 & $^{3}_{3}$  & 48 & $^{3}_{2}$  & 1.2 & $^{0.5}_{0.6}$ \\
Sk $-67^\circ$167 & 5.88 & $^{0.03}_{0.03}$  & 38500 & $^{1000}_{1000}$  & 3.45 & $^{0.10}_{0.07}$  & 19.8 & $^{0.6}_{0.6}$  & 7.80 & $^{0.15}_{0.25}$  & 9.05 & $^{0.15}_{0.30}$  & 9.00 & $^{0.05}_{0.55}$  & 0.145 & $^{0.035}_{0.020}$  & 175 & $^{25}_{35}$  & 43 & $^{8}_{5}$  & 56 & $^{2}_{2}$  & 59 & $^{2}_{2}$  & 2.4 & $^{0.1}_{0.1}$ \\
ST 92-4-18 & 5.67 & $^{0.04}_{0.08}$  & 41500 & $^{1250}_{2500}$  & 3.73 & $^{0.10}_{0.30}$  & 13.4 & $^{0.6}_{0.4}$  & 8.30 & $^{0.30}_{0.50}$  & 8.60 & $^{0.35}_{0.15}$  & 8.05 & $^{0.95}_{0.45}$  & 0.195 & $^{0.035}_{0.045}$  & 240 & $^{25}_{20}$  & 39 & $^{7}_{15}$  & 44 & $^{3}_{3}$  & 46 & $^{3}_{3}$  & 2.4 & $^{0.4}_{0.3}$ \\
Sk $-67^\circ$69 & 5.72 & $^{0.05}_{0.04}$  & 41500 & $^{1750}_{1000}$  & 3.58 & $^{0.17}_{0.07}$  & 14.2 & $^{0.5}_{0.5}$  & 7.90 & $^{0.40}_{0.40}$  & 8.50 & $^{0.30}_{0.15}$  & 7.60 & $^{0.30}_{0.35}$  & 0.145 & $^{0.035}_{0.030}$  & 115 & $^{20}_{15}$  & 28 & $^{10}_{3}$  & 49 & $^{2}_{2}$  & 51 & $^{3}_{2}$  & 2.3 & $^{0.2}_{0.3}$ \\
Farina-88 & 5.69 & $^{0.04}_{0.04}$  & 43750 & $^{1000}_{250}$  & 3.73 & $^{0.15}_{0.03}$  & 12.3 & $^{0.6}_{0.6}$  & 8.10 & $^{0.10}_{0.15}$  & 8.55 & $^{0.20}_{0.30}$  & 7.75 & $^{0.10}_{0.20}$  & 0.140 & $^{0.025}_{0.040}$  & 100 & $^{20}_{20}$  & 30 & $^{10}_{3}$  & 50 & $^{1}_{1}$  & 51 & $^{1}_{1}$  & 1.9 & $^{0.1}_{0.2}$ \\
LMCe 078-1 & 5.55 & $^{0.05}_{0.03}$  & 37750 & $^{1500}_{500}$  & 3.42 & $^{0.25}_{0.05}$  & 14.0 & $^{0.4}_{0.5}$  & 8.25 & $^{0.30}_{0.25}$  & 8.95 & $^{0.25}_{0.35}$  & 8.10 & $^{0.95}_{0.45}$  & 0.195 & $^{0.055}_{0.025}$  & 150 & $^{25}_{20}$  & 21 & $^{12}_{2}$  & 39 & $^{1}_{1}$  & 40 & $^{2}_{1}$  & 3.0 & $^{0.2}_{0.2}$ \\
Sk $-67^\circ$111 & 5.74 & $^{0.05}_{0.03}$  & 35000 & $^{1500}_{750}$  & 3.33 & $^{0.20}_{0.07}$  & 20.3 & $^{0.6}_{0.7}$  & 8.10 & $^{0.30}_{0.45}$  & 8.95 & $^{0.35}_{0.30}$  & 8.30 & $^{0.65}_{0.65}$  & 0.145 & $^{0.005}_{0.055}$  & 215 & $^{15}_{35}$  & 37 & $^{14}_{4}$  & 45 & $^{2}_{1}$  & 48 & $^{3}_{2}$  & 3.0 & $^{0.1}_{0.2}$ \\
N11-018 & 5.69 & $^{0.03}_{0.04}$  & 37000 & $^{250}_{1000}$  & 3.50 & $^{0.07}_{0.20}$  & 17.2 & $^{0.6}_{0.6}$  & 7.70 & $^{0.35}_{0.40}$  & 8.70 & $^{0.35}_{0.35}$  & 8.15 & $^{0.35}_{0.80}$  & 0.135 & $^{0.030}_{0.045}$  & 95 & $^{30}_{10}$  & 35 & $^{5}_{11}$  & 43 & $^{1}_{1}$  & 45 & $^{1}_{1}$  & 3.0 & $^{0.1}_{0.1}$ \\
Sk $-69^\circ$50 & 5.48 & $^{0.02}_{0.02}$  & 34500 & $^{250}_{250}$  & 3.33 & $^{0.03}_{0.05}$  & 15.5 & $^{0.4}_{0.4}$  & 7.60 & $^{0.65}_{0.05}$  & 8.95 & $^{0.05}_{0.20}$  & 8.10 & $^{0.95}_{0.30}$  & 0.140 & $^{0.030}_{0.030}$  & 185 & $^{30}_{10}$  & 21 & $^{1}_{2}$  & 33 & $^{0}_{0}$  & 35 & $^{0}_{0}$  & 3.8 & $^{0.1}_{0.1}$ \\
Sk $-71^\circ$50 & 5.47 & $^{0.02}_{0.04}$  & 36250 & $^{250}_{1250}$  & 3.42 & $^{0.12}_{0.10}$  & 14.0 & $^{0.4}_{0.4}$  & 7.65 & $^{0.25}_{0.25}$  & 8.55 & $^{0.05}_{0.70}$  & 8.55 & $^{0.10}_{0.45}$  & 0.120 & $^{0.040}_{0.030}$  & 255 & $^{20}_{25}$  & 24 & $^{5}_{3}$  & 33 & $^{1}_{1}$  & 35 & $^{1}_{1}$  & 3.6 & $^{0.2}_{0.1}$ \\
Sk $-69^\circ$104 & 5.88 & $^{0.04}_{0.02}$  & 34500 & $^{1000}_{250}$  & 3.38 & $^{0.20}_{0.07}$  & 24.5 & $^{0.7}_{0.7}$  & 7.45 & $^{0.40}_{0.10}$  & 7.95 & $^{0.65}_{1.10}$  & 8.95 & $^{0.10}_{0.60}$  & 0.095 & $^{0.015}_{0.005}$  & 195 & $^{15}_{20}$  & 57 & $^{24}_{6}$  & 53 & $^{2}_{1}$  & 57 & $^{2}_{1}$  & 2.7 & $^{0.1}_{0.1}$ \\
Sk $-68^\circ$16 & 5.65 & $^{0.03}_{0.04}$  & 37000 & $^{750}_{1000}$  & 3.62 & $^{0.23}_{0.12}$  & 16.3 & $^{0.6}_{0.6}$  & 7.70 & $^{0.30}_{0.35}$  & 8.35 & $^{0.35}_{0.55}$  & 8.05 & $^{0.35}_{0.40}$  & 0.095 & $^{0.030}_{0.005}$  & 110 & $^{20}_{20}$  & 42 & $^{23}_{9}$  & 41 & $^{1}_{1}$  & 43 & $^{1}_{1}$  & 3.1 & $^{0.1}_{0.1}$ \\
LH 9-34 & 5.72 & $^{0.04}_{0.04}$  & 32250 & $^{1000}_{1000}$  & 3.27 & $^{0.03}_{0.23}$  & 23.5 & $^{0.8}_{0.8}$  & 7.75 & $^{0.70}_{0.05}$  & 7.15 & $^{1.35}_{1.10}$  & 8.30 & $^{0.50}_{0.55}$  & 0.110 & $^{0.050}_{0.020}$  & 175 & $^{25}_{25}$  & 42 & $^{3}_{14}$  & 42 & $^{2}_{1}$  & 45 & $^{2}_{2}$  & 3.4 & $^{0.1}_{0.1}$ \\
Sk $-66^\circ$171 & 5.80 & $^{0.03}_{0.07}$  & 32250 & $^{750}_{1750}$  & 3.20 & $^{0.47}_{0.12}$  & 25.8 & $^{1.1}_{0.9}$  & 8.25 & $^{0.65}_{0.30}$  & 7.85 & $^{0.30}_{1.00}$  & 8.70 & $^{0.15}_{0.95}$  & 0.105 & $^{0.060}_{0.015}$  & 125 & $^{10}_{25}$  & 40 & $^{68}_{7}$  & 46 & $^{2}_{3}$  & 49 & $^{2}_{3}$  & 3.2 & $^{0.2}_{0.1}$ \\
Sk $-71^\circ$41 & 5.65 & $^{0.04}_{0.05}$  & 31000 & $^{1000}_{1250}$  & 3.12 & $^{0.23}_{0.07}$  & 23.5 & $^{0.9}_{0.8}$  & 8.65 & $^{0.20}_{0.55}$  & 7.15 & $^{0.60}_{0.75}$  & 8.75 & $^{0.30}_{1.45}$  & 0.105 & $^{0.040}_{0.015}$  & 115 & $^{10}_{20}$  & 29 & $^{15}_{3}$  & 38 & $^{2}_{2}$  & 41 & $^{2}_{2}$  & 3.7 & $^{0.2}_{0.1}$ \\
Sk $-67^\circ$5 & 5.40 & $^{0.04}_{0.04}$  & 27500 & $^{1000}_{1000}$  & 2.98 & $^{0.10}_{0.15}$  & 22.3 & $^{0.7}_{0.7}$  & 8.40 & $^{0.30}_{0.40}$  & 8.10 & $^{0.10}_{0.85}$  & 8.75 & $^{0.30}_{1.45}$  & 0.095 & $^{0.025}_{0.005}$  & 110 & $^{10}_{15}$  & 19 & $^{3}_{4}$  & 28 & $^{1}_{1}$  & 29 & $^{1}_{1}$  & 5.1 & $^{0.2}_{0.2}$ \\
\\[-10.0pt]
\multicolumn{8}{p{0.2\textwidth}}{\textit{Signs of binarity}} \\ 
LH~114-7 & 5.88 & $^{0.07}_{0.04}$  & 59000 & $^{3250}_{1750}$  & 4.22 & $^{0.17}_{0.07}$  & 8.4 & $^{0.2}_{0.3}$  & 8.50 & $^{0.35}_{0.50}$  & 8.75 & $^{0.50}_{0.25}$  & 8.00 & $^{0.45}_{0.25}$  & 0.185 & $^{0.080}_{0.045}$  & 120 & $^{45}_{30}$  & 44 & $^{16}_{5}$  &  \multicolumn{2}{c}{$-$}  &  \multicolumn{2}{c}{$-$}  &  \multicolumn{2}{c}{$-$} \\
VFTS-267 & 5.91 & $^{0.04}_{0.04}$  & 42500 & $^{1250}_{1250}$  & 3.75 & $^{0.15}_{0.15}$  & 16.8 & $^{0.6}_{0.6}$  & 7.70 & $^{0.30}_{0.50}$  & 7.85 & $^{0.40}_{0.90}$  & 8.20 & $^{0.25}_{1.10}$  & 0.095 & $^{0.010}_{0.005}$  & 155 & $^{25}_{45}$  & 60 & $^{20}_{14}$  & 61 & $^{3}_{2}$  & 64 & $^{3}_{3}$  & 2.0 & $^{0.2}_{0.1}$ \\
Sk $-71^\circ$46 & 5.88 & $^{0.04}_{0.02}$  & 37750 & $^{1250}_{250}$  & 3.73 & $^{0.28}_{0.28}$  & 20.6 & $^{0.6}_{0.6}$  & 7.95 & $^{0.30}_{0.60}$  & 8.05 & $^{0.45}_{0.25}$  & 8.65 & $^{0.40}_{0.60}$  & 0.095 & $^{0.025}_{0.005}$  & 175 & $^{30}_{40}$  & 86 & $^{60}_{36}$  & 59 & $^{2}_{2}$  & 63 & $^{3}_{2}$  & 2.3 & $^{0.1}_{0.1}$ \\
Sk $-67^\circ$108 & 5.97 & $^{0.03}_{0.04}$  & 43500 & $^{750}_{1000}$  & 3.88 & $^{0.03}_{0.17}$  & 17.3 & $^{0.6}_{0.5}$  & 7.90 & $^{0.50}_{0.10}$  & 8.40 & $^{0.15}_{0.60}$  & 8.65 & $^{0.30}_{0.35}$  & 0.095 & $^{0.020}_{0.005}$  & 185 & $^{15}_{20}$  & 85 & $^{5}_{24}$  & 66 & $^{2}_{2}$  & 71 & $^{2}_{2}$  & 1.8 & $^{0.1}_{0.1}$ \\
Sk $-71^\circ$19 & 5.17 & $^{0.04}_{0.04}$  & 39750 & $^{250}_{500}$  & 3.90 & $^{0.03}_{0.17}$  & 8.2 & $^{0.4}_{0.3}$  & 7.75 & $^{0.30}_{0.20}$  & 8.05 & $^{0.45}_{0.30}$  & 9.00 & $^{0.05}_{0.60}$  & 0.140 & $^{0.020}_{0.025}$  & 325 & $^{20}_{15}$  & 24 & $^{2}_{6}$  & 29 & $^{0}_{1}$  & 29 & $^{0}_{0}$  & 2.3 & $^{0.3}_{0.2}$ \\
Sk $-70^\circ$115 & 6.02 & $^{0.04}_{0.04}$  & 34750 & $^{1000}_{1000}$  & 3.27 & $^{0.10}_{0.12}$  & 28.4 & $^{0.9}_{0.9}$  & 8.20 & $^{0.75}_{0.35}$  & 7.80 & $^{0.30}_{1.25}$  & 8.55 & $^{0.35}_{0.55}$  & 0.095 & $^{0.015}_{0.005}$  & 125 & $^{10}_{40}$  & 58 & $^{11}_{11}$  & 64 & $^{3}_{2}$  & 70 & $^{3}_{3}$  & 2.4 & $^{0.1}_{0.1}$ \\
BI~173 & 5.64 & $^{0.05}_{0.02}$  & 34500 & $^{1500}_{250}$  & 3.58 & $^{0.20}_{0.10}$  & 18.6 & $^{0.5}_{0.6}$  & 8.00 & $^{0.30}_{0.35}$  & 7.20 & $^{0.65}_{0.60}$  & 8.45 & $^{0.50}_{0.30}$  & 0.095 & $^{0.030}_{0.005}$  & 180 & $^{25}_{25}$  & 51 & $^{21}_{8}$  & 38 & $^{1}_{1}$  & 40 & $^{2}_{1}$  & 3.4 & $^{0.1}_{0.2}$ \\
Sk $-68^\circ$155 & 5.81 & $^{0.03}_{0.02}$  & 31500 & $^{750}_{250}$  & 3.20 & $^{0.20}_{0.05}$  & 27.3 & $^{0.8}_{0.8}$  & 8.15 & $^{0.35}_{0.20}$  & 8.00 & $^{0.35}_{0.50}$  & 6.85 & $^{1.85}_{0.65}$  & 0.135 & $^{0.060}_{0.005}$  & 155 & $^{20}_{5}$  & 47 & $^{20}_{3}$  & 47 & $^{1}_{1}$  & 51 & $^{1}_{1}$  & 3.1 & $^{0.1}_{0.1}$ \\
 \hline 
 \multicolumn{27}{p{21.4cm}}{\small \textbf{Notes.} All values in this table are results from the combined optical and UV run, except for $\varv \sin i$ and $y_\mathrm{He}$, which were fixed in the combined analysis to the best fit values from the optical-only run (the optical-only parameters are listed here for $\varv \sin i$ and $y_\mathrm{He}$; a complete overview can be found on \href{https://zenodo.org/records/15013513}{Zenodo}). Evolutionary mass $M_\mathrm{evol}$, initial mass $M_\mathrm{ini}$ and age are derived using \textsc{Bonnsai} (see text). Surface gravity values quoted here are not corrected for centrifugal acceleration, but this is considered when computing the spectroscopic mass, which is defined as $M_\mathrm{spec} = g_e  R_*^2/G$, with $g_e = g + (v_{\mathrm{eq}}\sin i)^2/R_*$ the gravitational acceleration at the surface corrected for centrifugal acceleration, and $G$ the gravitational constant. Baseline abundances in the LMC are $x_\mathrm{C} = 8.01$, $x_\mathrm{N} = 7.03$, $x_\mathrm{O} = 8.40$ \citep{2022MNRAS.515.4130C,Vinkinprep}.  }
    \end{tabular}
\end{sidewaystable*}

\begin{sidewaystable*}[]
    \centering
    \small 
    \parbox{20.2cm}{\caption{Best fit values of wind parameters, Eddington factors and ionising fluxes. \label{tab:bestfitswind}}}
\begin{tabular}{l r@{$\pm$}l  r@{$\pm$}l  r@{$\pm$}l  r@{$\pm$}l  r@{$\pm$}l  r@{$\pm$}l  r@{$\pm$}l  r@{$\pm$}l  r@{$\pm$}l  c  c  c}
\hline \hline
Source  &  \multicolumn{2}{c}{$\log \dot{M}$} &  \multicolumn{2}{c}{$\varv_\infty$} &  \multicolumn{2}{c}{$\varv_\mathrm{windturb}$} &  \multicolumn{2}{c}{$\beta$} &  \multicolumn{2}{c}{$f_\mathrm{cl}$} &  \multicolumn{2}{c}{$f_\mathrm{ic}$} &  \multicolumn{2}{c}{$f_\mathrm{vel}$} &  \multicolumn{2}{c}{$\varv_\mathrm{cl,start}$} &  \multicolumn{2}{c}{$\Gamma_\mathrm{Edd,e}$} & $\log Q_0$ & $\log Q_1$ \\
 &  \multicolumn{2}{c}{[$\mathrm{M}_\odot$~yr$^{-1}$]} &  \multicolumn{2}{c}{[km~s$^{-1}$]} &  \multicolumn{2}{c}{[$\varv_\infty$]} &  \multicolumn{2}{c}{} &  \multicolumn{2}{c}{} &  \multicolumn{2}{c}{} &  \multicolumn{2}{c}{} &  \multicolumn{2}{c}{[$\varv_\infty$]} &  \multicolumn{2}{c}{} &  [s$^{-1}$] &  [s$^{-1}$] \\
\hline
\\[-10.0pt]
\multicolumn{8}{p{0.2\textwidth}}{\textit{Presumed single}} \\ 
W61-28-23 & $-6.25$ & $^{0.25}_{0.10}$  & $3125$ & $^{125}_{150}$  & $0.04$ & $^{0.04}_{0.04}$  & $1.80$ & $^{0.30}_{0.50}$  & $(2$ & $^{13}_{2})$  & $(0.29$ & $^{0.15}_{0.26})$  & $(0.40$ & $^{0.60}_{0.30})$  & $(0.07$ & $^{0.23}_{0.02})$  & $0.22$ & $^{0.02}_{0.03}$  & $49.4$ & $48.7$\\
Sk $-67^\circ$167 & $-5.66$ & $^{0.13}_{0.08}$  & $2375$ & $^{100}_{225}$  & $0.12$ & $^{0.09}_{0.02}$  & $0.90$ & $^{0.25}_{0.15}$  & $42$ & $^{5}_{19}$  & $0.09$ & $^{0.09}_{0.04}$  & $0.40$ & $^{0.24}_{0.10}$  & $0.07$ & $^{0.04}_{0.02}$  & $0.36$ & $^{0.03}_{0.03}$  & $49.6$ & $48.7$\\
ST 92-4-18 & $-6.08$ & $^{0.05}_{0.18}$  & $2350$ & $^{100}_{125}$  & $0.07$ & $^{0.06}_{0.02}$  & $1.50$ & $^{0.45}_{0.35}$  & $32$ & $^{19}_{10}$  & $(0.46$ & $^{0.07}_{0.45})$  & $(0.38$ & $^{0.64}_{0.16})$  & $0.18$ & $^{0.06}_{0.06}$  & $0.28$ & $^{0.03}_{0.05}$  & $49.4$ & $48.6$\\
Sk $-67^\circ$69 & $-6.18$ & $^{0.03}_{0.23}$  & $2550$ & $^{75}_{125}$  & $0.07$ & $^{0.05}_{0.02}$  & $1.65$ & $^{0.40}_{0.30}$  & $11$ & $^{18}_{6}$  & $0.42$ & $^{0.11}_{0.31}$  & $0.56$ & $^{0.38}_{0.26}$  & $0.08$ & $^{0.05}_{0.02}$  & $0.28$ & $^{0.04}_{0.03}$  & $49.5$ & $48.7$\\
Farina-88 & $-6.35$ & $^{0.06}_{0.15}$  & $2750$ & $^{100}_{25}$  & $0.08$ & $^{0.01}_{0.05}$  & $1.60$ & $^{0.25}_{0.05}$  & $10$ & $^{8}_{4}$  & $0.11$ & $^{0.14}_{0.00}$  & $0.44$ & $^{0.10}_{0.16}$  & $0.06$ & $^{0.01}_{0.04}$  & $0.26$ & $^{0.03}_{0.02}$  & $49.5$ & $48.8$\\
LMCe 078-1 & $-6.01$ & $^{0.18}_{0.08}$  & $2325$ & $^{125}_{200}$  & $0.13$ & $^{0.04}_{0.08}$  & $1.65$ & $^{0.20}_{0.30}$  & $50$ & $^{1}_{33}$  & $0.14$ & $^{0.23}_{0.07}$  & $0.74$ & $^{0.28}_{0.22}$  & $0.28$ & $^{0.03}_{0.05}$  & $0.24$ & $^{0.03}_{0.02}$  & $49.3$ & $48.4$\\
Sk $-67^\circ$111 & $-6.01$ & $^{0.15}_{0.08}$  & $2000$ & $^{125}_{100}$  & $0.08$ & $^{0.04}_{0.04}$  & $1.70$ & $^{0.75}_{0.30}$  & $31$ & $^{14}_{15}$  & $0.28$ & $^{0.23}_{0.23}$  & $(0.82$ & $^{0.18}_{0.46})$  & $0.14$ & $^{0.08}_{0.03}$  & $0.32$ & $^{0.04}_{0.02}$  & $49.4$ & $48.3$\\
N11-018 & $-6.03$ & $^{0.15}_{0.08}$  & $2300$ & $^{75}_{125}$  & $0.10$ & $^{0.04}_{0.03}$  & $1.75$ & $^{0.25}_{0.50}$  & $48$ & $^{3}_{24}$  & $0.38$ & $^{0.14}_{0.24}$  & $0.92$ & $^{0.10}_{0.30}$  & $0.29$ & $^{0.01}_{0.05}$  & $0.30$ & $^{0.02}_{0.03}$  & $49.4$ & $48.4$\\
Sk $-69^\circ$50 & $-6.16$ & $^{0.03}_{0.13}$  & $1925$ & $^{25}_{150}$  & $0.14$ & $^{0.04}_{0.01}$  & $0.95$ & $^{0.10}_{0.05}$  & $35$ & $^{8}_{4}$  & $0.23$ & $^{0.06}_{0.02}$  & $(0.04$ & $^{0.04}_{0.02})$  & $0.14$ & $^{0.03}_{0.01}$  & $0.23$ & $^{0.01}_{0.01}$  & $49.1$ & $47.9$\\
Sk $-71^\circ$50 & $-6.43$ & $^{0.13}_{0.05}$  & $1875$ & $^{50}_{125}$  & $0.14$ & $^{0.06}_{0.01}$  & $1.25$ & $^{0.30}_{0.30}$  & $43$ & $^{2}_{18}$  & $0.32$ & $^{0.01}_{0.23}$  & $0.52$ & $^{0.30}_{0.04}$  & $0.14$ & $^{0.04}_{0.04}$  & $0.23$ & $^{0.01}_{0.02}$  & $49.1$ & $48.1$\\
Sk $-69^\circ$104 & $-5.93$ & $^{0.13}_{0.15}$  & $2700$ & $^{225}_{200}$  & $0.20$ & $^{0.08}_{0.10}$  & $0.75$ & $^{0.25}_{0.15}$  & $44$ & $^{8}_{24}$  & $0.05$ & $^{0.03}_{0.02}$  & $0.28$ & $^{0.20}_{0.10}$  & $0.03$ & $^{0.05}_{0.02}$  & $0.37$ & $^{0.03}_{0.02}$  & $49.4$ & $48.3$\\
Sk $-68^\circ$16 & $-6.22$ & $^{0.13}_{0.15}$  & $2400$ & $^{250}_{75}$  & $0.14$ & $^{0.16}_{0.10}$  & $1.70$ & $^{0.35}_{0.40}$  & $45$ & $^{7}_{17}$  & $0.03$ & $^{0.04}_{0.02}$  & $0.50$ & $^{0.20}_{0.20}$  & $0.29$ & $^{0.01}_{0.04}$  & $0.28$ & $^{0.02}_{0.03}$  & $49.3$ & $48.3$\\
LH 9-34 & $-5.80$ & $^{0.15}_{0.30}$  & $1500$ & $^{75}_{100}$  & $0.09$ & $^{0.07}_{0.01}$  & $1.15$ & $^{0.60}_{0.15}$  & $6$ & $^{11}_{4}$  & $0.38$ & $^{0.02}_{0.17}$  & $0.92$ & $^{0.08}_{0.22}$  & $0.10$ & $^{0.14}_{0.04}$  & $0.33$ & $^{0.03}_{0.03}$  & $49.2$ & $47.7$\\
Sk $-66^\circ$171 & $-6.18$ & $^{0.55}_{0.08}$  & $1850$ & $^{75}_{100}$  & $0.17$ & $^{0.02}_{0.06}$  & $1.60$ & $^{0.70}_{0.30}$  & $(41$ & $^{1}_{35})$  & $(0.08$ & $^{0.07}_{0.05})$  & $(0.90$ & $^{0.08}_{0.20})$  & $(0.06$ & $^{0.18}_{0.04})$  & $0.36$ & $^{0.03}_{0.06}$  & $49.3$ & $47.9$\\
Sk $-71^\circ$41 & $-6.55$ & $^{0.18}_{0.05}$  & $1850$ & $^{75}_{175}$  & $0.18$ & $^{0.06}_{0.11}$  & $2.90$ & $^{0.65}_{0.25}$  & $30$ & $^{19}_{8}$  & $(0.25$ & $^{0.19}_{0.23})$  & $0.76$ & $^{0.20}_{0.12}$  & $0.04$ & $^{0.07}_{0.01}$  & $0.31$ & $^{0.03}_{0.04}$  & $49.1$ & $47.4$\\
Sk $-67^\circ$5 & $-6.64$ & $^{0.15}_{0.13}$  & $1425$ & $^{75}_{75}$  & $0.14$ & $^{0.06}_{0.04}$  & $2.25$ & $^{0.50}_{0.90}$  & $34$ & $^{17}_{8}$  & $0.18$ & $^{0.25}_{0.10}$  & $0.82$ & $^{0.10}_{0.68}$  & $0.10$ & $^{0.07}_{0.06}$  & $0.23$ & $^{0.03}_{0.02}$  & $48.4$ & $45.6$\\
\\[-10.0pt]

\multicolumn{8}{p{0.2\textwidth}}{\textit{Signs of binarity}} \\ 
LH~114-7 & $-6.46$ & $^{0.10}_{0.20}$  & $3425$ & $^{350}_{375}$  & $0.27$ & $^{0.04}_{0.14}$  & $1.00$ & $^{0.25}_{0.20}$  & $24$ & $^{28}_{5}$  & $0.36$ & $^{0.16}_{0.33}$  & $0.42$ & $^{0.54}_{0.26}$  & $0.01$ & $^{0.04}_{0.01}$  &  0.47 & $_{0.03}^{0.12}$  & $49.8$ & $49.3$\\
VFTS-267 & $-6.15$ & $^{0.18}_{0.03}$  & $2700$ & $^{200}_{75}$  & $0.04$ & $^{0.07}_{0.03}$  & $1.70$ & $^{0.40}_{0.40}$  & $51$ & $^{1}_{21}$  & $0.05$ & $^{0.07}_{0.02}$  & $0.26$ & $^{0.40}_{0.24}$  & $0.10$ & $^{0.04}_{0.01}$  & $0.35$ & $^{0.04}_{0.04}$  & $49.7$ & $48.9$\\
Sk $-71^\circ$46 & $-5.68$ & $^{0.10}_{0.10}$  & $2575$ & $^{100}_{100}$  & $0.06$ & $^{0.04}_{0.01}$  & $1.70$ & $^{0.20}_{0.30}$  & $29$ & $^{16}_{7}$  & $(0.50$ & $^{0.02}_{0.49})$  & $0.96$ & $^{0.06}_{0.18}$  & $0.29$ & $^{0.01}_{0.04}$  & $0.34$ & $^{0.03}_{0.02}$  & $49.6$ & $48.7$\\
Sk $-67^\circ$108 & $-6.37$ & $^{0.10}_{0.15}$  & $2675$ & $^{125}_{125}$  & $0.03$ & $^{0.07}_{0.04}$  & $1.35$ & $^{0.15}_{0.40}$  & $36$ & $^{5}_{13}$  & $0.10$ & $^{0.00}_{0.06}$  & $0.02$ & $^{0.06}_{0.02}$  & $0.04$ & $^{0.01}_{0.02}$  & $0.37$ & $^{0.03}_{0.03}$  & $49.7$ & $49.0$\\
Sk $-71^\circ$19 & $-7.20$ & $^{0.04}_{0.33}$  & $1575$ & $^{225}_{275}$  & $0.20$ & $^{0.07}_{0.20}$  & $0.75$ & $^{0.65}_{0.10}$  & $40$ & $^{9}_{19}$  & $0.01$ & $^{0.01}_{0.00}$  & $0.02$ & $^{0.06}_{0.02}$  & $0.10$ & $^{0.06}_{0.04}$  & $0.13$ & $^{0.01}_{0.01}$  & $48.8$ & $48.0$\\
Sk $-70^\circ$115 & $-5.73$ & $^{0.08}_{0.13}$  & $2150$ & $^{100}_{125}$  & $0.04$ & $^{0.04}_{0.03}$  & $1.25$ & $^{0.15}_{0.45}$  & $33$ & $^{17}_{5}$  & $0.14$ & $^{0.13}_{0.06}$  & $0.08$ & $^{0.18}_{0.08}$  & $0.05$ & $^{0.03}_{0.04}$  & $0.43$ & $^{0.04}_{0.04}$  & $49.7$ & $48.6$\\
BI 173 & $-6.71$ & $^{0.18}_{0.13}$  & $2725$ & $^{475}_{250}$  & $0.03$ & $^{0.28}_{0.03}$  & $3.50$ & $^{0.05}_{0.50}$  & $(11$ & $^{36}_{11})$  & $(0.08$ & $^{0.06}_{0.05})$  & $(0.22$ & $^{0.36}_{0.22})$  & $(0.28$ & $^{0.03}_{0.07})$  & $0.30$ & $^{0.04}_{0.02}$  & $49.1$ & $47.8$\\
Sk $-68^\circ$155 & $-5.85$ & $^{0.13}_{0.10}$  & $1600$ & $^{50}_{50}$  & $0.08$ & $^{0.01}_{0.04}$  & $1.65$ & $^{0.35}_{0.15}$  & $17$ & $^{10}_{7}$  & $0.22$ & $^{0.31}_{0.15}$  & $0.90$ & $^{0.10}_{0.10}$  & $0.26$ & $^{0.05}_{0.05}$  & $0.36$ & $^{0.03}_{0.02}$  & $49.3$ & $47.6$\\
\hline 
 \multicolumn{22}{p{19.5cm}}{\small \textbf{Notes.} All values in this table are results from the combined optical and UV run. The electron scattering Eddington parameter is defined as  $\Gamma_{\mathrm{e}} = L \kappa_{e} / 4\pi cGM$, where we adopt $\kappa_e = 0.34$ \citep{2015AA...580A..20S} and $M=M_\mathrm{evol}$; for LH~114-7 $M_\mathrm{evol}$ could not be derived and we use $M=M_\mathrm{spec}$. The H and \hei ionising fluxes $Q_0$, $Q_1$ are defined as $Q_x = q_x 4 \pi R_*^2$, with $q_x$ the ionising radiation (number of photons) per unit surface area per second and $x \in \{0,1\}$. The values that appear within brackets have $2\sigma$ uncertainties that span the full parameter space; they are not considered for the wind structure parameter analysis (\cref{section:wind_structure}).} 
    \end{tabular}

\end{sidewaystable*}

\renewcommand{\arraystretch}{1.2}

\clearpage 

\section{Radial velocities and multiplicity \label{app:RVs}}

We measure the radial velocities with a twofold purpose. First, to shift the wavelengths to rest wavelength in order to compare the data with stellar atmosphere models, and second, to reveal the presence of possible companions. For each source we have two to six epochs.  

The radial velocities of the optical spectra are measured by fitting Gaussian functions to the line centres of the following lines: 
He\textsc{~ii}~$\lambda$4200, He\textsc{~i}~$\lambda$4471, He\textsc{~ii}~$\lambda$4541, He\textsc{~i}~$\lambda$4713, He\textsc{~i}~$\lambda$4921, He\textsc{~ii}~$\lambda$5412, He\textsc{~i}~$\lambda$5875,  N\textsc{~iv}~$\lambda$6380, O\textsc{~iii}~$\lambda$5592, He\textsc{~ii}~$\lambda$6406, He\textsc{~i}~$\lambda$6678, and He\textsc{~ii}~$\lambda$6683. For each source we inspect all Gaussian fits by eye, remove lines that have a bad fit, or where two diagnostics are blended\footnote{For stars in the intermediate temperature regime, the  He\textsc{~i}~$\lambda$6678 and He\textsc{~ii}~$\lambda$6683 lines are blended and thus removed from the selection.}, and then adopt the mean velocity for correcting the full optical spectrum. We adopted the standard deviation on this mean as our 1$\sigma$ uncertainty. We note that the barycentric velocity correction applied to the spectra released in eDR1 had the wrong sign; we corrected for this before carrying out this radial-velocity analysis. 
For the radial-velocity determination we used the co-added 1D spectra; we checked that this does not pose a problem for the two sources with multiple epochs; indeed, we find no radial velocity differences between the epochs. 
The radial velocities of the UV spectra were obtained in the normalisation process (see \cref{sec:UVnorm}). Uncertainties were determined using the $\chi^2$-value of the fit with the iron pseudo-continuum, following the method described in Sec. 3.4 of \citet{brands2022}. We obtained radial velocities of all settings (gratings) available, also if these gratings were not used for the spectroscopic analysis. 

The radial-velocity measurements for each source are presented in \cref{tab:rv_measurements}. In order to identify binaries we use the statistical test described in \citet{2013AA...550A.107S}. We compare any two radial-velocity measurements of a given source and assess the statistical significance of rejection of the null hypothesis of constant radial velocity as following: 
\begin{equation}
    \sigma_\textrm{RV} = \textrm{max}\left(\frac{|\varv_i - \varv_j|}{\sqrt{\sigma_i^2 + \sigma_j^2}} \right),
\end{equation}
where $\varv_i$ and $\varv_j$ are the measured radial velocities at epochs $i$ and $j$, with $\sigma_i$ and $\sigma_j$ the $1\sigma$ uncertainties on the velocity measurements. The cut-off value of $\sigma_\textrm{RV} > 4$ as  adopted by \citet{2013AA...550A.107S} results in only 1 out of 1000 false positives; a less strict limit of for example $\sigma_\textrm{RV} > 2$ would result in 1 out of 20 false positives. We note that the HLSP ULLYSES spectra are corrected for heliocentric velocity, whereas the optical XshootU are corrected for the barycentric velocity; the difference between the two is very small and this thus has no significant influence on our radial-velocity comparison. 

The $ \sigma_\textrm{RV}$ we find for each source is shown in the last column of \cref{tab:rv_measurements}. We detect radial-velocity variability at $\sigma_\textrm{RV} > 4$ for three sources: Sk~$-71^{\circ}$46, BI~173, and BI~272. Two other sources with signs of binarity in their spectra also have a relatively high $\sigma_\textrm{RV}$. The other three sources we marked as likely binaries do not show significant radial-velocity variations. Of the sources we marked as presumed single, three sources have $1 < \sigma_\textrm{RV} < 2$. We do not see obvious signs of binarity in their spectra.
We note that a radial-velocity variability analysis is not guaranteed to detect all binaries in the sample; this is true in general (binaries with small radial-velocity variations may go undetected), but even more so since the observation strategy was not optimised for binary detection. In particular, some epochs are taken very close in time (in particular the observations of the COS/G130M and COS/G160M gratings, which are usually taken right after each other), and some UV observations result in large uncertainties on the radial velocity (in particular observations obtained with FUSE and those with STIS/E230M).

\clearpage  

\begin{table*}
\caption{Radial velocities of stars in the sample, measured in all available gratings per star. \label{tab:rv_measurements}}
\begin{tabular}{l l l l l l l l l l l}
\hline \hline
Source & $\varv_{\rm FUSE}$ & $\varv_{\rm G130M}$ & $\varv_{\rm G160M}$ & $\varv_{\rm E140M}$ & $\varv_{\rm E140H}$ & $\varv_{\rm E230M}$ & $\varv_{\rm E230H}$ & $\varv_{\rm XSh}$ & $\langle \varv_{\rm rad}\rangle$ & $\sigma_{\rm RV}$  \\
& [km~s$^{-1}$] & [km~s$^{-1}$] & [km~s$^{-1}$] & [km~s$^{-1}$] & [km~s$^{-1}$] &[km~s$^{-1}$] & [km~s$^{-1}$] & [km~s$^{-1}$] &  [km~s$^{-1}$] & \\ \hline
\\[-10.0pt]\\[-10.0pt]\\[-10.0pt]
\multicolumn{2}{p{0.2\textwidth}}{\textit{Presumed single}} \\[-10.0pt] \\ 
W61-28-23 & $325\pm^{15}_{10}$ & $370\pm^{10}_{10}$ & $370\pm^{10}_{0}$ & $\cdots$ & $\cdots$ & $\cdots$ &$\cdots$ &$350\pm^{17}_{17}$ & $353\pm18$ & 1.5 \\
Sk -67$^\circ$167 & $255\pm^{45}_{35}$ & $\cdots$ & $\cdots$ & $310\pm^{15}_{10}$ & $\cdots$ & $\cdots$ &$\cdots$ &$312\pm^{16}_{16}$ & $292\pm26$ & 0.7 \\
ST 92-4-18 & $\cdots$ & $190\pm^{10}_{20}$ & $190\pm^{20}_{20}$ & $\cdots$ & $\cdots$ & $\cdots$ &$\cdots$ &$196\pm^{19}_{19}$ & $192\pm2$ & 0.2 \\
Sk -67$^\circ$69 & $\cdots$ & $\cdots$ & $\cdots$ & $315\pm^{5}_{5}$ & $\cdots$ & $\cdots$ &$\cdots$ &$307\pm^{14}_{14}$ & $311\pm3$ & 0.4 \\
Farina-88 & $\cdots$ & $250\pm^{5}_{5}$ & $260\pm^{5}_{10}$ & $\cdots$ & $\cdots$ & $\cdots$ &$\cdots$ &$253\pm^{9}_{9}$ & $254\pm4$ & 0.5 \\
LMCE 078-1 & $\cdots$ & $320\pm^{20}_{10}$ & $320\pm^{20}_{10}$ & $\cdots$ & $\cdots$ & $\cdots$ &$\cdots$ &$340\pm^{15}_{15}$ & $326\pm9$ & 0.7 \\
Sk -67$^\circ$111 & $225\pm^{45}_{40}$ & $\cdots$ & $\cdots$ & $250\pm^{20}_{15}$ & $\cdots$ & $\cdots$ &$\cdots$ &$246\pm^{16}_{16}$ & $240\pm11$ & 0.3 \\
N11-018 & $345\pm^{60}_{45}$ & $300\pm^{5}_{10}$ & $305\pm^{5}_{5}$ & $\cdots$ & $\cdots$ & $\cdots$ &$\cdots$ &$303\pm^{10}_{10}$ & $313\pm18$ & 0.4 \\
Sk -69$^\circ$50 & $200\pm^{100}_{50}$ & $\cdots$ & $\cdots$ & $235\pm^{20}_{20}$ & $\cdots$ & $\cdots$ &$\cdots$ &$214\pm^{16}_{16}$ & $216\pm14$ & 0.5 \\
Sk -71$^\circ$50 & $450\pm^{0}_{205}$ & $\cdots$ & $\cdots$ & $245\pm^{35}_{30}$ & $\cdots$ & $230\pm^{70}_{75}$ & $\cdots$ &$239\pm^{21}_{21}$ & $291\pm91$ & 0.7 \\
Sk -69$^\circ$104 & $280\pm^{20}_{45}$ & $\cdots$ & $\cdots$ & $285\pm^{5}_{5}$ & $\cdots$ & $275\pm^{15}_{10}$ & $\cdots$ &$261\pm^{19}_{19}$ & $275\pm8$ & 1.7 \\
Sk -68$^\circ$16 & $\cdots$ & $\cdots$ & $\cdots$ & $285\pm^{5}_{5}$ & $\cdots$ & $\cdots$ &$\cdots$ &$276\pm^{14}_{14}$ & $280\pm4$ & 0.7 \\
LH 9-34 & $\cdots$ & $\cdots$ & $\cdots$ & $295\pm^{10}_{15}$ & $\cdots$ & $\cdots$ &$\cdots$ &$287\pm^{33}_{33}$ & $291\pm3$ & 0.2 \\
Sk -66$^\circ$171 & $410\pm^{40}_{70}$ & $\cdots$ & $\cdots$ & $410\pm^{5}_{10}$ & $\cdots$ & $405\pm^{20}_{25}$ & $\cdots$ &$407\pm^{16}_{16}$ & $408\pm2$ & 0.1 \\
Sk -71$^\circ$41 & $305\pm^{40}_{35}$ & $\cdots$ & $\cdots$ & $220\pm^{15}_{10}$ & $\cdots$ & $\cdots$ &$\cdots$ &$256\pm^{5}_{5}$ & $260\pm34$ & 1.4 \\
Sk -67$^\circ$5 & $360\pm^{45}_{40}$ & $\cdots$ & $\cdots$ & $300\pm^{15}_{15}$ & $250\pm^{40}_{30}$ & $300\pm^{25}_{25}$ & $255\pm^{35}_{45}$ & $296\pm^{15}_{15}$ & $293\pm36$ & 1.0 \\
\\[-10.0pt]\\[-10.0pt]\\[-7.0pt]
\multicolumn{2}{p{0.2\textwidth}}{\textit{Signs of binarity}} \\[-10.0pt] \\
LH 114-7 & $\cdots$ & $\cdots$ & $\cdots$ & $320\pm^{15}_{15}$ & $\cdots$ & $\cdots$ &$\cdots$ &$298\pm^{22}_{22}$ & $309\pm10$ & 0.7 \\
VFTS-267 & $\cdots$ & $265\pm^{15}_{25}$ & $265\pm^{15}_{20}$ & $\cdots$ & $\cdots$ & $\cdots$ &$\cdots$ &$259\pm^{23}_{23}$ & $263\pm2$ & 0.2 \\
Sk -71$^\circ$46 & $\cdots$ & $310\pm^{15}_{20}$ & $335\pm^{10}_{5}$ & $\cdots$ & $\cdots$ & $\cdots$ &$\cdots$ &$191\pm^{31}_{31}$ & $278\pm62$ & 5.7 \\
Sk -67$^\circ$108 & $350\pm^{45}_{40}$ & $\cdots$ & $\cdots$ & $345\pm^{10}_{10}$ & $\cdots$ & $\cdots$ &$\cdots$ &$293\pm^{19}_{19}$ & $329\pm25$ & 1.9 \\
Sk -71$^\circ$19 & $190\pm^{260}_{40}$ & $270\pm^{85}_{75}$ & $290\pm^{40}_{45}$ & $\cdots$ & $\cdots$ & $\cdots$ &$\cdots$ &$252\pm^{34}_{34}$ & $250\pm37$ & 0.4 \\
Sk -70$^\circ$115 & $260\pm^{40}_{50}$ & $\cdots$ & $\cdots$ & $245\pm^{5}_{5}$ & $235\pm^{25}_{25}$ & $110\pm^{30}_{20}$ & $170\pm^{75}_{20}$ & $247\pm^{20}_{20}$ & $211\pm53$ & 2.3 \\
BI 173 & $285\pm^{35}_{60}$ & $\cdots$ & $\cdots$ & $315\pm^{5}_{10}$ & $\cdots$ & $300\pm^{35}_{25}$ & $\cdots$ &$224\pm^{17}_{17}$ & $281\pm34$ & 4.1 \\
Sk -68$^\circ$155 & $370\pm^{80}_{220}$ & $220\pm^{30}_{35}$ & $230\pm^{10}_{10}$ & $\cdots$ & $\cdots$ & $230\pm^{35}_{40}$ & $\cdots$ &$227\pm^{23}_{23}$ & $255\pm57$ & 0.7 \\
BI 272 & $\cdots$ & $\cdots$ & $\cdots$ & $380\pm^{10}_{15}$ & $\cdots$ & $\cdots$ &$\cdots$ &$251\pm^{17}_{17}$ & $315\pm64$ & 4.1 \\
\hline
\end{tabular}
\end{table*}

\FloatBarrier

\newpage 
\clearpage 

\section{Wind-embedded shocks and estimating $f_X$\label{app:sec:X-rays}}

For the wind-embedded shocks and resulting X-rays we use the parameterisation as implemented in \fastwind by \citet{2016A&A...590A..88C}. In this parameterisation, the X-ray properties of the wind are described by five different parameters, where the X-ray volume filling fraction $f_X$, and the maximum jump velocity of the shocks $u_\infty$, have the most profound influence on the X-ray output; we discuss the values we adopt for these two parameters below. For the other parameters we adopt canonical values for all stars: $\gamma_X = 0.75$, $m_X = 30$, and $R_\mathrm{min}^\mathrm{input} = 1.45$ (following \citealt{brands2022}). 

For the maximum jump velocity we adopt $u_\infty = 0.3\varv_\infty$ (also following \citealt{brands2022}), where we estimate $\varv_\infty$ prior to the \kiwiGA run by reading off the velocity at the  blue edge of \CIVline, or in a few cases where this line is weak, from the blue edge of the Ly-$\alpha$ corrected \nvuvline. 
We want to choose the value of $f_X$ such that the typical output X-ray luminosity matches observational constraints; typically, a value of $L_X/L \sim 10^{-7}$ is observed \citep{1997A&A...322..167B,2022MNRAS.515.4130C}. In previous work where X-rays were included in \kiwiGA fits, $f_X$ was estimated based on an observed relation with $\dot{M}/\varv_\infty$ \citep{1996rftu.proc....9K}. With this relation $f_X$ could be predicted with an accuracy of about 1.0~dex \citep[see][their Appendix G.2.]{brands2022}. 

\begin{figure}[h]
\centering 
    \includegraphics[width=0.46\textwidth]{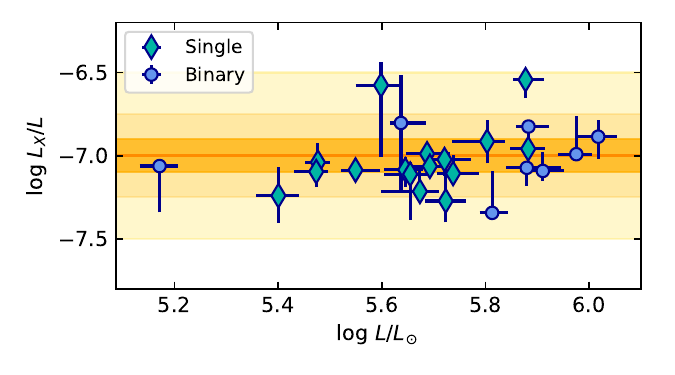}
     \caption{Adopted X-ray output luminosities as a function of total luminosity (diamonds and circles indicate single stars and binaries, respectively). Shaded areas indicate deviations of 0.1, 0.25, and 0.5 dex from $\log (L_X/L) = -7$. The adopted X-ray luminosity lies within 0.25~dex of the desired $\log (L_X/L) = -7$ for 80\% of the sample stars, and within 0.5~dex for all stars. \label{app:fig:xraycheck}}
\end{figure}

In this work we find an improved relation between wind parameters, $f_X$ and the X-ray luminosity. We base this relation on 1920 \fastwind models that are randomly sampled in the O-star parameter space -- including randomly sampled values of $f_X$  
\footnote{Parameters that were varied: $f_X$, \teff, \logg, \mdot, \yhe, \vinf, \vmicro, $\beta$, \fcl, \fic, \fvel, \vclonset, \vwindturb, and C, N, and O abundance; here $f_X$ was sampled in log space, with $\log f_X \ \epsilon \ [-4, 1.2]$, \yhe was sampled in the range $y_\mathrm{He} \ \epsilon \ [0.09, 0.40]$, and the other parameters could assume values in the typical parameter range used for the fitting; see the fitness plots in \cref{fig:examplefit:farina88} and on \href{https://zenodo.org/records/15013513}{Zenodo}.}. 
For these models both $f_X$ as well as the output X-ray luminosity is known. In order to obtain a relation, we define a measure for the wind density,  $\rho_\mathrm{wind}$: 
\begin{equation}\label{eq:winddensity}
    \rho_\mathrm{wind} = \dot{M}/(4 \pi R_*^2 \varv_\infty).
\end{equation}
Then we compare $f_X$ and $\rho_\mathrm{wind}$ of each model with the output X-ray luminosity, in other words, we fit a quadratic function of $\rho_\mathrm{wind}$  and $\log (L_X/L)$ through all 1920 points, to obtain a relation that expresses $f_X$ in terms of the wind density and the desired X-ray output luminosity: 
\begin{equation}
\begin{split}\label{eq:xrayfull}
     \log f_X & = 3.9785 + 0.4225 \log (L_X/L) - 0.0321 \log^2 (L_X/L)   +  1.2442 \log \rho_\mathrm{wind} + 0.0851 \log ^2\rho_\mathrm{wind},
\end{split}
\end{equation}
with $\rho_\mathrm{wind}$ expressed in cgs-units. As we desire all our models to have an output X-ray luminosity of $\log (L_X/L) = -7$, we substitute this into \cref{eq:xrayfull}, and adopt a value of $f_X$ using the following relation:
\begin{equation}\label{eq:fx_final_eq}
     \log f_X = -0.5541 + 1.2442 \log \rho_\mathrm{wind} + 0.0851 \log ^2\rho_\mathrm{wind}.
\end{equation}
This relation is implemented in \kiwiGA and used for all our fits. Using this relation, we obtain $\log (L_X/L) = -7$ within 0.1~dex for 50\% of our sample stars, within 0.25~dex for 80\% of the stars, and never are we more off than 0.5~dex (see \cref{app:fig:xraycheck}).  We emphasise that in the fitting process $f_X$ is not fitted, but  given by  \cref{eq:winddensity,eq:fx_final_eq}.

\clearpage 

\section{Spectral type versus \teff and \logg \label{app:spectraltype}}

\begin{figure}[h]
    \centering
    \includegraphics[width=0.85\textwidth]{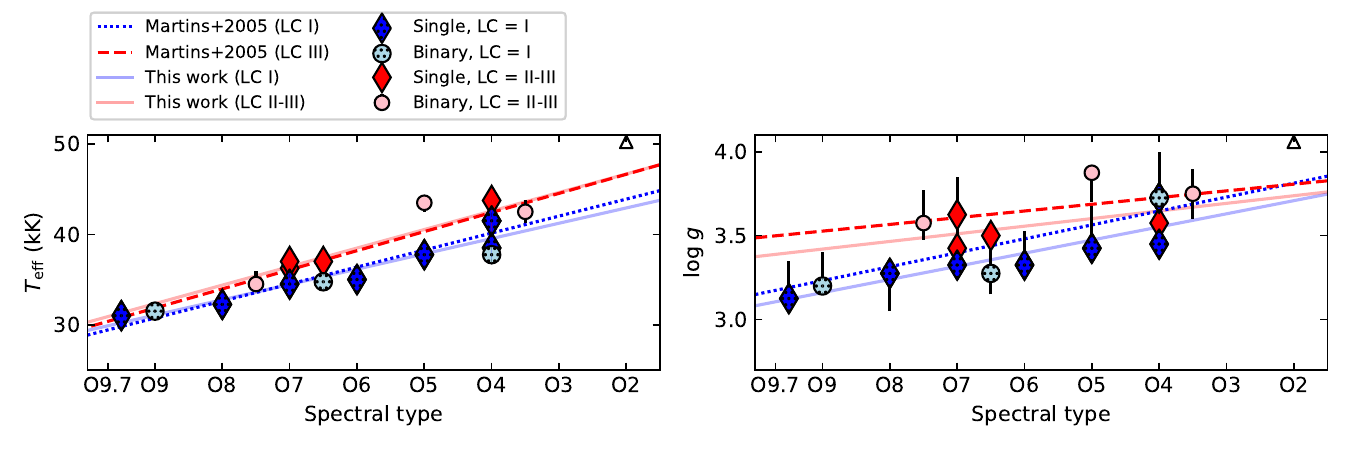}
    \caption{The relation between spectral sub-type and \teff and $\log g$ for O-type stars of luminosity class I (blue shaded markers), and II and III (red markers). \label{fig:spectype_calibration}}
\end{figure}

\Cref{fig:spectype_calibration} shows \teff and \logg as a function of spectral sub-type.  For \teff the fit to our single stars (diamonds) is in excellent agreement with the observed Galactic relations from \citet{2005AA...436.1049M}. For \logg, we find a small offset $<0.1$~dex; smaller than the typical scatter around the relation.

\FloatBarrier

\section{Comparison to literature values\label{app:literature}}

\begin{figure*}[h!]
    \centering
    \includegraphics[width=\textwidth]{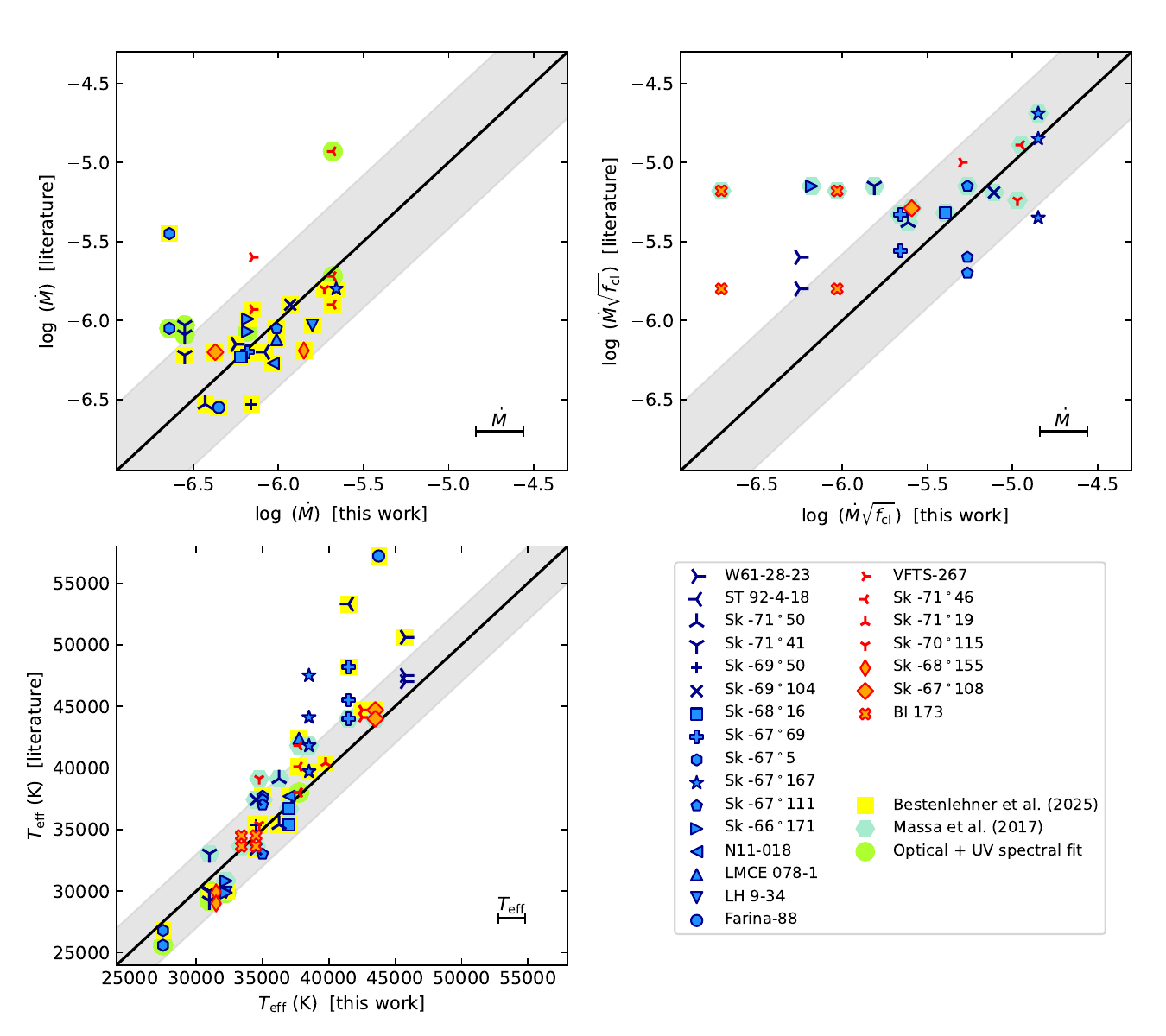}
    \caption{Values of \mdot (top left), $\log (\dot{M}\sqrt{f_\mathrm{cl}})$ (top right) and \teff (lower left) derived in this work compared to values found in literature. Studies that include clumping (either by fitting a clumping factor or assuming it) are shown in the \mdot-plot; studies that do not consider clumping are shown in the $\log (\dot{M}\sqrt{f_\mathrm{cl}})$-plot. The \teff-plot contains all values. Blue symbols correspond to the presumed single stars, orange-red symbols to the stars with signs of binarity; various markers indicate different sources. In the corner of each panel the typical error bar of values derived in this work is shown. The majority of the literature values come from \citet{2017MNRAS.470.3765M} and \citet{Bestenlehnerinprep}; these are marked in the plot (light blue hexagons and yellow squares, respectively). References to other studies included in this plot can be found in \Cref{table:literature,table:literature2}. 
    Values derived using quantitative optical and UV spectroscopy are marked with green circles. The black diagonal line indicates where the literature values equal ours, and the grey area corresponds to the systematic uncertainty that can be expected when analysing O-star spectra with different codes and methods \citep[see][]{Sanderinprep}. 
    }
    \label{fig:literature}
\end{figure*}  

\Cref{table:literature,table:literature2} contain an overview of literature values of \teff and \mdot for our sample stars. The tables also include values for \fcl in case a value was either adopted or fitted, $\dot{M}\sqrt{f_\textrm{cl}}$ (assuming $f_{\rm cl} = 1$ in case clumping was not considered), and a brief description of the methodology of each analysis. 
We only included analyses in which both mass-loss rate and effective temperature were derived (for LH~114-7 there were no such studies). 
Few stars from our sample are analysed before in as much detail as we do in this paper. 
Most studies use spectral type to determine \teff, several quantitative spectroscopy (but not always including UV); for the determination of \mdot most studies use spectroscopy, except for one, which derives its \mdot from the observed mid-IR excess (i.e., photometry; \citealt{2017ApJ...837..122M}).
Quantitative combined optical and UV spectroscopy was only carried out for W61-28-23 \citep{2005ApJ...627..477M}, Sk~$-67^\circ5$ \citep{Alkousainprep}, Sk~$-71^\circ$46 \citep{2018AA...609A...7R},  Sk$~-71^\circ41$ \citep{2018AA...615A..40R},  Sk$~-66^\circ171$ (\citealt{Sanderinprep}, \citealt{Alkousainprep}) and Sk~$-69^\circ50$ \citep{Sanderinprep}\footnote{Sk~$-66^\circ171$ and Sk~$-69^\circ50$ are special cases, which \citet{Sanderinprep} analysed with different stellar atmosphere codes and fitting strategies with the aim to investigate the parameter spread due to different fitting approaches. As \citet{Sanderinprep} provide an in depth discussion of these comparisons, and because the values of this paper are also used in their study, we do not include the values of the \citet{Sanderinprep}-analyses in our literature study. }. 

In \cref{fig:literature} a comparison of \teff, \mdot and $\dot{M}\sqrt{f_\textrm{cl}}$ is shown. For stars where clumping was  considered (i.e., a clumping factor was either adopted or fitted) we compare \mdot directly, for the other stars we compare $\dot{M}\sqrt{f_\textrm{cl}}$. Considering typical uncertainties on the derived values (indicated by the horizontal bar in the lower right corner of the $\dot{M}$-subplot of \cref{fig:literature}) and the systematic uncertainty that can be expected when analysing O-star spectra with different codes
and methods  (indicated by the shaded areas in \cref{fig:literature}; see \citealt{Sanderinprep}) the values of \mdot agree rather well. In particular the values derived by \citet{Bestenlehnerinprep} agree well with ours, while their \mdot analysis does not rely on UV line profile fitting (only for determination of \vinf) and their analysis is largely automated, allowing them to analyse the full ULLYSES sample. Two outliers in the \mdot-plot are Sk~$-67^\circ5$ and Sk~$-71^\circ46$, for which \citep{Bestenlehnerinprep} and \citep{2018AA...609A...7R}, respectively, find way higher mass-loss rates than we do. For Sk~$-67^\circ5$ the emission strength of \halpha is overestimated in the best fitting model of \citep{Bestenlehnerinprep}, though \heiiline is too deep in absorption. In our analysis the emission strength of \halpha matches better, as do the UV lines in our case, but the match with \heiiline is poor too. Considering the position of Sk~$-67^\circ5$ in the $\log$~\mdot-\logll plot (\cref{fig:mdot_vs_theory}) and the fit to \halpha, it seems that our value for \mdot is more likely for this star. 
For Sk~$-71^\circ$46 \citet{2018AA...609A...7R} find a mass-loss rate way higher than ours when adopting $f_\mathrm{cl} = 10$, though they also derive a second value assuming $f_\mathrm{cl} = 20$: in that case the mass-loss rates we derive are in agreement (we find $f_\mathrm{cl} = 29\pm^{16}_{7}$ for this source). Two more stars lie (just) outside the uncertainty range of \citet{Sanderinprep}. For Sk~$-71^\circ$41 \citet{2018AA...615A..40R} find a higher rate: looking at the fits we notice that their their model predicts a too strong \siivline, and has a poor fit to \halpha. However, we also notice that ours and their \halpha data are very different: in our data \halpha is in emission, in theirs strongly in absorption. Sk~$-67^\circ$5 also falls just outside of the uncertainty range of \citet{Sanderinprep} and was analysed by \citet{Alkousainprep}. Their fit is rather good, as is ours; possibly the differences arise in differing assumptions regarding the wind structure. 
For the analyses that do not consider clumping we compare $\dot{M}\sqrt{f_\textrm{cl}}$ and find good agreement for $\dot{M}\sqrt{f_\textrm{cl}} \gtrsim-5.7$, while for $\dot{M}\sqrt{f_\textrm{cl}} \lesssim-5.7$ the literature values are typically higher. With one exception, these studies do not employ UV wind lines and therefore may only have been able to derive an upper limit on the mass-loss rate. 

The last panel in \cref{fig:literature} shows a comparison of the effective temperatures. The agreement is rather good, especially for temperatures below 40000~K. At higher temperatures, the points of \citet{Bestenlehnerinprep} stand out, being all significantly higher than the values we find, in the most extreme case the difference exceeds 13000~K. In concerns the sources W61-28-23, ST 92-4-18, Farina-88, and Sk~$-67^\circ69$. For the latter three, the models of \citet{Bestenlehnerinprep} predict \nvopt lines where these are not visible in the data, implying that these models overestimate the temperature. An explanation for this is that the normalisation around these lines is not always good as they lie close to \heiiline, causing the pipeline of \citet{Bestenlehnerinprep} to decrease the weight of the data at these wavelengths. This way, the \nvopt lines hardly contributes to the $\chi^2$ value and the temperature can be overestimated.  For W61-28-23 the \nvopt lines do fit well for \citet{Bestenlehnerinprep}, where in our fit they are slightly too weak, implying that the higher temperature of \citet{Bestenlehnerinprep} might be more probable for this star. We conclude this section by stressing that \citet{Bestenlehnerinprep} do report uncertainties on their derived parameters, and that the \teff uncertainties are, as expected, rather large when too few lines were available for \teff constraints. For example the lower $1\sigma$ uncertainty on the \teff of Farina-88 was 11700~K, meaning that within uncertainties our \teff and theirs are actually in agreement. 

\clearpage

\begin{table*}[]
    \small 
    \caption{An overview of literature values of \teff, \mdot, and $\dot{M}\sqrt{f_\textrm{cl}}$ for the presumed single stars of our sample (see also \cref{table:literature2}). }
    \label{table:literature}
    \begin{tabular}{l l p{1.0cm} p{1.4cm} l l}
\hline\hline
\\[-10.0pt]
Source & \teff & $\log \dot{M}$ & $\log (\dot{M}\sqrt{f_\mathrm{cl}})$ & Reference & Comments\\ 
 & [K] & [$\mathrm{M}_\odot$~yr$^{-1}$] & [$\mathrm{M}_\odot$~yr$^{-1}$] &  & \\ 
\hline
W61-28-23           & 45750 & $-6.25$       & $-6.25$  & This work & $f_\mathrm{cl}$ not constrained \\
                    & 50600 & $-6.15$       & $-5.65$  & \citet{Bestenlehnerinprep}  &  Optical spectral fit, $\varv_\infty$ from UV,  $f_\mathrm{cl} = 10$  \\ 
                    & 47500 & $-5.60$       & $-5.60$  & \citet{2005ApJ...627..477M} &  Optical spectral fit, $\varv_\infty$ from UV \\ 
                    & 47000 & $-5.80$       & $-5.80$  & \citet{2012AA...543A..95R}  &  Optical spectral fit  \\ 
 Sk -67$^\circ$167  & $38500 $ & $-5.66 $   & $-4.84$  & This work & Best fitting clumping factor:  $f_\mathrm{cl} = 42$ \\
                     & 39700 & $-5.80$       & $-5.30$  & \citet{Bestenlehnerinprep}  &  Optical spectral fit, $\varv_\infty$ from UV,  $f_\mathrm{cl} = 10$  \\ 
                   & 41809 & $-4.69$       & $-4.69$  & \citet{2017MNRAS.470.3765M}  & \teff from spectral type, $\dot{M}$ from mid-IR excess\\
                    & 47500 & $-4.85$       & $-4.85$  & \citet{1996AA...305..171P}  & Optical spectral fit  \\
                    & 44100 & $-5.35$       & $-5.35$  & \citet{1988ApJ...334..626L} & \teff from spectral type, $\dot{M}$ from H$\alpha$ luminosity \\
ST 92-4-18 & $41500 $ & $-6.08 $ & $-5.33 $  & This work & Best fitting clumping factor:  $f_\mathrm{cl} = 32$. \\ 
& 53300 & $-6.20$  & $-5.70$  & \citet{Bestenlehnerinprep}  &  Optical spectral fit, $\varv_\infty$ from UV,  $f_\mathrm{cl} = 10$  \\
Sk -67$^\circ$69    & $41500 $  & $-6.18 $  & $-5.65$  & This work & Best fitting clumping factor:  $f_\mathrm{cl} = 11$ \\
                    & 48200 & $-6.20$  & $-5.70$  & \citet{Bestenlehnerinprep}  &  Optical spectral fit, $\varv_\infty$ from UV,  $f_\mathrm{cl} = 10$  \\                    & $ 43985$ & $-5.33 $   & $-5.33$  & \citet{2017MNRAS.470.3765M}  & \teff from spectral type, $\dot{M}$ from mid-IR excess \\
                    & $ 45500$ & $-5.56$    & $-5.56$  & \citet{1988ApJ...334..626L} & \teff from spectral type, $\dot{M}$ from H$\alpha$ luminosity \\
Farina-88 & $43750 $ & $-6.35 $  & $-5.85$ & This work & Best fitting clumping factor:  $f_\mathrm{cl} = 10$. \\ 
& 57200 & $-6.55$  & $-6.05$  & \citet{Bestenlehnerinprep}  &  Optical spectral fit, $\varv_\infty$ from UV,  $f_\mathrm{cl} = 10$  \\ 
LMCE 078-1 & $ 37750$ & $-6.01 $& $-5.16 $  & This work & Best fitting clumping factor:  $f_\mathrm{cl} = 50$. \\  
& 42400 & $-6.12$  & $-5.62$  & \citet{Bestenlehnerinprep}  &  Optical spectral fit, $\varv_\infty$ from UV,  $f_\mathrm{cl} = 10$  \\
Sk -67$^\circ$111   & $ 35000$ & $-6.01 $   & $-5.26$  & This work & Best fitting clumping factor:  $f_\mathrm{cl} = 31$ \\ 
& 37700 & $-6.05$  & $-5.55$  & \citet{Bestenlehnerinprep}  &  Optical spectral fit, $\varv_\infty$ from UV,  $f_\mathrm{cl} = 10$  \\ 
                    & $ 37415$ & $-5.15 $   & $-5.15$  & \citet{2017MNRAS.470.3765M} & \teff from spectral type, $\dot{M}$ from mid-IR excess \\ 
                    & $ 33000$ & $-5.70 $   & $-5.70$  & \citet{2000ApJ...538L..43F} & UV spectral fit \\ 
                    & $ 37000$ & $-5.60$    & $-5.60$  & \citet{1988ApJ...334..626L} & \teff from spectral type, $\dot{M}$ from H$\alpha$ luminosity \\
N11-018 & $ 37000$ & $-6.03 $  & $-5.19$ & This work & Best fitting clumping factor:  $f_\mathrm{cl} = 48$. \\  
    & 37700 & $-6.27$  & $-5.77$  & \citet{Bestenlehnerinprep}  &  Optical spectral fit, $\varv_\infty$ from UV,  $f_\mathrm{cl} = 10$  \\ 
Sk -$69^\circ50$    & $ 34500$ & $-6.16 $   & $-5.39$  & This work & Best fitting clumping factor:  $f_\mathrm{cl} = 35$ \\
& 35400 & $-6.53$  & $-6.03$  & \citet{Bestenlehnerinprep}  &  Optical spectral fit, $\varv_\infty$ from UV,  $f_\mathrm{cl} = 10$  \\ 
Sk -$71^\circ50$    & $36250 $ & $-6.43 $   & $-5.61$  & This work & Best fitting clumping factor:  $f_\mathrm{cl} = 43$ \\ 
    & 35400 & $-6.53$  & $-6.03$  & \citet{Bestenlehnerinprep}  &  Optical spectral fit, $\varv_\infty$ from UV,  $f_\mathrm{cl} = 10$  \\ 
                    & $39121$ & $-5.38 $    & $-5.38$  & \citet{2017MNRAS.470.3765M} & \teff from spectral type, $\dot{M}$ from mid-IR excess \\ 
Sk -69$^\circ$104   & $ 34500$ & $-5.93$   & $-5.11$  & This work & Best fitting clumping factor:  $f_\mathrm{cl} = 44$ \\ 
    & 33400 & $-5.90$  & $-5.40$  & \citet{Bestenlehnerinprep}  &  Optical spectral fit, $\varv_\infty$ from UV,  $f_\mathrm{cl} = 10$  \\ 
                    & $ 37415$ & $-5.19$    & $-5.19$  & \citet{2017MNRAS.470.3765M} & \teff from spectral type, $\dot{M}$ from mid-IR excess \\ 
Sk -$68^\circ16$    & $37000 $ & $-6.22$    & $-5.39$  & This work & Best fitting clumping factor:  $f_\mathrm{cl} = 45$ \\ 
    &  35400 & $-6.23$  & $-5.73$  & \citet{Bestenlehnerinprep}  &  Optical spectral fit, $\varv_\infty$ from UV,  $f_\mathrm{cl} = 10$  \\ 
                    & $36689$ & $-5.32$     & $-5.32$  & \citet{2017MNRAS.470.3765M} & \teff from spectral type, $\dot{M}$ from mid-IR excess \\ 
LH 9-34 & $32250 $ & $-5.80$  & $-5.41$ & This work & Best fitting clumping factor:  $f_\mathrm{cl} = 6$. \\ 
    &  29900 & $-6.03$  & $-5.53$  & \citet{Bestenlehnerinprep}  &  Optical spectral fit, $\varv_\infty$ from UV,  $f_\mathrm{cl} = 10$  \\ 

Sk -$66^\circ171$   & $32250 $ & $-6.18 $   & $-6.18$  & This work & Clumping parameters not constrained \\
    &  29900 & $-5.99$  & $-5.49$  & \citet{Bestenlehnerinprep}  &  Optical spectral fit, $\varv_\infty$ from UV,  $f_\mathrm{cl} = 10$  \\ 
                    & $30824$ & $-5.15$     & $-5.15$  & \citet{2017MNRAS.470.3765M} & \teff from spectral type, $\dot{M}$ from mid-IR excess \\ 
                    & $29854$ & $-6.07$     & $-5.33$  & \citet{Alkousainprep} & Optical + UV spectral fit,  $f_\mathrm{cl} = 30$ \\
Sk -$71^\circ41$    & $ 31000$ & $-6.55 $   & $-5.81$  & This work & Best fitting clumping factor:  $f_\mathrm{cl} = 30$ \\ 
    & 29900  & $-6.22$  & $-5.72$  & \citet{Bestenlehnerinprep}  &  Optical spectral fit, $\varv_\infty$ from UV,  $f_\mathrm{cl} = 10$  \\ 
                    & $33021$ & $-5.15$     & $-5.15$  & \citet{2017MNRAS.470.3765M} & \teff from spectral type, $\dot{M}$ from mid-IR excess \\ 
                    & $30000$ & $-6.09$     & $-5.59$  & \citet{2018AA...615A..40R} & Optical + UV spectral fit,  $f_\mathrm{cl} = 10$ \\ 
                    & $29200$ & $-6.03$     & $-5.53$  & \citet{Alkousainprep} & Optical + UV spectral fit,  $f_\mathrm{cl} = 10$ \\ 
Sk -$67^\circ5$ & $27500 $ & $-6.64 $  & $-5.87$ & This work & Best fitting clumping factor:  $f_\mathrm{cl} = 34$ \\  
    & 26800  & $-5.95$  & $-5.45$  & \citet{Bestenlehnerinprep}  &  Optical spectral fit, $\varv_\infty$ from UV,  $f_\mathrm{cl} = 10$  \\ 
                & $25600$ & $-6.05 $  & $-5.55$ & \citet{Alkousainprep} & Optical + UV spectral fit,  $f_\mathrm{cl} = 10$  \\  
\hline 
\multicolumn{6}{p{16.4cm}}{\textbf{Notes.} In this overview we include only analyses in which both mass-loss rate and effective temperature were derived. For computing \mdot$\sqrt{f_\textrm{cl}}$, we assume $f_\textrm{cl} = 1$ unless another clumping factor was adopted or derived from the fitting process.  } \\
    \end{tabular}
\end{table*}

\begin{table*}[h!]
    \small 
    \caption{As \cref{table:literature} but for the stars with signs of binarity. }
    \label{table:literature2}
    \begin{tabular}{l l p{1.0cm} p{1.4cm} l l}
\hline\hline
\\[-10.0pt]
Source & \teff & $\log \dot{M}$ & $\log (\dot{M}\sqrt{f_\mathrm{cl}})$ & Reference & Comments\\ 
 & [K] & [$\mathrm{M}_\odot$~yr$^{-1}$] & [$\mathrm{M}_\odot$~yr$^{-1}$] &  & \\ 
\hline     
VFTS-267            & $42500$   & $-6.15$   & $-5.30$  & This work & Best fitting clumping factor:  $f_\mathrm{cl} = 51$ \\ 
    & 44700  & $-5.93$  & $-5.43$  & \citet{Bestenlehnerinprep}  &  Optical spectral fit, $\varv_\infty$ from UV,  $f_\mathrm{cl} = 10$  \\ 
                    & $44100$   & $-5.00$   & $-5.00$  & \citet{2017AA...600A..81R} & Optical spectral fit \\ 
                    & $44700$   & $-5.60$   & $-5.10$  & \citet{2014AA...570A..38B} & Optical spectral fit,  $f_\mathrm{cl} = 10$  \\ 
Sk -71$^\circ$46    & $37750$   & $-5.68$   & $-4.95$  & This work & Best fitting clumping factor:  $f_\mathrm{cl} = 29$ \\ 
    & 40100  & $-5.90$  & $-5.40$  & \citet{Bestenlehnerinprep}  &  Optical spectral fit, $\varv_\infty$ from UV,  $f_\mathrm{cl} = 10$  \\ 
                    & $38000$   & $-4.93$   & $-4.43$  & \citet{2018AA...609A...7R} & Optical + UV spectral fit, $f_\mathrm{cl} = 10$ \\ 
                    & $38000$   & $-5.72$   & $-5.06$  & \citet{2018AA...609A...7R} & Optical + UV spectral fit, $f_\mathrm{cl} = 20$ \\ 
                    & $41809$   & $-4.89$   & $-4.89$  &  \citet{2017MNRAS.470.3765M} & \teff from spectral type, $\dot{M}$ from mid-IR excess \\ 
Sk -$67^\circ108$   & $43500$   & $-6.37$   & $-5.59$  & This work & Best fitting clumping factor:  $f_\mathrm{cl} = 36$ \\ 
    & 44700  & $-6.20$  & $-5.70$  & \citet{Bestenlehnerinprep}  &  Optical spectral fit, $\varv_\infty$ from UV,  $f_\mathrm{cl} = 10$  \\ 
                    & $43985$   & $-5.29$   & $-5.29$  & \citet{2017MNRAS.470.3765M} & \teff from spectral type, $\dot{M}$ from mid-IR excess \\ 
  Sk -$71^\circ19$  & $39750$   & $-7.20$   & $-6.39$  & This work & Best fitting clumping factor:  $f_\mathrm{cl} = 40$ \\ 
    & 40400  & $-7.15$  & $-6.65$  & \citet{Bestenlehnerinprep}  &  Optical spectral fit, $\varv_\infty$ from UV,  $f_\mathrm{cl} = 10$  \\ 
  
  Sk -$70^\circ115$ & $34750$   & $-5.73$   & $-4.97$  & This work & Best fitting clumping factor:  $f_\mathrm{cl} = 33$ \\ 
    & 35400  & $-5.80$  & $-5.30$  & \citet{Bestenlehnerinprep}  &  Optical spectral fit, $\varv_\infty$ from UV,  $f_\mathrm{cl} = 10$  \\ 
                    & $39121$   & $-5.24$   & $-5.24$  & \citet{2017MNRAS.470.3765M} & \teff from spectral type, $\dot{M}$ from mid-IR excess \\ 
  BI 173            & $34500$   & $-6.71$   & $-6.71$  & This work & $f_\mathrm{cl}$ not constrained \\
    & 33400  & $-6.53$  & $-6.03$  & \citet{Bestenlehnerinprep}  &  Optical spectral fit, $\varv_\infty$ from UV,  $f_\mathrm{cl} = 10$  \\ 
                    & $33639 $  & $-5.18$   & $-5.18$  & \citet{2017MNRAS.470.3765M} & \teff from spectral type, $\dot{M}$ from mid-IR excess \\ 
                    & $34500 $  & $-5.80$   & $-5.80$  & \citet{2009ApJ...692..618M} &  Optical spectral fit,  \\ 
Sk -$68^\circ155$ & $31500$ & $-5.85$ & $-5.23$ & This work & Best fitting clumping factor:  $f_\mathrm{cl} = 17$ \\ 
    &  29900 & $-6.19$  & $-5.69$  & \citet{Bestenlehnerinprep}  &  Optical spectral fit, $\varv_\infty$ from UV,  $f_\mathrm{cl} = 10$  \\ 
                 & $29000$ & $-6.19$ & $-5.45$ & \citet{Alkousainprep} & Optical + UV spectral fit,  $f_\mathrm{cl} = 30$  \\  
\hline 
\multicolumn{6}{p{16.4cm}}{\textbf{Notes.} In this overview we include only analyses in which both mass-loss rate and effective temperature were derived. For LH 114-7 there were no such studies. For computing \mdot$\sqrt{f_\textrm{cl}}$, we assume $f_\textrm{cl} = 1$ unless another clumping factor was adopted or derived from the fitting process.  } \\

    \end{tabular}
\end{table*}

\FloatBarrier

\end{appendix}

\end{document}